\documentclass[11pt,a4paper,logo,copyright]{agentsociety}

\usepackage[utf8]{inputenc} 
\usepackage[T1]{fontenc}    
\usepackage{hyperref}       
\usepackage{url}            
\usepackage{booktabs}       
\usepackage{amsfonts}       
\usepackage{amsmath}
\usepackage{nicefrac}       
\usepackage{microtype}      
\usepackage{xcolor}         
\usepackage{graphicx}
\usepackage{subcaption}
\usepackage{multirow}
\usepackage{tabularx}
\usepackage{makecell}
\usepackage{wrapfig}

\usepackage[most]{tcolorbox}
\usepackage{xcolor}
\usepackage{enumitem}

\usepackage{array}
\usepackage{multirow}
\usepackage{booktabs}
\usepackage{rotating}
\usepackage{amssymb}
\usepackage{graphicx}

\newcommand{\sstar}{\raisebox{0.03ex}{\scalebox{0.72}{$\bigstar$}}}
\newcommand{\stars}[1]{%
  \makebox[4.8em][c]{%
    \ifcase#1
    \or \sstar
    \or \sstar\kern0.12em\sstar
    \or \sstar\kern0.12em\sstar\kern0.12em\sstar
    \or \sstar\kern0.12em\sstar\kern0.12em\sstar\kern0.12em\sstar
    \or \sstar\kern0.12em\sstar\kern0.12em\sstar\kern0.12em\sstar\kern0.12em\sstar
    \fi
  }%
}

\newcolumntype{M}[1]{>{\raggedright\arraybackslash}m{#1}}

\title{AgentSociety 2: An Integrated Research Environment for Executable Social Science}

\author[1\dagger]{Jinghua Piao}
\author[1\dagger]{Jun Zhang}
\author[1]{Haoyu Huang}
\author[1]{Keming Zhang}
\author[1]{Jing Yi Wang}
\author[1]{Xinran Zhao}
\author[1]{Songwei Li}
\author[1]{Boyuan Sun}
\author[1]{Jiayi Chang}
\author[1]{Fengli Xu}
\author[2]{Chunyan Wang}
\author[3]{Fang Zhang}
\author[4]{Ke Rong}
\author[3]{Jun Su}
\author[4,5]{Tianguang Meng}
\author[2]{Yi Liu}
\author[3,5]{Qingguo Meng}
\author[1,5]{Yu Wang}
\author[1,5\ddagger]{Yong Li}
\affiliation[1]{Department of Electronic Engineering, Tsinghua University}
\affiliation[2]{School of Environment, Tsinghua University}
\affiliation[3]{School of Public Policy and Management, Tsinghua University}
\affiliation[4]{School of Social Sciences, Tsinghua University}
\affiliation[5]{Laboratory of Computational Social Science and State Governance, Tsinghua University}
\contribution[\dagger]{Equal Contribution and Project Leads}
\contribution[\ddagger]{Corresponding Author.}

\abstract{AI scientist systems are beginning to automate parts of scientific research, but social science poses a distinct challenge: the objects of inquiry are not merely datasets or laboratory protocols, but integrated social processes involving situated participants, interaction contexts, interventions, and the outcomes they produce. Yet a critical link is missing: existing systems either assist isolated research tasks or simulate agents as experimental subjects, leaving the research workflow and the simulated society decoupled. Here we introduce \textit{AgentSociety$^2$}, an Integrated Research Environment (IRE) for executable social science. \textit{AgentSociety$^2$} couples two roles of LLM agents in the same research runtime: AI social scientists that coordinate literature grounding, hypothesis generation, experiment design, simulation execution, result interpretation, and manuscript drafting; and silicon participants that generate behavioral responses within configurable social environments. This dual-role design turns social-science hypotheses into auditable agent behaviors, environment rules, interventions, and measurements, thereby supporting an end-to-end social science research workflow. Across seven illustrative studies spanning micro-level social-science laboratory experiments, meso-level dynamics in social media, and macro-level urban scenarios, we demonstrate the platform's capacity to support diverse disciplinary research questions, reproduce major qualitative patterns from prior studies, identify informative deviations, and enable large-scale simulations through optimized agent-environment interactions. By preserving human researchers' high-level agency while delegating procedural orchestration to agentic systems, \textit{AgentSociety$^2$} provides a human-in-the-loop and controllable infrastructure for next-generation computational social science, with broader applications in scalable computational social experimentation and AI-enabled social governance platforms.}

\date{\monthyeardate\today}
\resource[\faEnvelope]{\email{liyong07@tsinghua.edu.cn}}
\resource[\faGithub]{\href{https://github.com/tsinghua-fib-lab/AgentSociety}{tsinghua-fib-lab/AgentSociety}}
\resourceline[\faGlobe]{\textbf{Online Platform:} \href{https://agentsociety2.fiblab.net/}{https://agentsociety2.fiblab.net/}}
\keywords{large language model agents, AI scientist, social simulation, computational social science}

\begin{document}

\maketitle

\newpage
\begingroup
\setlength{\baselineskip}{1.25\baselineskip}
\tableofcontents
\endgroup

\newpage

\section{Introduction}

\begin{figure}[b!]
    \centering
    \includegraphics[width=\linewidth]{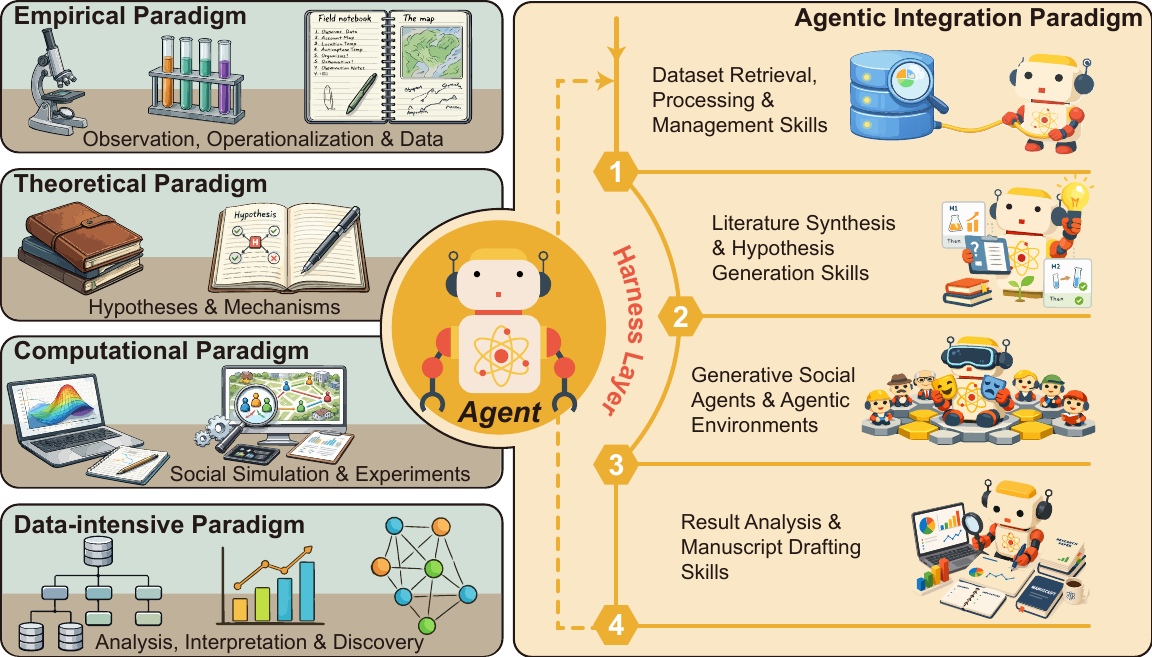}
    \caption{\textit{AgentSociety$^2$} as an agentic integration paradigm for computational social science. Empirical, theoretical, computational, and data-intensive paradigms have expanded how social scientists observe, explain, simulate, and infer from complex social systems. \textit{AgentSociety$^2$} integrates these modes of inquiry in a unified agentic framework by coordinating specialized teams of AI social scientists.}
    \label{fig:overview}
\end{figure}

Scientific inquiry has often evolved alongside technological progress; each major technological leap acts as a catalyst, reshaping not only the instruments of science but the very paradigms of knowledge production~\cite{hey2009fourth,kuhn1970structure}. From Galileo’s systematic observations and Newton’s mathematical laws to computational simulation and data-intensive discovery, technological advances have repeatedly expanded how scientists observe, theorize, compute, and infer~\cite{hacking1983representing,winsberg2019science,lazer2009computational,salganik2019bit}. These advances are especially salient in social science, where empirical observation and theoretical modeling provide the foundations for characterizing social phenomena and explaining their underlying mechanisms, while computational social science extends this foundation to large-scale social systems through data-intensive analysis and social simulation~\cite{lazer2009computational,macy2002factors,epstein2012generative}. Yet as these modes of inquiry converge in the same research workflow, social scientists face a growing integration burden: theory, data, simulation, and analysis remain distributed across separate tools, datasets, models, and experimental environments. This fragmentation makes it difficult to connect empirical evidence, theoretical mechanisms, executable experiments, and scientific interpretation in a single research process.


Rapid advances in large language models (LLMs) and autonomous agent systems are profoundly restructuring human workflow~\cite{brown2020language,achiam2023gpt,deng2023mind2web}. This restructuring is already visible in software engineering and agent-native platforms: natural-language-first ``vibe coding'' shifts human work from code construction to intent specification and iterative steering, while OpenClaw shows how collective interaction and content production can be reorganized around agents at scale~\cite{shi2022natural,chen2021evaluating,github2022quantifying,steinberger2026openclaw}. In scientific inquiry, the implications extend deeper, beginning to reshape not only how research tasks are executed, but how research workflows are organized. As AI scientists, agents have evolved from conversational assistants into tool-using experts and, increasingly, autonomous orchestrators capable of planning, acting, observing feedback, and revising actions over multiple steps~\cite{wei2022chain,wang-etal-2023-plan,yao2022react,deng2023mind2web,patilberkeley}. Building on this progression, recent AI scientist systems have begun to assist a broad range of research tasks, including literature synthesis, idea generation, experimental execution, result analysis, and scientific writing~\cite{boiko2023autonomous,swanson2025virtual,lu2026towards,gottweis2026accelerating,ghareeb2026multi}, suggesting that agents can undertake substantial portions of the cognitive and procedural labor traditionally coordinated by human researchers~\cite{lu2026towards,shao2025omniscientist,li2026autosota,feng2026internagent,aygun2026ai}.

However, social science poses a distinct challenge for existing AI scientist systems. Unlike the machine-learning, biological, and chemical domains that these systems have primarily targeted, social science concerns objects of inquiry that are not merely datasets, benchmarks, molecules, experimental protocols, or measurable outputs, but integrated social processes shaped by situated participants, interaction contexts, interventions, and the outcomes they produce. To make such social-science inquiry experimentally tractable, a complementary line of work uses generative social agents as silicon participants to pilot surveys, behavioral experiments, and policy interventions in controlled, repeatable environments~\cite{aher2023using,argyle2023out,horton2023large,dillion2023can,demszky2023using,park2023generative,piao2025agentsociety}. Yet these two lines of progress remain largely disconnected: AI scientist systems mainly assist isolated research tasks, while generative-agent simulations mainly provide experimental subjects and environments for piloting social experiments. As a result, the research workflow and the simulated society remain decoupled, with hypotheses, experimental designs, social environments, interventions, measurements, and interpretations rarely represented and executed in the same runtime. This gap motivates an \textbf{``agentic integration paradigm''} for computational social science, in which LLM agents assume dual roles as AI social scientists and simulated participants, coupling scientific reasoning with executable social simulation in a unified, human-in-the-loop research workflow~\cite{lu2026towards,shao2025omniscientist,li2026autosota}.


Inspired by this shift, we introduce \textit{AgentSociety$^2$}, an Integrated Research Environment (IRE) for executable computational social science. The central idea is to couple two roles of LLM agents in the same research runtime: AI social scientists that coordinate literature grounding, hypothesis generation, experiment design, simulation execution, result interpretation, and manuscript drafting; and silicon participants that simulate behavioral responses within configurable social environments. By connecting these two roles, \textit{AgentSociety$^2$} turns social-science hypotheses into auditable agent behaviors, environment rules, interventions, and measurements, thereby linking the research workflow with the simulated society it seeks to study. To operationalize this dual-role design, \textit{AgentSociety$^2$} introduces an agent harness layer that binds domain-specific skills, subagents, tool interfaces, and auxiliary scripts to an explicit staged workflow. This design organizes empirical, theoretical, computational, and data-intensive modes of social inquiry into a unified and auditable research process. Specifically, we first design dataset skills to collect, process, and manage empirical data  from real-world social systems, grounding research questions and computational social experiments in observed social structures and human behaviors. Second, we introduce literature synthesis and hypothesis formulation skills that draw on a curated interdisciplinary corpus to distill prior social-scientific knowledge into testable hypotheses and mechanism specifications for computational experiments. Third, we build social simulation and experiment skills by redesigning and extending the agent-environment architecture of \textit{AgentSociety-1}. These skills enable AI social scientist agents to translate empirical conditions, theoretical mechanisms, intervention designs, and measurement procedures into executable computational experiments over generative social agents and configurable environments. Finally, result analysis and writing skills turn simulation outputs into interpretable claims and manuscript drafts, closing the loop from empirical grounding and theoretical mechanism formulation to computational experimentation and social-scientific explanation. Taken together, the harness layer and these skills transform general-purpose AI assistants into an AI social scientist, coordinating empirical grounding, theoretical formulation, computational experimentation, and data-intensive synthesis in an integrated research workflow.

Importantly, this integration is not intended to make social science fully autonomous. Meaningful scientific insight remains grounded in human judgement: researchers define concepts, assess construct validity, specify interventions, interpret deviations, and decide which claims are warranted. The IRE therefore keeps human researchers in the loop while delegating procedural orchestration to AI social scientists. Rather than fully delegating scientific reasoning to conversational agents or requiring researchers to engineer complex simulation pipelines from scratch, \textit{AgentSociety$^2$} provides a structured and interactive workspace in which researchers can inspect intermediate artifacts, revise agent outputs, approve key workflow transitions, and translate early-stage scientific ideas into executable agentic experiments. In this workspace, researchers can implement diverse social-science methods, including surveys, interactive interviews, behavioral laboratory experiments, and field-like interventions in simulated societies spanning micro-level individual behavior, meso-level collective dynamics, and macro-level social systems. This breadth supports research across psychology, economics, political science, communication, environmental studies, urban science, and computational social science.

To demonstrate the generality and usability of \textit{AgentSociety$^2$}, we present seven illustrative practices together with platform-level performance evaluations. At the micro level, the platform supports individual psychological surveys and classic behavioral laboratory experiments; at the meso level, it simulates networked social dynamics such as algorithmic echo chambers, opinion polarization, and inequalities in information consumption; and at the macro level, it extends simulation to urban living and disaster-response behaviors in physical space. Across these studies, \textit{AgentSociety$^2$} supports diverse disciplinary research questions, reproduces major qualitative patterns from prior social-science studies, identifies informative deviations, and enables large-scale simulations through optimized agent-environment interactions. Specifically, the platform supports simulations involving up to 10,000 agents and 100 million interactions, while reproducing more than 95\% of qualitative patterns reported in prior studies. It also achieves strong quantitative alignment with empirical observations in macro-level scenarios, reproducing population mobility responses across two disaster scenarios and three crisis stages with low RMSE values ranging from 0.0073 to 0.0188. Importantly, the few deviations from prior findings are not merely reproduction failures, but informative inconsistencies that suggest new hypotheses about behavioral assumptions, mechanism specification, and contextual boundary conditions. To support such scalable experimentation, \textit{AgentSociety$^2$} further optimizes the agent-environment architecture, improving interaction reliability by up to 31.6\% over the best baseline. Pre-generation and caching handle more than 70\% of routine requests and reduce LLM calls by 66.5\%, substantially lowering token overhead. Overall, these results show that \textit{AgentSociety$^2$} provides a scalable, controllable, and empirically grounded IRE for advancing computational social science.


\section{AgentSociety$^2$: Design Principles}



\begin{figure}[t]
    \centering
    \includegraphics[width=\linewidth]{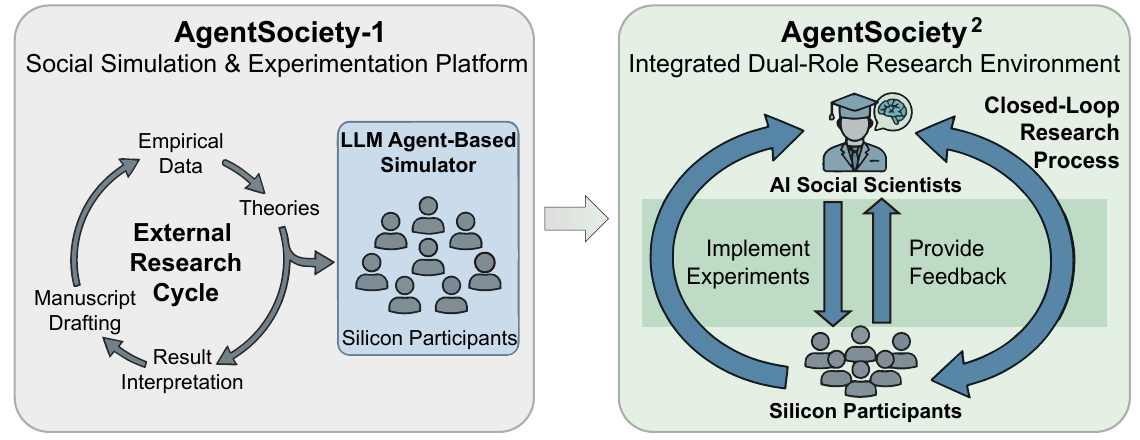}
    \caption{\textbf{From social simulation platform to integrated dual-role research environment.} \textit{AgentSociety-1} provides a social simulation and experimentation platform in which LLM agents mainly serve as silicon participants, while the broader research cycle, including empirical data collection, theory development, result interpretation, and manuscript drafting, remains external to the simulator. \textit{AgentSociety$^2$} extends this foundation into an integrated dual-role research environment, where AI social scientists coordinate the research workflow and silicon participants provide experimental feedback, enabling a closed-loop process from empirical grounding and theory formulation to experiment implementation, interpretation, and research presentation.}
    \label{fig:dp1}
\end{figure}

As discussed above, \textit{AgentSociety$^2$} is designed to support an agentic integration paradigm for computational social science. This goal raises three design tensions that the platform must address: how to integrate agents as silicon participants with agents as AI social scientists, how to balance human judgement with agentic assistance, and how to align general-purpose AI capabilities with the methodological demands of social science. Accordingly, we organize the design of \textit{AgentSociety$^2$} around three principles: an \textbf{integrated dual-role research environment}, which brings simulated experimental populations and agentic research workflows into the same platform; a \textbf{human-steered IRE}, which preserves researchers' conceptual agency while delegating routine orchestration to AI social scientists; and \textbf{generative simulation as experimental substrate}, which turns hypotheses into configurable agent behaviors, environmental rules, and intervention procedures for mechanism-oriented social inquiry.

\paragraph{Design Principle 1: From Simulation Platform to Integrated Dual-Role Research Environment.}
The first design principle concerns the transition from \textit{AgentSociety-1} to \textit{AgentSociety$^2$} (Figure~\ref{fig:dp1}). \textit{AgentSociety-1} was developed as a social simulation and experimentation platform, in which LLM agents primarily served as silicon participants within an agent-based simulator. This design provided a controllable experimental substrate for generating behavioral trajectories, testing theories, and studying social interaction and collective dynamics. However, the broader research cycle remained external to the simulator: empirical data collection, theory development, result interpretation, and manuscript drafting still had to be conducted across separate tools and human-managed workflows.

\textit{AgentSociety$^2$} extends this foundation into an integrated dual-role research environment. Beyond silicon participants, the platform introduces AI social scientists that coordinate the research workflow through the harness layer and domain-specific skills. These AI social scientists support empirical grounding, literature synthesis, theory and hypothesis formulation, experiment implementation, result interpretation, and manuscript generation, while silicon participants provide behavioral responses and experimental feedback within simulated environments. The two roles are therefore coupled within the same closed-loop research process: research questions and theories are translated into executable experiments, implemented over silicon participants, evaluated through generated behavioral trajectories and emergent patterns, and refined against real-world observations. This design transforms \textit{AgentSociety$^2$} from a social simulation platform into an integrated research environment for agentic computational social science.

\begin{figure}[t]
    \centering
    \includegraphics[width=\linewidth]{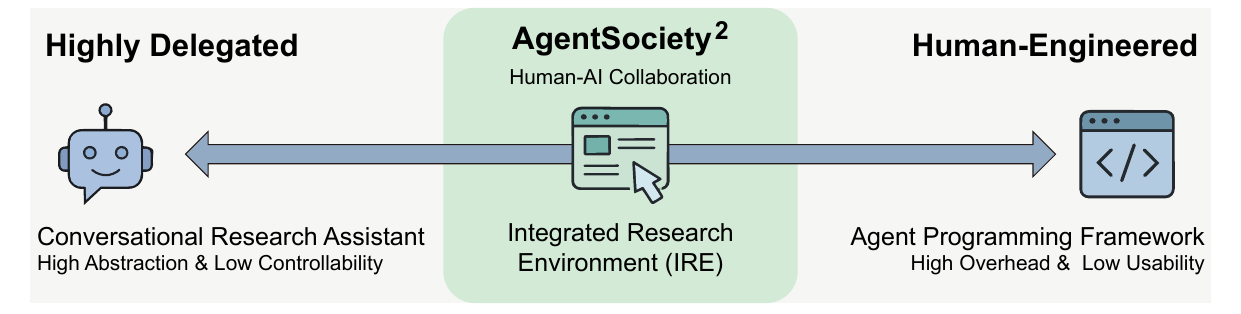}
    \caption{\textit{AgentSociety$^2$} as an IRE for human-AI collaboration. The IRE is positioned between highly delegated conversational assistants, which offer high abstraction but limited controllability, and low-level agent programming frameworks, which offer fine-grained control but impose high engineering overhead.}
    \label{fig:dp2}
\end{figure}

\begin{figure}[bth]
    \centering
    \includegraphics[width=0.85\linewidth]{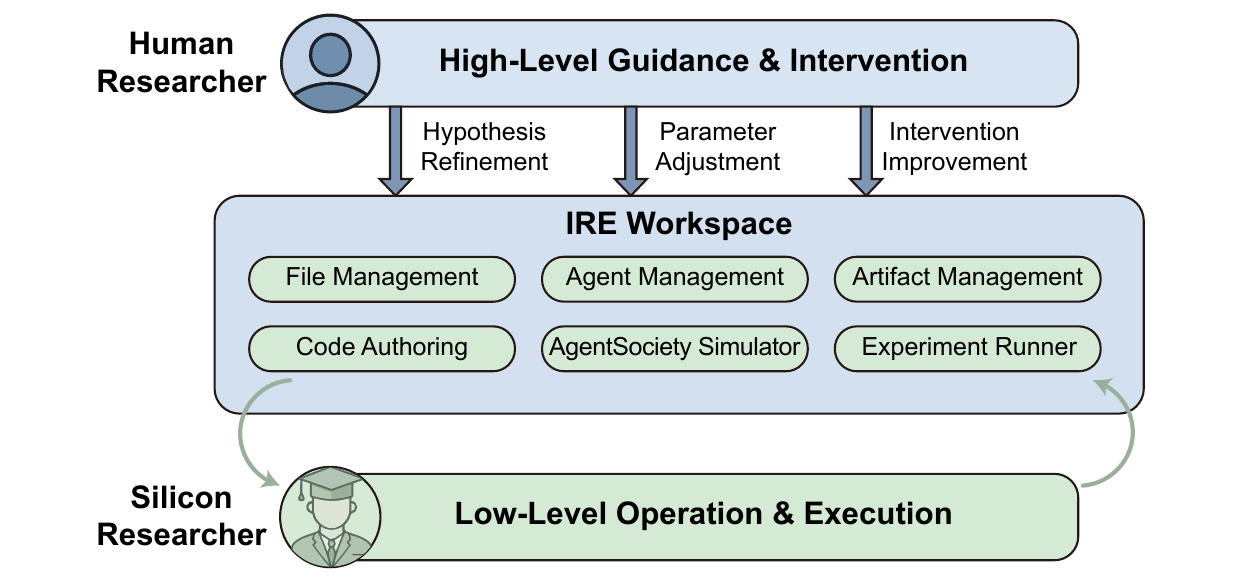}
    \caption{\textit{AgentSociety$^2$} structures human-steered collaboration with AI social scientists inside the IRE. Human researchers guide and intervene in key research decisions, while AI social scientists support low-level operation and execution through shared workspace components.}
    \label{fig:dp22}
\end{figure}


\paragraph{Design Principle 2: IRE for Human-Agent Collaboration.} While the dual-role architecture brings silicon participants and AI social scientists into the same platform, meaningful scientific progress still depends on how human researchers guide, inspect, and revise the research process. Existing agent-based research tools often fall along a spectrum between two extremes (Figure~\ref{fig:dp2}): highly delegated conversational assistants can help with simple tasks, but they often hide methodological structure behind natural-language interaction; low-level simulation frameworks expose detailed control over agents, environments, and interventions, but require researchers to engineer complex experimental pipelines from scratch. \textit{AgentSociety$^2$} adopts a different interaction paradigm by positioning the IRE between these two extremes.

Rather than treating AI social scientists as opaque autonomous executors, the IRE provides a structured and transparent workspace in which human researchers collaborate with them across the research workflow (Figure~\ref{fig:dp22}). In this IRE, the harness layer and domain-specific skills enable general-purpose AI assistants to act as AI social scientists, supporting the full research workflow from idea generation, literature retrieval, and dataset preparation to experiment design, simulation execution, result analysis, and manuscript drafting. These capabilities, however, are embedded in a human-steered process with explicit approval gates. Human researchers must confirm key transitions and judgement-intensive artifacts, including hypothesis approval, construct-to-variable mapping, intervention specification, validity interpretation, and claim release. Around these gates, researchers can inspect intermediate artifacts, revise agent outputs, adjust parameters, rerun simulations, and intervene whenever the proposed workflow departs from theoretical assumptions, empirical constraints, or ethical requirements. In this sense, \textit{AgentSociety$^2$} is designed as an AI social scientist environment rather than an autonomous replacement for researchers. By placing human judgement and agentic orchestration in the same interactive workspace, the IRE reduces the engineering burden of constructing agent-based studies while preserving the methodological visibility and fine-grained controllability required for social-scientific inquiry.


\begin{figure}[tb]
    \centering
    \includegraphics[width=\linewidth]{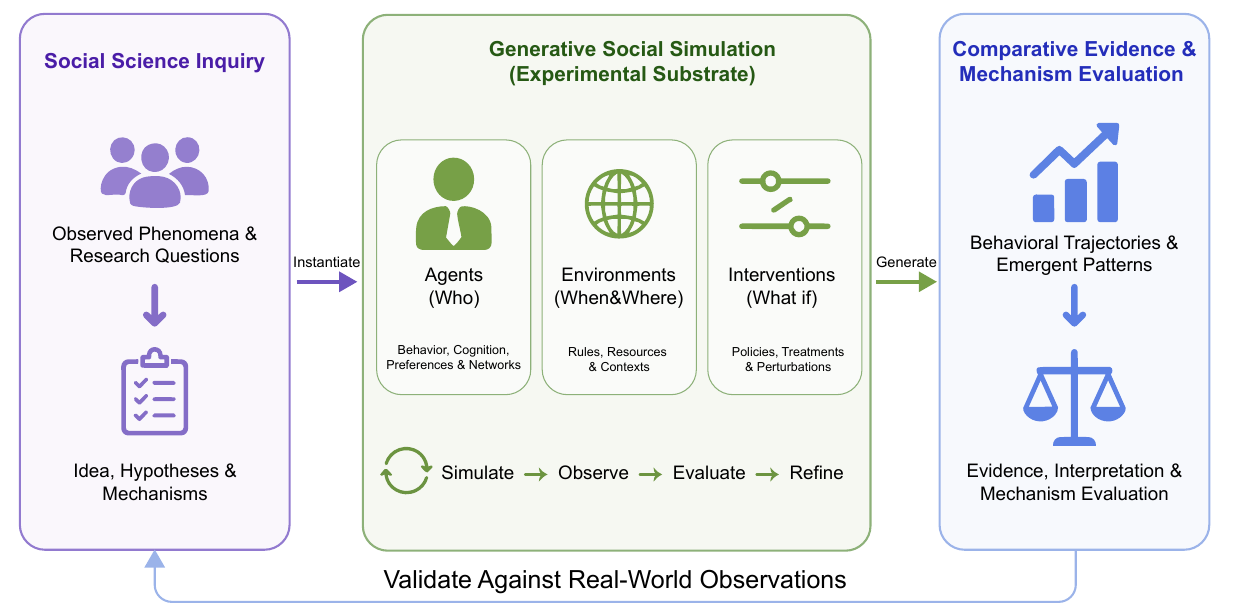}
    \vspace{-5mm}
    \caption{Generative social simulation as an experimental substrate for social science. \textit{AgentSociety$^2$} translates research questions, ideas, and hypotheses into configurable agent behaviors, environmental rules, intervention procedures, and measurement designs. Behavioral trajectories and emergent patterns generated by silicon participants provide evidence for mechanism evaluation and iterative hypothesis refinement.}
    \label{fig:dp3}
\end{figure}

\paragraph{Design Principle 3: Generative Social Simulation as an Experimental Substrate for Social Science.} Beyond architectural integration and interaction design, \textit{AgentSociety$^2$} is shaped by the methodological needs of social science. Many AI-for-science systems focus on well-defined optimization or prediction tasks, where structured inputs are mapped to measurable outputs. Social science inquiry, by contrast, often begins with complex collective phenomena and asks how individual cognition, social interaction, institutional conditions, and environmental contexts jointly generate them. The central challenge is therefore not only to predict outcomes, but to formulate, instantiate, and test plausible mechanisms that link micro-level behavior to macro-level patterns.

This motivates the use of generative social simulation as an experimental substrate for social science (Figure~\ref{fig:dp3}). In \textit{AgentSociety$^2$}, research questions, ideas, and hypotheses are translated into configurable agent behaviors, environmental rules, intervention procedures, and measurement designs. Building on \textit{AgentSociety-1}, the platform redesigns the agent–environment architecture so that generative social agents, configurable environments, and experimental procedures can be flexibly composed across research contexts. Environments are no longer passive simulation backdrops, but structured experimental containers in which mechanisms can be implemented, interventions can be applied, and behavioral trajectories can be collected as evidence for mechanism evaluation.

To make this experimental substrate reusable and extensible, \textit{AgentSociety$^2$} further introduces a skill-based repository for accumulating social-scientific research experience. Environment capabilities, experimental procedures, analysis routines, and domain-specific methods can be encapsulated as modular skills, which are then invoked by AI social scientists. As researchers build new agents, environments, interventions, and evaluation procedures, the corresponding skills can be inherited, adapted, and shared across projects. In this sense, \textit{AgentSociety$^2$} is not only a platform for running individual computational experiments, but also a shared base of reusable social-science skills, supporting a community-oriented infrastructure in which methodological experience can accumulate over time.

\section{Executable Social Experiments} 
The core of executable computational social science is to turn social-science hypotheses into runnable experiments. In \textit{AgentSociety$^2$}, this requires two coupled components: social generative agents and agentic environments. Social generative agents provide programmable silicon participants that can perceive, reason, remember, act, and respond to experimental tasks. Agentic environments provide the executable contexts in which these participants interact, receive resources, encounter rules, and experience interventions. Together, they instantiate hypotheses and mechanisms as controllable simulations that produce behavioural trajectories, emergent patterns, and measurable outcomes. This section introduces these two components in turn: Section~\ref{sec:social-generative-agents} describes how we redesign social generative agents into skill-based, workspace-grounded experimental participants, and Section~\ref{sec:agentic-environments} describes how agentic environments make social contexts, interaction protocols, interventions, and measurements reusable and executable.

\subsection{Social Generative Agents}
\label{sec:social-generative-agents}


\begin{figure}[tb]
    \centering
    \includegraphics[width=\linewidth]{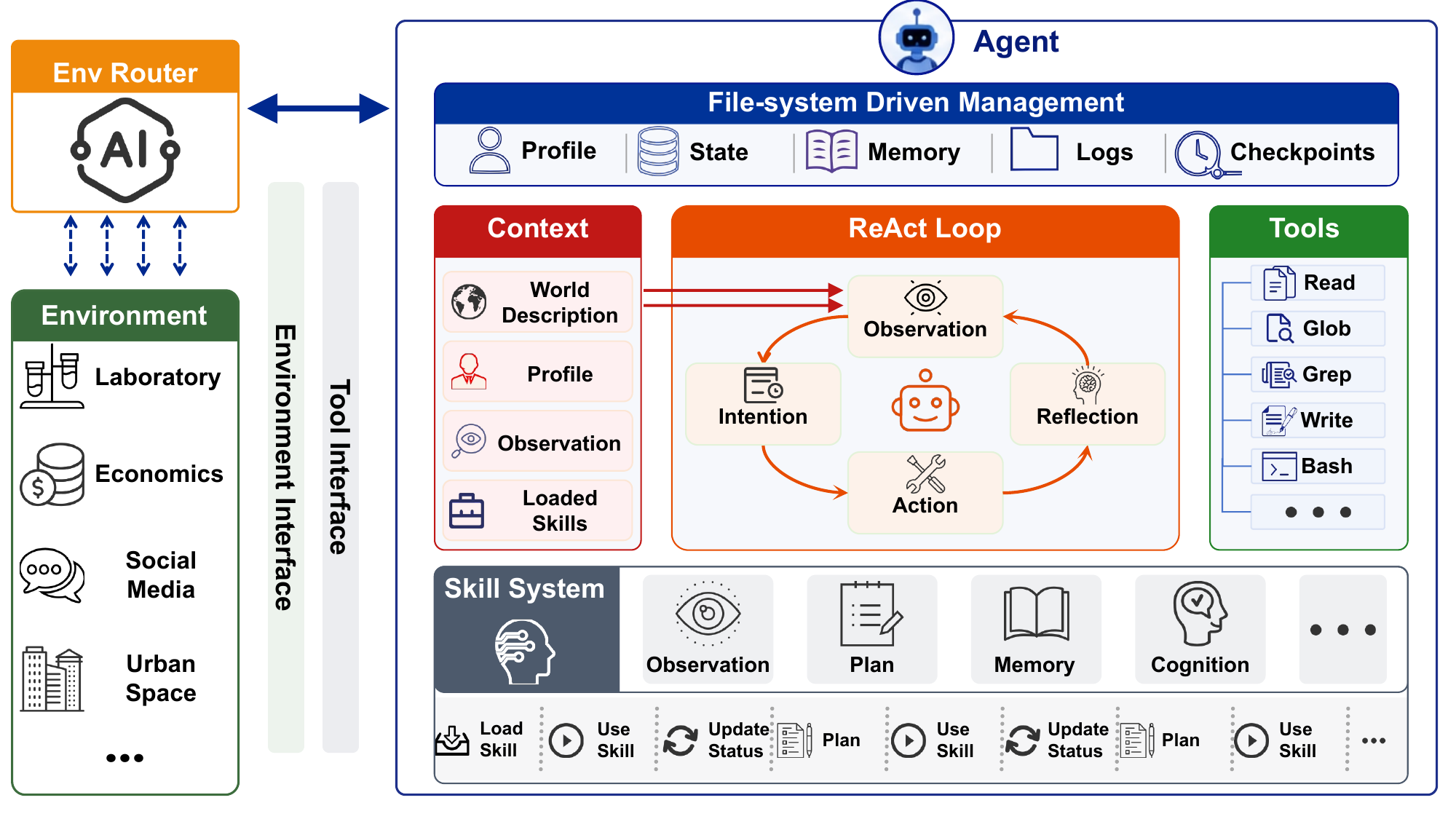}
    \caption{Overview of the Social Generative Agent.  Each agent maintains a persistent workspace that stores its profile, state, memory, logs, and checkpoints. At each simulation tick, the agent reasons over its context and may act directly, use tools, or activate skills, with actions and updates written back to the workspace for continuity, traceability, and reproducibility.}
    \label{fig:social-generative-agents}
\end{figure}

\subsubsection{Overview}





As LLM agents advance in natural language understanding, reasoning, tool use, and interactive decision making, they are becoming an important technical approach for social simulation \citep{park2023generative,mou2026individual}. Unlike traditional agents based on rules or manually specified parameters, LLM based agents can interpret experimental tasks, role descriptions, environmental observations, interaction histories, and tool instructions in natural language. They can then generate context sensitive responses, decisions, or actions. Existing systems, such as \textit{AgentSociety-1}  \citep{piao2025agentsociety}, OASIS \citep{yang2024oasis}, and YuLan-OneSim \citep{wang2025yulan}, have demonstrated the feasibility of using LLM based agents for social simulation. They organize agents within social environments such as urban spaces, social media platforms, and group interaction settings, and support simulations with a certain degree of scale and complexity.

However, existing work mainly shows that LLM based agents can participate in social simulations. It does not fully address a more fundamental design problem: how agent behavioural capabilities should be organized, reused, and controlled when social experiments become more diverse and complex. Different social experiments often involve distinct behavioural rules, interaction protocols, task objectives, and evaluation requirements \citep{gao2024large}. If agent capabilities are mainly implemented through prompts, platform specific code interfaces, or fixed workflows, then each new experimental setting requires researchers to translate experimental logic into system executable configurations, scripts, or prompt templates. This raises the barrier to LLM based social science research. It also introduces an additional translation cost between social scientific intent and executable agent behaviour. More importantly, such a design makes it difficult for agent capabilities to accumulate as composable and extensible modules, which limits the system's ability to support heterogeneous social experiments.

At the same time, prompt driven workflows introduce a substantial runtime burden in complex experiments. As experimental complexity increases, each agent decision may depend on a growing amount of information, including role background, experimental rules, environmental states, available tools, interaction histories, and stage specific objectives. These elements are often repeatedly inserted into prompts or passed across multiple prompt templates, making behaviour generation highly dependent on the quality of context organization. Even when a decision only requires a small subset of the available information, the agent may still need to carry a large amount of content that is not directly relevant to the current task. As the context further expands, key information may become diluted, local rules may be overlooked, and response quality and decision consistency may decline \citep{jin2025long,liu2024longgenbench,liu2024lost}. In long running or large scale simulations, excessively long contexts can also increase token cost, inference latency, and engineering risks such as timeouts, failed calls, or interrupted runs. These factors add further operational burden to experimental execution.

To address these problems, we propose skill based social generative agents. The core idea is to decouple agent behavioural capabilities from a single bulky prompt or a fixed workflow, and to represent them explicitly as a set of modular skills. Each skill encapsulates specific procedural knowledge and behavioural capability, including its conditions of use, input and output constraints, execution steps, tool use, state updates, and logging rules. At runtime, the agent activates relevant skills according to the current task, environmental state, and interaction feedback. A ReAct \citep{yao2022react} style loop connects observation, reasoning, action, tool use, and feedback updates, without requiring the agent to load the full experimental logic at every decision step. Through this design, new experimental capabilities can be introduced, extended, and composed as skills. Researchers can focus more on research questions, behavioural mechanisms, and experimental objectives, while reducing the burden of repeatedly translating experimental logic into platform specific code or fixed workflows. Meanwhile, on demand skill activation reduces contextual redundancy. The agent only needs to attend to the rules, states, and tool information that are directly relevant to the current task, thereby improving the runtime efficiency of complex experiments.

\subsubsection{Skill-based Agent Architecture}



Recent engineering practice in agent systems such as Claude Code and Codex suggests that skills are becoming a practical abstraction for organizing agent capabilities. Anthropic defines Agent Skills as modular capabilities that package instructions, metadata, and optional resources, such as scripts and templates, which Claude can use when they are relevant to a task \citep{anthropic2025agentskills}. OpenAI Codex adopts a similar mechanism, where a skill packages task-specific instructions, resources, and optional scripts into a reusable capability so that Codex can follow repeated workflows more reliably \citep{openai2026codexskills}. Inspired by this design pattern, we adapt skills from programming and productivity agents to social simulation, where they are used to organize the cognitive, social, and experimental behavioural capabilities of social generative agents. We design a skill-centered behavioural generation framework for social generative agents. The framework consists of agent workspaces, built-in tools, modular skills, and a ReAct-style execution loop.

We organize agent behavioural capabilities that were previously embedded in global prompts or fixed workflows into independent skill packages. Each skill package may describe general cognitive and behavioural mechanisms required by a social actor, such as observation, memory, cognition and planning. It may also describe task-specific behavioural guidance for a particular experimental setting, such as interaction rules, decision constraints, and output requirements. Each skill package follows a layered structure. The first layer is a skill header. It exposes only lightweight information, including the skill name, a short description, applicable scenarios, and a concise summary of expected inputs and outputs. This layer is used for skill selection: the agent can decide whether a skill is relevant to the current task without reading its full implementation details. The second layer is the full skill specification. It is loaded only when the skill is selected for execution. This layer defines how the skill should operate, including what information should be read from the agent state or environment, what steps should be followed, which tools may be used, what constraints must be respected, and what outputs should be produced. In this way, the specification acts as the executable behavioural guideline of the skill. The third layer provides optional supporting resources for skills that require more structured or repeatable execution. Instead of encoding everything in the main instruction, this layer can contain reusable materials such as example cases, output templates, domain-specific checklists, or small utility scripts. These resources help the agent perform complex behaviours more consistently while keeping the main skill specification concise.

This layered organization enables progressive disclosure. At the discovery stage, the agent only reads lightweight skill headers to identify potentially relevant capabilities. At the execution stage, it loads the full skill specification only for the selected skill, obtaining the necessary context requirements, procedural rules, tool-use constraints, and output format. When the task requires more structured or repeatable execution, the agent can additionally access optional supporting resources, such as examples, templates, checklists, or utility routines. Thus, a skill package separates capability discovery, execution guidance, and auxiliary support. By avoiding the repeated insertion of all skill details and supporting materials into the context, this design reduces contextual redundancy and improves runtime efficiency.

Skill-based modularization provides a structured interface for extending agent capabilities. For a new experimental setting, researchers no longer need to redesign the entire behavioural workflow. Instead, they can encode experiment-specific behavioural rules, interaction protocols, and task constraints as standardized skill packages, which are loaded and executed by a unified agent runtime. Skill-generation tools, such as Anthropic’s skill-creator \citep{anthropic_skill_creator}, can further assist in creating these packages by transforming high-level task requirements into structured skill specifications, thereby lowering the cost of translating experimental logic into executable agent behaviour.

We also introduce a skill repository mechanism to support organization, reuse, and sharing. Researchers can reuse skills from external repositories or contribute their own skills as transferable capability modules. This decouples agent capabilities from individual experiments and enables them to be reused, extended, and composed across experimental settings and research groups.




\subsubsection{Workspace-based Execution}
To isolate agents from one another, the system assigns an independent directory to each agent at runtime, which we call the agent workspace. The workspace is the private working directory of an agent during simulation. It stores the agent's relatively stable profile, dynamic state, memory, log files, and intermediate artifacts produced during skill execution. By storing this information in separate directories, each agent can maintain its own state across multiple rounds of interaction without mixing its states or histories with those of other agents. The workspace also prevents agent state from relying entirely on the temporary context carried in prompts. Instead, agent information is stored as persistent files that can be read, updated, and inspected. When needed, the agent can use tools to read relevant states or memories, and after execution it can write new observations, plans, memories, or logs back to the workspace.

However, the workspace itself only serves as a storage substrate for agent states, memories, intermediate artifacts, and execution traces. It does not determine how the agent should access, update, or act upon the information stored inside it. To operationalize the workspace, we provide a set of built-in tools that expose workspace operations and environment interactions through explicit interfaces. These tools cover file access, workspace navigation, content search, script execution, environment interaction, and execution control. Through these tool interfaces, the agent can retrieve information such as its profile, current state, and memory; write new observations, plans, and intermediate results back to the workspace; query environmental feedback; and record each execution step as a traceable record. In this way, the workspace provides persistent and transparent storage, while the built-in tools define how the agent can use that storage during reasoning and action.

\begin{table}[t]
\centering
\small
\caption{Built-in tools for workspace operation and environment interaction.}
\label{tab:builtin-tools}
\begin{tabular}{
    >{\centering\arraybackslash}p{0.24\textwidth}
    >{\centering\arraybackslash}p{0.20\textwidth}
    p{0.48\textwidth}
}
\toprule
\textbf{Tool} & \textbf{Category} & \textbf{Function} \\
\midrule
\texttt{workspace\_read}, \texttt{workspace\_write}, \texttt{workspace\_list}
& Workspace access
& Reads, writes, and lists workspace files, including agent states, memories, observations, plans, and intermediate artifacts. \\
\midrule
\texttt{grep}, \texttt{glob}
& Search
& Searches file contents and matches workspace files according to path patterns. \\
\midrule
\texttt{bash}
& Script execution
& Executes auxiliary commands or scripts when deterministic computation or batch processing is required. \\
\midrule
\texttt{codegen}
& Environment interaction
& Interacts with environment modules through executable instructions. \\
\midrule
\texttt{done}
& Execution control
& Signals the completion of a skill execution. \\
\bottomrule
\end{tabular}
\end{table}

With an independent workspace, built-in tools, and modular skills, the main execution loop connects these components through a ReAct-style execution process. In this process, the workspace provides a persistent substrate for states, memories, and execution traces. Tools provide executable operations. Skills provide high-level behavioural guidance. At each step, the agent first observes and reasons based on the current task, environmental feedback, and relevant states from the workspace. It then determines what behavioural capability is needed for the current decision and selects potentially relevant skills based on their headers. When necessary, the agent loads the full skill specification and follows the execution logic defined by the selected skill. It then calls the corresponding tools to interact with the environment or update its state, and writes the returned results back to the workspace. In this way, agent behaviour generation is not a one-off text response. It is a stateful execution process around an independent workspace, organized as a cycle of reasoning, action, feedback, and update.



\subsubsection{Context Management for Long-Horizon Simulations}

Based on the above design, an agent may perform multiple rounds of ReAct-style execution within each simulation step and may activate skills and call tools several times during this process. Although progressive skill disclosure avoids loading all skill details at every decision step, tool results, environmental feedback, historical reasoning, and state update records still accumulate in the conversation thread. As simulation steps proceed, the thread context continues to grow. Without a dedicated context management mechanism, the agent would not only be constrained by the model context window, but may also lose focus on the key states that are most relevant to the current decision.

To address this issue, we design a budget-aware context management mechanism for agents. The system first estimates the context utilization of the current thread based on the model context window and token counters. When the utilization reaches predefined thresholds, compression is triggered. The compression process follows a tiered strategy. Under light pressure, the system removes duplicated tool results and discards low-value historical messages according to priority. Under medium pressure, the system uses an LLM to compress old thread content into a structured summary. Under high pressure, the system further maintains a rolling summary so that long-term history can be represented more compactly. This allows the system to compress context progressively according to context pressure, rather than simply truncating the full history.

Context compression does not mean that historical information is discarded. Before each compression, the system writes the complete thread history into the runtime log directory of the workspace, so that it remains available as an external historical record. The compressed current context is then reconstructed from a structured summary, key states from the workspace, persistent memories, and the most recent interactions. Through this design, the workspace preserves the full trace and searchable history, while the current prompt only carries the summary, key states, and recent feedback needed for the current decision. This enables the agent to control context size during long-running simulations and improves the feasibility of long-running or large-scale execution.



\subsubsection{Built-in Skills Adapted from AgentSociety-1}
\textit{AgentSociety-1}  provides a theory-informed design for LLM-driven social generative agents. It represents each agent with a relatively stable profile and a dynamic status, where the status includes psychological state, economic state, and social relationships. It further models mental processes through emotion, needs, and cognition, and connects these internal states with social behaviours such as mobility, social interaction, employment, and consumption through memory.

\textit{AgentSociety$^2$}  inherits this behavioural modelling perspective, but reorganizes it into a skill-based architecture. Instead of using a fixed global workflow, \textit{AgentSociety$^2$}  provides four built-in skills: \texttt{observation}, \texttt{cognition}, \texttt{plan}, and \texttt{memory}. These skills implement the basic mind-memory-action loop of social agents: observing the current situation, updating internal state, forming executable actions, and retaining experience.

This design preserves the core behavioural structure of \textit{AgentSociety-1}  while making it modular in \textit{AgentSociety$^2$} . The built-in skills serve as default behavioural components that can be reused across experiments, replaced when different behavioural assumptions are required, or composed with additional skills for specific experimental settings. Table~\ref{tab:builtin-skills} summarizes these built-in skills.

\begin{table}[t]
\centering
\small
\caption{Built-in skills in \textit{AgentSociety$^2$} .}
\label{tab:builtin-skills}
\begin{tabular}{p{0.22\linewidth} p{0.68\linewidth}}
\toprule
\textbf{Skill} & \textbf{Function} \\
\midrule
Observation & Perceives the current environment and extracts information relevant to the agent's current task. \\
Cognition & Interprets observations and updates the agent's internal state, such as emotion, needs, cognition, or intention. \\
Plan & Translates intentions, goals, and constraints into executable actions or short-term plans. \\
Memory & Retains important experiences and makes them available for future decisions. \\
\bottomrule
\end{tabular}
\end{table}

\subsection{Agentic Environments}
\label{sec:agentic-environments}

\subsubsection{Overview}
\begin{figure*}[ht]
\centering
\includegraphics[width=\linewidth]{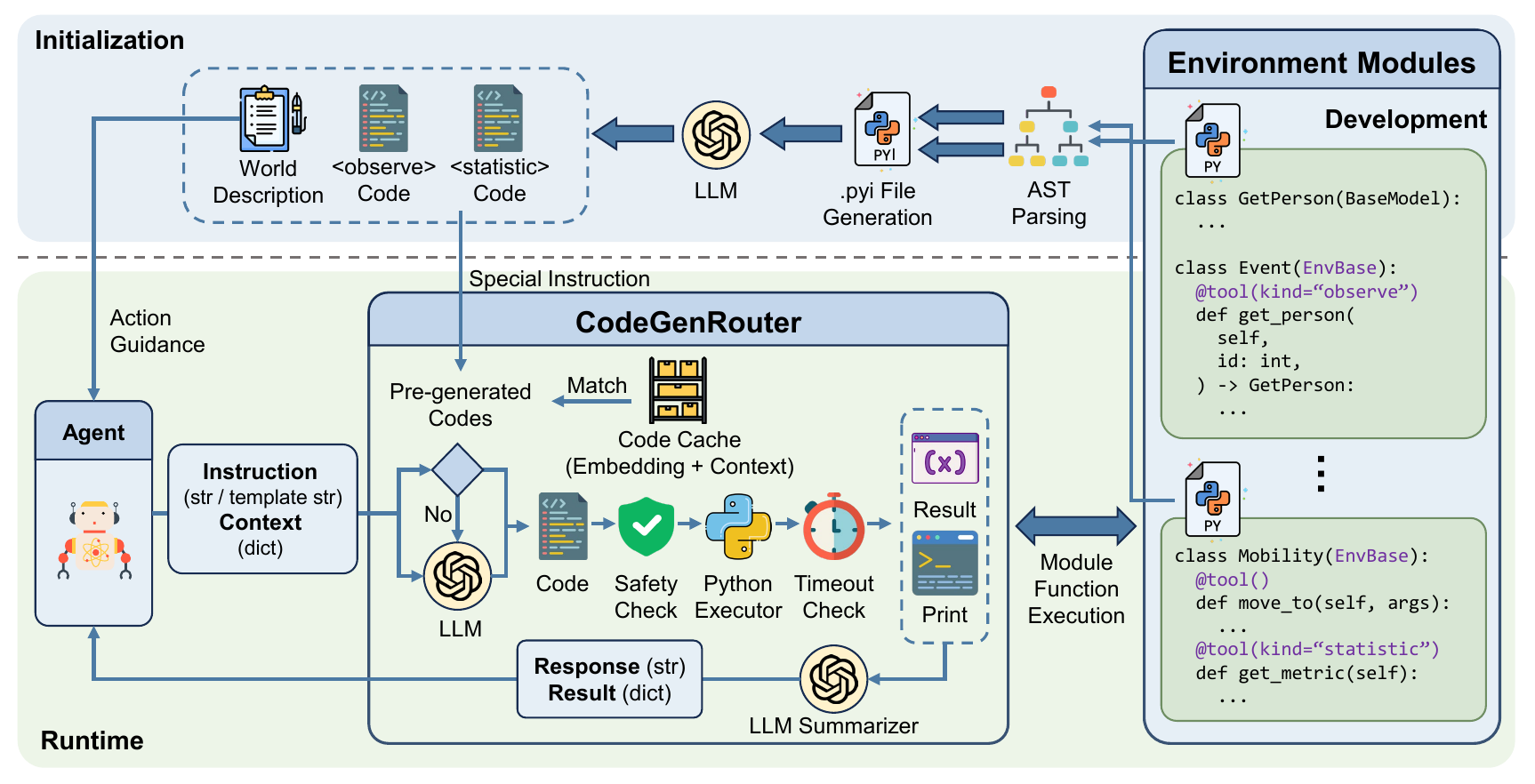}
\caption{The overview of the agentic environments.}
\label{fig:env}
\end{figure*}

In LLM agent simulations for social-science experiments, environments are not only the background in which agents act, but also the substrate through which research questions are translated into executable experiments.
They instantiate social contexts and interaction mechanisms, and provide observers with measurable behavioral and system-level outcomes.
AgentSociety-1 has demonstrated this role by combining LLM agents with city-scale mobility, social, and economic simulation environments, enabling researchers to model complex urban scenarios as computational testbeds for analyzing social phenomena and validating research concepts.

However, extending such environments to broader computational social science experiments raises a central challenge.
Existing simulation environments are often tightly coupled with specific agent implementations, environment schemas, and interaction protocols.
As a result, extending the platform to a new research domain typically requires substantial engineering effort.
Researchers must also manually adapt existing simulators, compose multiple environments, or repackage environment functions into interfaces callable by LLM agents.
This coupling limits the accumulation of reusable simulation modules and makes it difficult for researchers to rapidly construct experimental settings for different research questions.

To address this challenge, \textit{AgentSociety$^2$}  introduces a generalized agentic environment architecture, shown in Figure~\ref{fig:env}.
The contribution of this environment layer is threefold.
First, it provides a unified agent--environment interface that decouples LLM agents from environment-specific implementations while supporting both natural-language flexibility and structured execution precision.
Second, it offers a lightweight integration mechanism that allows existing simulators and newly generated custom environments to be registered as reusable modules with minimal code changes.
Third, it introduces \textbf{CodeGenRouter}, an LLM-assisted routing mechanism that translates agent intentions into validated environment operations and reduces interaction overhead through code pre-generation and semantic caching.
These designs transform heterogeneous social simulation environments into composable, reusable, and agent-accessible experimental modules.

The remainder of this section explains how this architecture is realized.
We first describe the unified agent--environment interface, then explain how environment modules are integrated or generated, and finally discuss CodeGenRouter and its efficiency optimizations.

\begin{figure*}[t]
\centering
\includegraphics[width=\linewidth]{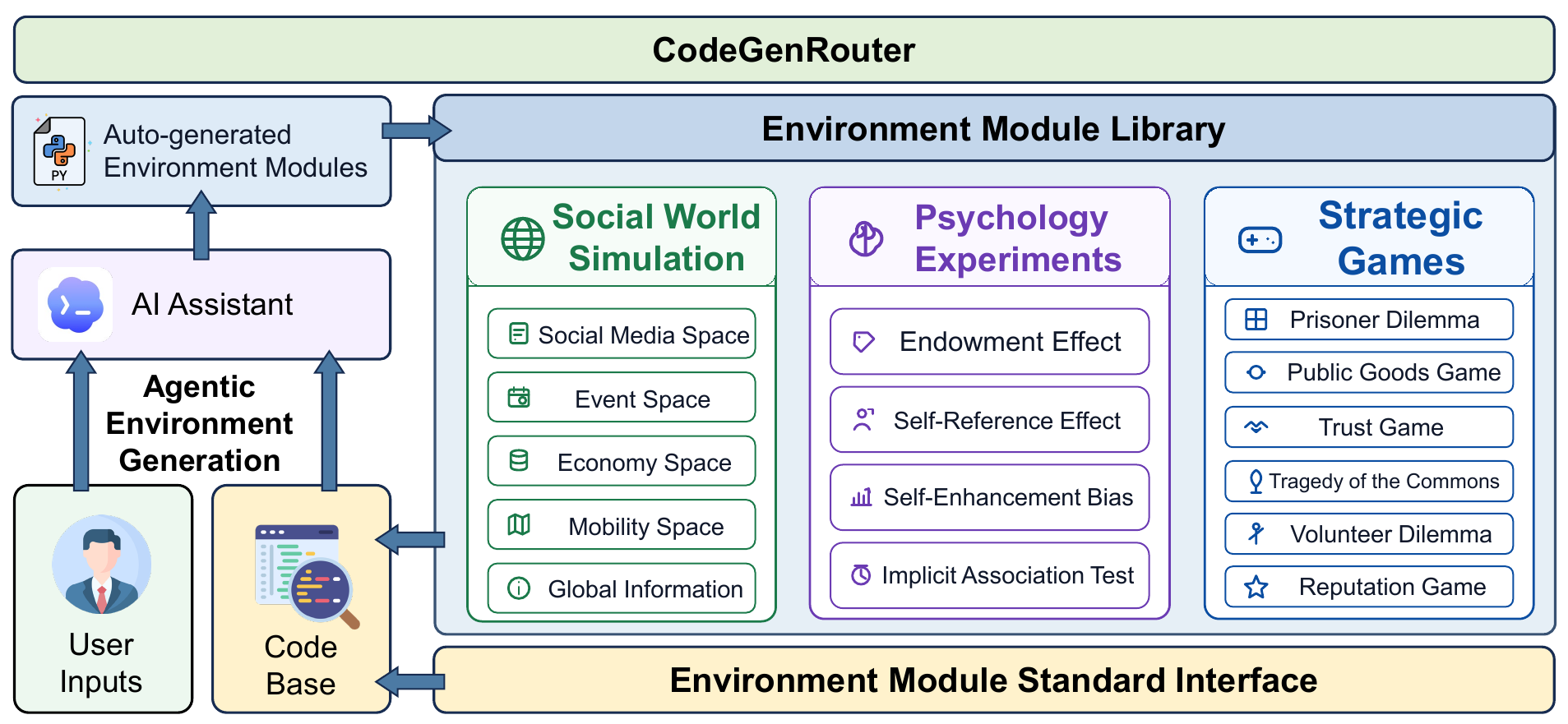}
\caption{Environment modules and the agentic environment generation pipeline implemented in \textit{AgentSociety$^2$} .}
\label{fig:existing_env}
\end{figure*}

Based on these designs, the architecture supports diverse experimental environments through flexible composability.
As shown in Figure~\ref{fig:existing_env}, extending the urban simulations from AgentSociety-1, we have expanded the platform to include social network environments modeling information diffusion and opinion dynamics, laboratory environments implementing classic behavioral games such as the Prisoner's Dilemma and Public Goods Game, among others.
Each module can be used independently or combined with others to study cross-domain interactions.

\subsubsection{Unified Agent--Environment Interface}

In conventional agent-based frameworks, the interface typically takes the form of structured function calls defined by the environment schema.
While this approach ensures reliable execution, it imposes rigid coupling between agents and environments.
Agents must invoke specific functions with correctly formatted arguments, and environments must expose functionality through predetermined schemas.
This coupling complicates experimentation across diverse scenarios, as researchers must modify agent implementations when switching environments or restructure environments to accommodate new protocols.

\textit{AgentSociety$^2$}  addresses this challenge through a dual-channel interface that simultaneously supports textual and structured data exchange.
On the input side, agents provide a \textit{textual instruction} describing their intended action, optionally supplemented by a \textit{structured context} containing parameter values and state information.
The textual instruction allows agents to express complex intentions without knowing available functions, while the structured context enables precise parameter passing and template-based optimizations.
On the output side, environments return both a \textit{textual response} and a \textit{structured result} object, accommodating diverse agent architectures.
This decoupling of agent reasoning from environmental implementation details allows the same agent architecture to operate across different environments.

\subsubsection{Environment Module Integration}

To realize a composable ecosystem, we provide a straightforward mechanism for incorporating existing simulation code.
Our integration system minimizes overhead by allowing existing code to be adapted through lightweight annotations rather than structural rewrites.
The process relies on two core components: environment classes inherit from a base \texttt{EnvBase} class providing common functionality, and developers use the \texttt{@tool()} decorator to annotate methods exposed to agents.
This decorator accepts optional parameters including a \texttt{kind} attribute specifying semantic categories: \texttt{kind="observe"} for state queries, \texttt{kind="statistics"} for aggregated metrics, and the default for state modifications or interventions.

Type annotations and docstrings serve three audiences simultaneously: human researchers, the LLM synthesizing invocation code, and the automated system generating interface descriptions.
During initialization, we parse each module using AST analysis to extract signatures, types, and documentation, then synthesize this information into a compact \texttt{.pyi}-style stub representation that is held in memory.
These stubs serve as semantic descriptions in LLM prompts, providing sufficient information for correct code generation while minimizing token consumption.
Adapting an existing simulator typically involves only adding \texttt{EnvBase} inheritance, decorating public methods with \texttt{@tool()}, and ensuring complete type annotations.
The core simulation logic remains unchanged.

\paragraph{From manual integration to agentic environment generation.}

Although the \texttt{EnvBase} inheritance and \texttt{@tool()} annotation mechanism enables existing simulators to be incorporated with minimal modification, many novel research scenarios lack pre-existing source code suitable for integration.
In such cases, researchers must construct environment modules from scratch, specifying state variables, tool interfaces, and persistence semantics that are consistent with the experimental hypothesis.
To address this gap, the system introduces an environment creation skill that assists AI social scientist in constructing custom environment modules on top of the platform abstraction according to experimental requirements.

The skill organizes custom environment development as a two-stage transformation: \emph{natural-language requirements} $\rightarrow$ \emph{structured design specification} $\rightarrow$ \emph{code}.
A planner subagent first produces a \textbf{DesignSpec}, a JSON contract that specifies class identity, global and per-agent state, tool signatures, initialization fields, and persistence semantics derived from the experimental hypothesis.
An implementer subagent then generates a single \texttt{custom/envs/<module>.py} file from this specification, after which a reviewer subagent audits the implementation against the specification under a fresh context window.
This separation of roles eliminates the blind spots of monolithic generation, in which a single context both produces the code and reviews its own output, and renders design intent and implementation independently traceable.

After code generation, the skill invokes a platform validator that loads the module to verify a unique \texttt{EnvBase} subclass is exported, checks interface compliance, runs a smoke test that mounts the module through \texttt{CodeGenRouter}, and confirms registry visibility.
When any check fails, the skill returns to the upstream stage that produced the failure rather than regenerating the module from scratch: a specification inconsistency sends it back to the design stage, whereas an implementation error sends it back to code generation.
Once all checks pass, the module is registered through the same scanner mechanism that exposes built-in environments, allowing AI social scientist to transform research-level environmental definitions into executable, registrable, and reusable environment modules that share the runtime semantics of manually integrated counterparts.

\subsubsection{CodeGenRouter}

The \textbf{CodeGenRouter} acts as the central coordinator bridging the unified interface with the modular environment system.
It translates agent textual instructions and structured context into environment function calls by generating executable Python code.
This process addresses three key challenges: understanding agent intent, invoking functions with appropriate arguments, and ensuring robustness against malformed or unsafe code.
CodeGenRouter operates through initialization and runtime phases.

During initialization, the router processes all registered environment modules, extracts their \texttt{.pyi}-style stubs, and constructs a system prompt describing available functionality including signatures, types, and documentation.
The system prompt serves as the World Description for LLM agents to understand the simulation environment.
The router also binds two special commands to executable paths: \texttt{<observe>} requests are routed to a built-in runner that iterates over methods tagged with \texttt{kind="observe"}, while \texttt{<statistics>} requests trigger an LLM-driven code pre-generation step whose output is validated and cached for direct execution.
At runtime, when an agent issues a request, the router first checks whether the input matches a pre-generated command or cached template.
If a cache hit occurs, the stored code (or the built-in runner, for \texttt{<observe>}) executes directly.
Otherwise, the router constructs an LLM prompt containing the agent's instruction, structured context, environment descriptions, and output constraints.
It then tasks the LLM with generating Python code that extracts context information, invokes environment functions in the correct order, stores results in a designated variable, and optionally prints intermediate progress.

Security is ensured through sandboxed code execution.
We selectively disable certain Python advanced syntax features and prohibit potentially dangerous built-in functions, allowing only imports from a trusted package whitelist that covers numerical primitives (e.g., \texttt{numpy}) and a curated set of standard pure-Python libraries.
Execution is subject to strict timeout limits, and code runs in an isolated context preventing access to framework internals or other agents' state.
After execution completes, the router collects the \texttt{result} variable along with standard output and error messages, passing them to an LLM summarization step that synthesizes a natural-language response for the agent.
The final output combines this textual response with the structured result object, completing the unified interface contract.

\subsubsection{Efficiency Improvement}\label{sec:env:eff}

Generative code execution incurs computational overhead through LLM inference for code generation.
While flexibility benefits warrant this cost for many applications, large-scale simulations require careful attention to efficiency.
We implement three complementary optimizations that substantially reduce interaction overhead.

\textbf{AST-based interface generation.}
We minimize overhead by generating compact \texttt{.pyi}-style stubs through AST parsing.
These stubs contain only essential information: function signatures, type annotations, and concise docstrings, formatted in standard Python syntax.
This approach reduces input token consumption while improving consistency between LLM inputs and outputs, as generated code directly mirrors the interface structure.

\textbf{Code pre-generation for common patterns.}
Agent-environment interactions in large-scale social simulations exhibit regular patterns.
The most frequent interaction is observation, where agents periodically query local state or external users retrieve global statistics.
For \texttt{<observe>} requests, we bind a built-in runner that iterates over all methods tagged with \texttt{kind="observe"} without invoking the LLM at request time.
For \texttt{<statistics>} requests, we ask the LLM once during initialization to generate, validate, and cache a code fragment that aggregates all methods tagged with \texttt{kind="statistics"}.
When agents issue requests matching these patterns, the bound executable path runs immediately, handling over 70\% of interaction requests in typical simulations.

\textbf{Semantic caching for template-based requests.}
Beyond fixed patterns, agents often issue structurally similar requests differing only in specific parameters.
Our caching system exploits this regularity by storing previously generated code snippets indexed by both semantic similarity and structural template.
When a new request arrives, the router computes embedding similarity between instructions and checks whether structured context has compatible keys and types.
If both conditions pass a configurable threshold, cached code executes with current parameters substituted.
This mechanism enables code reuse across agents and time, reducing LLM calls by approximately 66.5\% in large-scale scenarios.

\section{AI Social Scientist: Harness Engineering for Social Sciences}
\subsection{Overview}

\begin{figure}[!t]
    \centering
    \includegraphics[width=\linewidth]{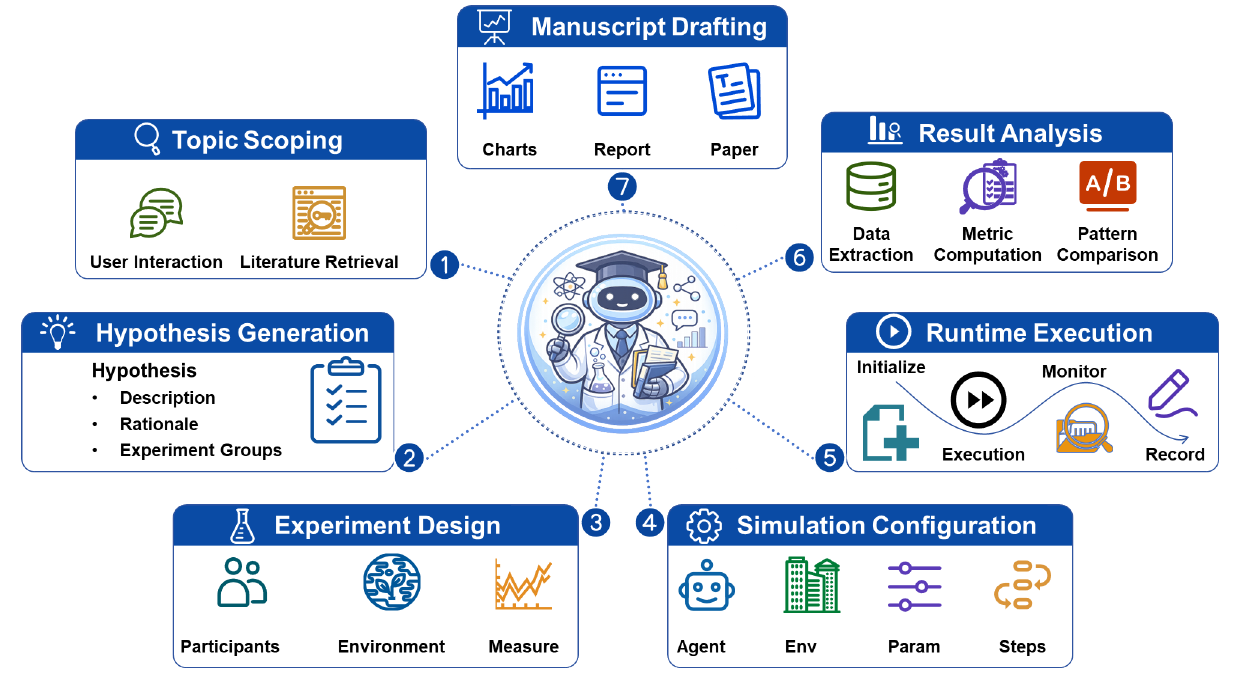}
    \caption{Overview of the AI social scientist workflow. The system supports the research process from topic scoping and hypothesis generation to simulation execution, result analysis, and report generation.}
    \label{fig:ai_scientist}
\end{figure}


The previous sections present \textit{AgentSociety$^2$}  as a unified LLM-based platform for agent simulation. Its agents, environment modules, and execution procedure can all be configured by users who can write the necessary code and configuration files. Yet moving from a substantive social-science question to a runnable experiment still involves substantial translation work. Abstract constructs must be turned into agent specifications, interventions into configuration files, and analyses into executable scripts. Because this kind of engineering work lies outside the standard training of many social-science researchers, the paradigm has not yet seen broad adoption in the field.


Recent autonomous research systems suggest a different approach. Rather than asking researchers to assemble the workflow themselves, these systems equip a general-purpose AI assistant such as Claude Code or Codex with a library of domain skills and let the assistant carry the workflow forward. Systems such as the AI Scientist, ResearchAgent, and related autonomous research agents \cite{lu2024ai,baek2025researchagent,boiko2023autonomous,zhou2025autonomous,shao2025omniscientist,feng2026internagent} already operate in this style, and agent-skill interfaces \cite{zhang2026equipping} further support on-demand loading of domain skills. This shift expands what the assistant can do, but it also places methodological discipline inside the assistant's sequence of autonomous decisions rather than in an explicit workflow. The same assistant that helps generate useful research insights may also produce simulator-incompatible configurations, edit unintended code, skip methodologically necessary steps, or leave behind artifacts that are difficult to trace or reproduce. As the capabilities of the assistant grow, explicit boundaries on action become increasingly important.


AI social scientist introduces these boundaries as a harness engineering layer for social science. Rather than functioning as a closed automation system, it adds an external scaffold on top of \textit{AgentSociety$^2$}  that aligns the behavior of a general-purpose AI assistant with social-scientific research practice. This harness combines a workflow substrate that tracks stage progression with a library of domain-specific skills, subagents, and auxiliary scripts. At the project level, the workflow substrate records progression through literature search, hypothesis formulation, experiment configuration, simulation execution, analysis, and downstream paper preparation, while the analysis stack further refines its own work through gated phases with persisted state and provenance \cite{cohen2017scientific,chirigati2013es}. Together, these components supply the methodologically grounded capabilities needed at each stage, covering literature retrieval, hypothesis generation, experiment design, simulation execution, result analysis, and manuscript preparation. The remaining subsections describe these two elements in turn, beginning with their shared specification and orchestration and then moving through the research stages.

\subsection{Skill Specifications and State-Machine Workflow Management}


The harness layer of AI social scientist rests on two complementary design choices. Each research capability is declared through a structured skill specification that defines how the assistant discovers and activates it. Workflow progression is then managed through explicit stage tracking at the pipeline level and finer-grained gated phases where the implementation requires them. Together, these two elements define how the assistant enters each research capability and how the overall process advances, while keeping artifacts and state transitions visible to the researcher.

\begin{figure}[!t]
    \centering
    \includegraphics[width=\linewidth]{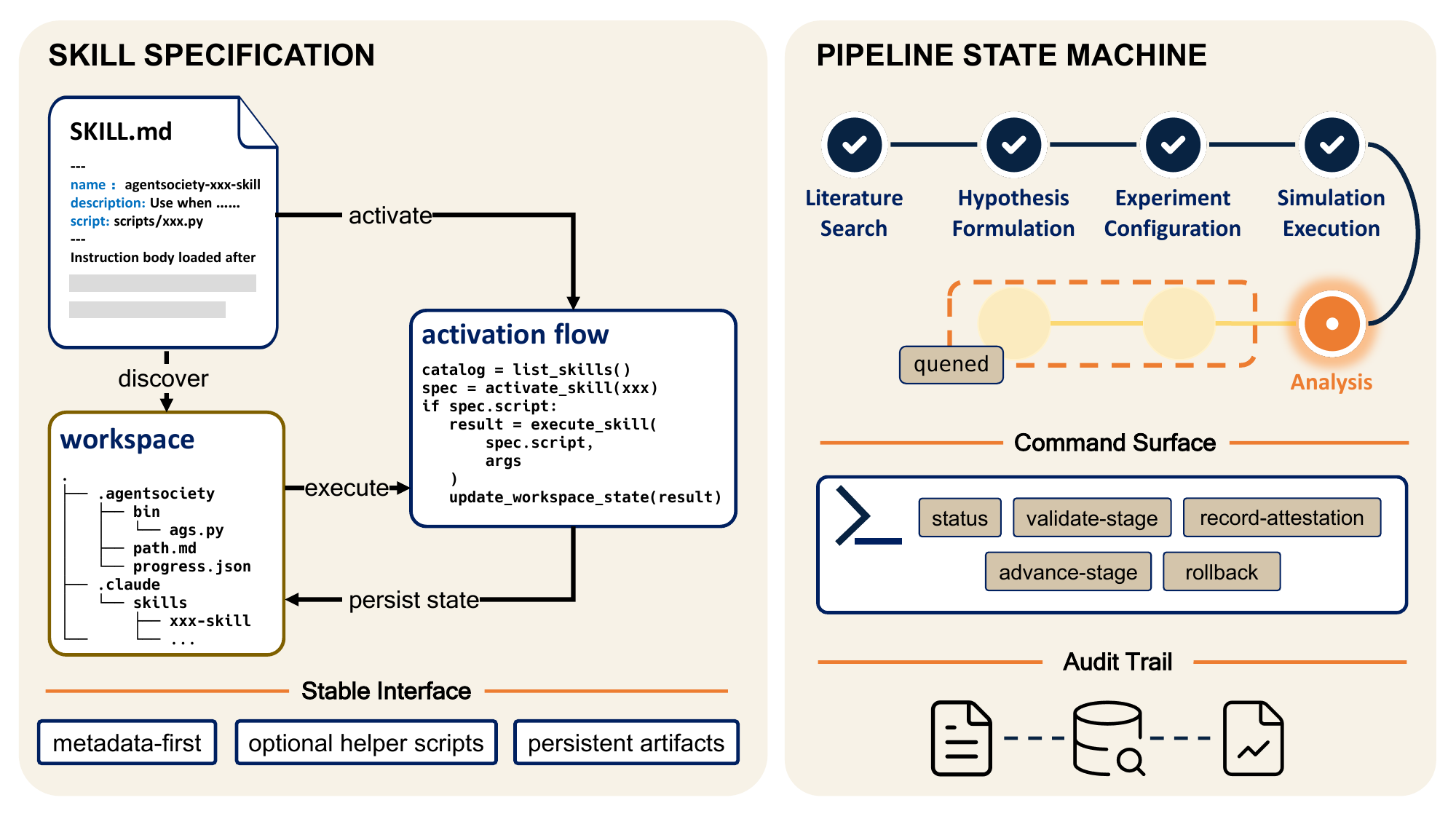}
    \caption{Harness contract for AI social scientist. Structured skill specifications provide a stable discovery-and-activation interface, while the shared workspace tracks explicit state transitions across the overall research pipeline from literature search to paper preparation.}
    \label{fig:scientist_harness}
\end{figure}


Skill specifications follow a lightweight uniform pattern. Each skill is packaged with a specification file whose structured frontmatter exposes discovery metadata, while the body provides the behavioral instructions loaded at activation time. Some skills are implemented as prompt-only guidance, whereas others also call deterministic helper scripts. The assistant therefore interacts with a stable discovery and activation interface, rather than rediscovering conventions by trial and error at each call. Extending the library then reduces to adding a new skill directory with its local specification and helper logic, without modifying the central skill registry.


Workflow management is split across two levels. At the project level, a pipeline tracker records progression through literature search, hypothesis formulation, experiment configuration, simulation execution, analysis, and paper preparation in the shared workspace state. Within the analysis stage, a dedicated harness persists machine-readable state and advances hypothesis-specific work through ordered phases from framing and exploration to claim formation, refinement, report production, and workspace-level synthesis. Paper preparation is handled as a downstream stage rather than by an internal paper-phase machine.


Within the analysis harness, a phase transition proceeds only after both gate layers have passed. Structural validation first checks deterministic preconditions such as the presence of required artifacts and the consistency of outputs from earlier phases. Assistant attestation then records a phase-specific rubric, including the touched artifacts, the key findings, and any remaining concern or blocking condition. Subagents support this process by returning selected outputs to the orchestrator, while the orchestrator remains responsible for gate evaluation and phase advancement. The workspace thereby accumulates a persistent record of analysis decisions, evidence, and outputs that can be inspected or audited later.


For the staged analysis workflow, state mutation is mediated through a documented command surface. This arrangement gives both the assistant and the researcher a stable operational interface to the persisted analysis state. The remaining subsections instantiate this architecture for the individual research stages. Each stage corresponds to one or more skills, and some stages introduce additional validators, rubrics, or persisted artifacts where tighter control is useful. This shared harness keeps the overall research workflow traceable end to end while allowing stage-specific mechanisms to remain local to the parts of the system that use them.

\subsection{Literature-Grounded Hypothesis Generation}

\begin{figure}[!t]
    \centering
    \includegraphics[width=\linewidth]{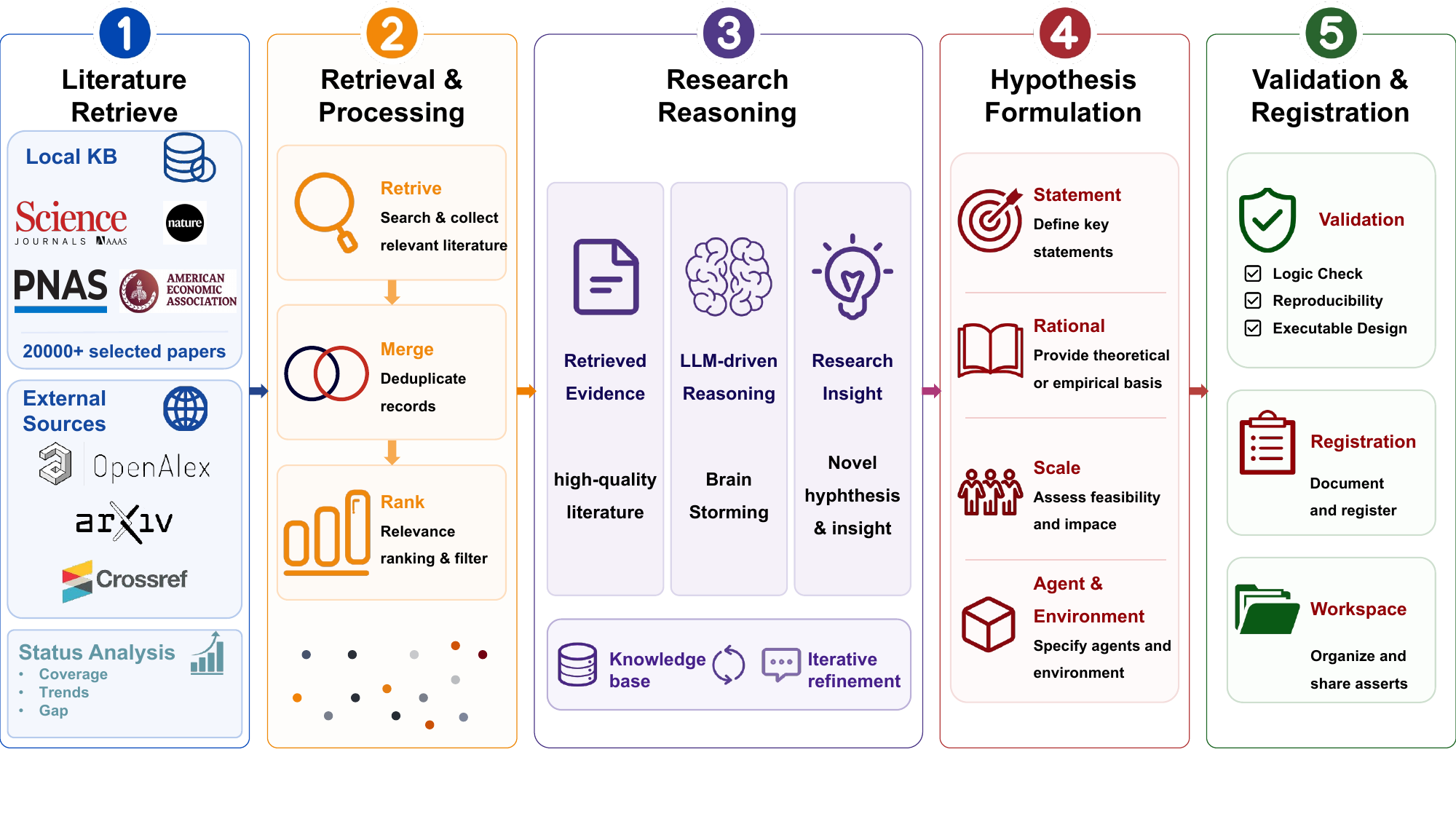}
    \caption{Overview of hypothesis generation workflow. Retrieved literature is processed into structured evidence, formulated into a hypothesis package, and registered in the workspace for subsequent experiment configuration.}
    \label{fig:hypothesis_generation}
\end{figure}


Hypothesis generation is a central step in social science research, as it connects theoretical reasoning, experimental design, and result interpretation. A meaningful hypothesis should not merely describe a potential empirical pattern, but should specify a theoretically motivated relationship between social conditions, behavioural outcomes, and underlying mechanisms. For an AI social scientist, this step is particularly important because it determines the research direction, the variables to be operationalized, the interventions to be designed, the data to be collected, and the framework under which simulation results are interpreted.


A straightforward approach is to ask LLMs to directly generate hypotheses from a given topic. However, such open-ended generation is insufficient for scientific research. Although LLMs can produce fluent and seemingly plausible research statements, their outputs may lack theoretical grounding, repeat well-established findings, rely on ambiguous constructs, or fail to specify experimentally testable mechanisms. Therefore, we argue that hypothesis generation should be grounded in external scholarly knowledge rather than relying solely on the parametric knowledge of the model.




To this end, we build a unified multi-source scholarly retrieval MCP service to support literature-grounded hypothesis generation. The service uses a curated local RAGFlow knowledge base as its authoritative and reproducible core, while further integrating external scholarly sources such as arXiv, CrossRef, and OpenAlex to improve coverage, timeliness, and cross-disciplinary recall. As shown in Table~\ref{tab:local_literature_database}, the local corpus contains 23,781 articles from leading journals and venues across economics, political science, sociology, management, behavioural science, environmental and urban studies, and computational science, covering publications from 2000 to 2025. By combining the high-quality local corpus with open scholarly retrieval sources, this MCP service provides the AI social scientist with a reliable and extensible scholarly foundation for retrieving established theories, empirical findings, variable definitions, experimental paradigms, and unresolved research gaps before generating hypotheses.

\begin{table}[!t]
\centering
\small
\caption{Composition of the Local Literature Database by Domain}
\label{tab:local_literature_database}
\resizebox{\textwidth}{!}{%
\begin{tabular}{llcc}
\toprule
\textbf{Domain} & \textbf{Journal / Venue} & \textbf{Article Count} & \textbf{Year Range} \\
\midrule

\multirow{4}{3.2cm}{Multidisciplinary Sciences}
 & Nature & 245 & 2000--2025 \\
 & Nature Communications & 1,494 & 2010--2025 \\
 & Scientific Data & 19 & 2018--2025 \\
 & Others, including Scientific Reports & 9,954 & 2011--2025 \\
\midrule

\multirow{5}{3.2cm}{Economic Sciences}
 & Quarterly Journal of Economics & 498 & 2015--2025 \\
 & American Economic Review & 1,222 & 2015--2025 \\
 & Econometrica & 538 & 2015--2024 \\
 & Journal of Political Economy & 771 & 2015--2025 \\
 & Review of Economic Studies & 859 & 2015--2025 \\
\midrule

\multirow{3}{3.2cm}{Management and Organizational Sciences}
 & Management Science & 1,437 & 2018--2025 \\
 & Academy of Management Journal & 809 & 2015--2025 \\
 & Academy of Management Review & 419 & 2015--2025 \\
\midrule

\multirow{6}{3.2cm}{Social, Political, and Behavioral Sciences}
 & Nature Human Behaviour & 321 & 2016--2025 \\
 & American Political Science Review & 776 & 2016--2025 \\
 & American Sociological Review & 445 & 2015--2025 \\
 & American Journal of Sociology & 225 & 2015--2025 \\
 & Journal of Personality and Social Psychology & 56 & 2015--2015 \\
 & Others, including Humanities and Social Sciences Communications & 2,801 & 2015--2025 \\
\midrule

\multirow{5}{3.2cm}{Environmental and Urban Sciences}
 & Nature Climate Change & 258 & 2011--2025 \\
 & Nature Sustainability & 255 & 2018--2025 \\
 & Nature Cities & 129 & 2023--2025 \\
 & Nature Ecology \& Evolution & 16 & 2017--2025 \\
 & Others, including npj Urban Sustainability & 199 & 2021--2025 \\
\midrule

\multirow{3}{3.2cm}{Computational and Physical Sciences}
 & Nature Physics & 4 & 2010--2023 \\
 & Nature Machine Intelligence & 12 & 2019--2024 \\
 & Nature Computational Science & 19 & 2021--2025 \\
\midrule

\textbf{Total} & & \textbf{23,781} & \textbf{2000--2025} \\
\bottomrule
\end{tabular}%
}
\end{table}


To complement the local corpus, the service further connects to external scholarly sources, including arXiv, OpenAlex, and CrossRef, which provide recent preprints, scholarly metadata, and publication records. The retrieval service normalizes heterogeneous outputs into a unified schema and applies post-processing steps such as duplicate merging, relevance filtering, venue-based weighting, and result ranking. Local documents are retrieved through vector search and aggregated at the article level, while external sources provide complementary metadata and related-work expansion. This pipeline provides the AI social scientist with traceable literature evidence that balances reliability, coverage, and recency.


Moreover, social science research is not a one-shot retrieval process, but an iterative process of knowledge accumulation, comparison, and revision. Therefore, the literature service supports both retrieval and ingestion: it retrieves papers from the local knowledge base and external scholarly sources to ground hypothesis generation, and it can also ingest papers, summaries, research notes, experimental records, or raw text back into the RAGFlow knowledge base. In this sense, the Literature module functions as a research memory infrastructure rather than a standalone search tool.


Built upon this infrastructure, hypothesis generation is formulated as a literature-driven research pipeline. The system retrieves relevant studies around a target social phenomenon, extracts theoretical mechanisms, variable relationships, experimental settings, empirical findings, and open questions, and transforms them into hypotheses with explicit conditions, expected outcomes, underlying mechanisms, and measurable indicators. This makes hypothesis generation a scholarly evidence-constrained and mechanism-oriented research procedure, rather than an unconstrained language generation task.


At the implementation level, this pipeline is instantiated as a structured hypothesis-package generation process. Instead of directly treating an LLM-generated statement as a hypothesis, the system converts the literature-grounded research intention into a structured object containing a hypothesis statement, theoretical rationale, experimental group design, and optional simulation module intentions. The package is then validated, assigned a unique identifier, and registered in the workspace with corresponding hypothesis records, simulation settings, and experimental group entries. Thus, a generated hypothesis becomes a traceable, versioned, and simulation-addressable research unit for subsequent experiment configuration and execution.

\subsection{Experiment Design}


\begin{figure}[!t]
    \centering
    \includegraphics[width=\textwidth]{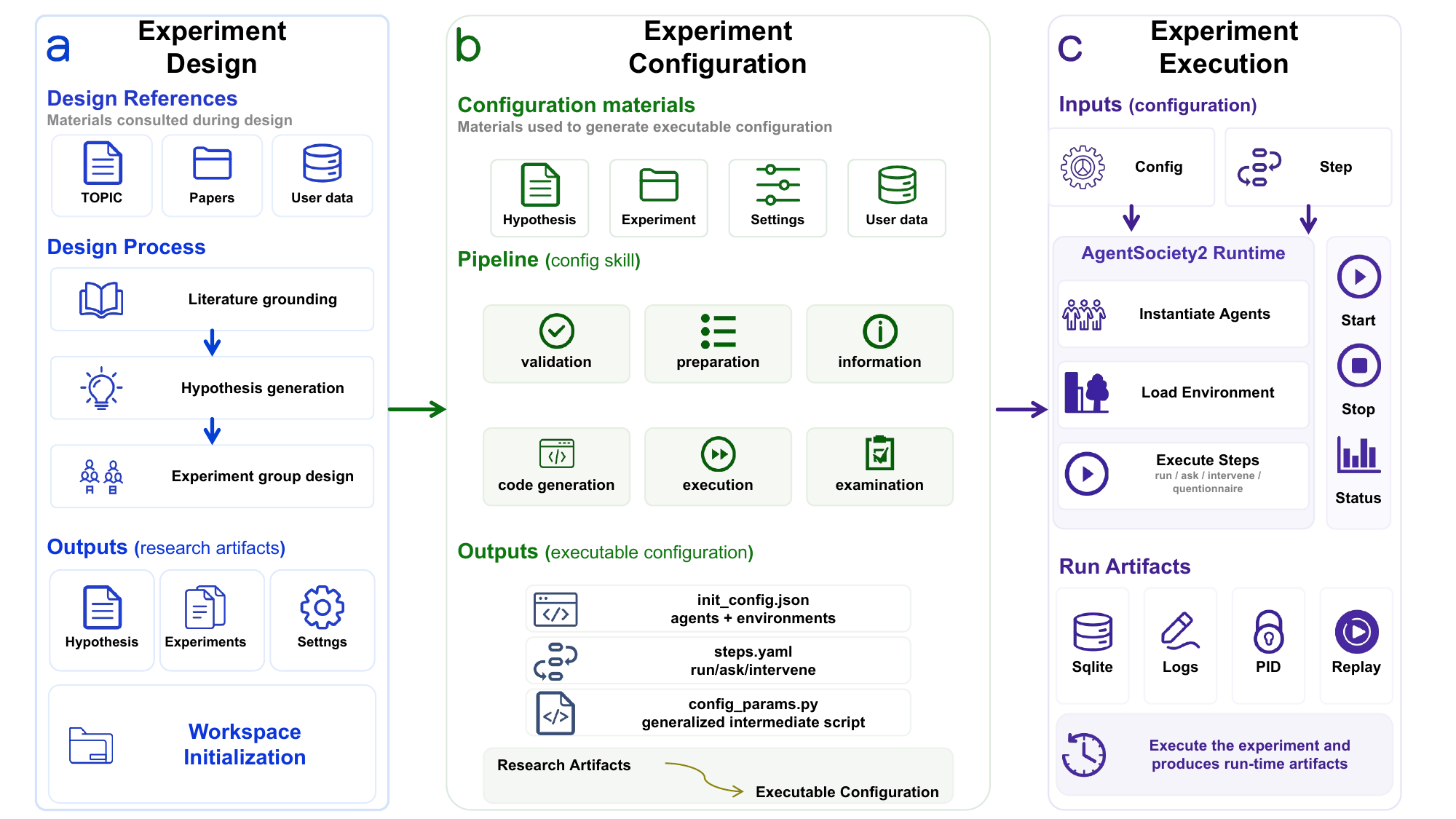}
    \caption{AI social scientist experiment workflow. The pipeline turns research ideas into executable experimental protocols, including experimental designs, simulation configurations, and run-time artifacts, through a structured design–configuration–execution process.}
\end{figure}


Following topic construction and hypothesis formalization, the system proceeds to the experiment design phase. The primary objective of this stage is to translate research ideas and theoretical constructs into executable experimental protocols, rather than merely generating simulation configuration files. Starting from the research description layer, the AI social co-scientist progressively specifies how hypotheses, experimental groups, agent populations, environment modules, intervention conditions, measurement targets, and execution requirements should be operationalized within a simulation. The current implementation therefore separates the research description layer from the executable configuration layer. The former organizes high-level semantic information, including hypotheses, construct definitions, experimental-group designs, selected agent types, environment modules, intervention logic, and measurement plans, while the latter materializes these decisions as strictly defined configuration files that can be loaded by the execution engine.


In the research description layer, each hypothesis is materialized as a structured workspace directory, typically containing \texttt{HYPOTHESIS.md}, \texttt{SIM\_SETTINGS.json}, and several \texttt{experiment\_\{id\}/} subdirectories. \texttt{HYPOTHESIS.md} records the hypothesis statement and theoretical rationale; \texttt{SIM\_SETTINGS.json} stores the selected agent classes and environment modules; and each \texttt{EXPERIMENT.md} under \texttt{experiment\_\{id\}/} describes the conditions, group type, and experimental setting of the corresponding group. These files express research semantics, but they are not directly executable simulation configurations.


In the executable configuration layer, an experiment is explicitly defined by \texttt{init\_config.json} and \texttt{steps.yaml}. The former specifies environment modules, the agent population, and router-related configurations, while the latter defines the experiment start time, temporal progression, and interaction steps, such as running the simulation, asking questions, applying interventions, and conducting questionnaires.


Therefore, the central task of experiment design is to establish a mapping between research semantics and executable configurations. This process is not an unconstrained automatic generation, but a template-guided and constraint-driven translation workflow with user-in-the-loop refinement. The system generates configuration templates and candidate structures from research descriptions, while constraining the results through workspace path conventions, module discovery, and executable configuration schemas.


Researchers or higher-level orchestration processes can then inspect, refine, and confirm the final configurations, ensuring that the experiment remains consistent with the research semantics while being executable by the simulation engine.

\subsection{Experiment Execution}


The execution phase translates the experimental protocol into a concrete simulation process, connecting research design, system configuration, and runtime records. In the current implementation, experiments are executed by a command-line runner that separates configuration generation from simulation execution. The research artifacts describe hypotheses and experiment groups, while the executable configuration files specify the environment modules, agent population, router settings, and ordered execution steps required by the simulation engine.


At startup, the runner loads and validates the core configuration files, including \texttt{init\_config.json} for initialization settings and \texttt{steps.yaml} for execution steps. The initialization configuration defines the environment modules, agents, and router options, whereas the step configuration specifies the simulation start time and the ordered runtime procedures. Before creating simulation objects, the runner checks whether the workspace contains a custom module directory and registers user-defined modules when available. Environment modules and agents are then instantiated from the module registry according to their configured types and initialization arguments.


Each execution is associated with a dedicated run directory for runtime state, replay data, and generated artifacts. The runner records process information, execution status, timestamps, and progress information in \texttt{pid.json}. In parallel, the system initializes a SQLite replay database, \texttt{sqlite.db}, through the replay writer. This database stores basic replay information and supports data tables dynamically registered by environment modules, allowing simulation-specific states, interactions, and measurements to be persistently recorded for later inspection and analysis.


The execution process is controlled by the ordered steps defined in the step configuration. A simulation step advances the agent society according to the configured number of steps and time interval. An ask step sends a natural language question to the society and saves the response as a Markdown file. An intervention step applies a natural language intervention and stores the corresponding output. A questionnaire step constructs a structured questionnaire and collects responses from the specified target agents, or from all agents when no target list is provided.


This execution design supports both behavioral simulation and interactive measurement. Quantitative states, interaction traces, and module-related records are stored in the replay database, while natural language and structured outputs produced by ask, intervention, and questionnaire steps are saved under the artifact directory. Typical outputs include the replay database, the process status file, and artifact files for ask responses, intervention outputs, and questionnaire results. When an explicit log file path is provided, execution logs can also be written into the run directory for debugging, auditing, and later analysis.

\subsection{Result Analysis}


\begin{figure}[!t]
    \centering
    \includegraphics[width=\textwidth]{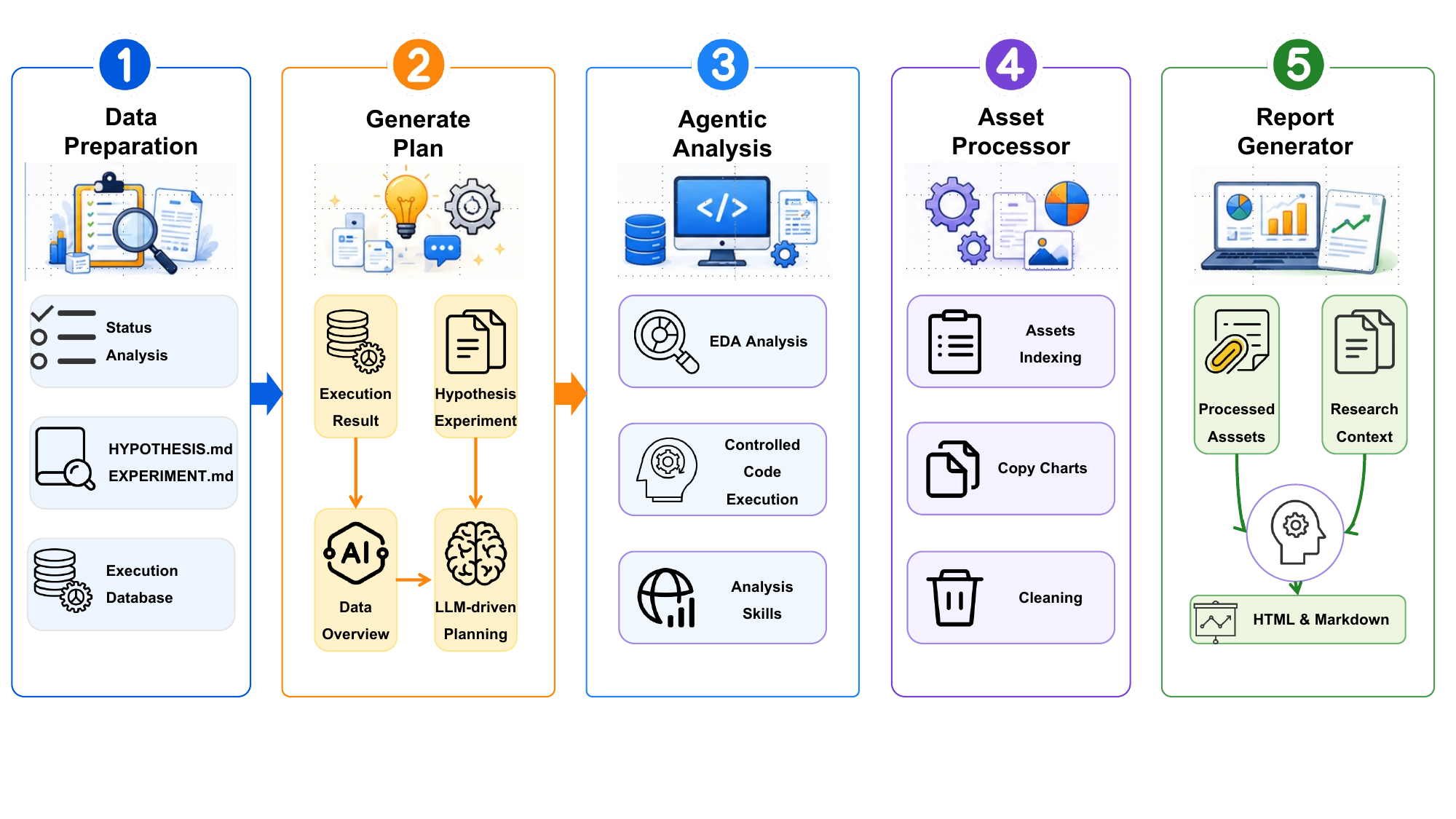}
    \caption{AI social scientist analysis and report-generation workflow. The pipeline prepares experimental data and research context, generates an analysis plan, executes agentic data analysis, processes generated assets, and produces a final HTML and Markdown report.}
    \label{fig:result-analysis}
\end{figure}


The analysis phase transforms raw execution records into interpretable research outputs. Rather than merely reading simulation results and producing a static report, it provides an integrated analysis pipeline that combines experiment context loading, data summarization, reasoning, tool execution, exploratory data analysis, report generation, and cross-experiment synthesis. \cite{lee2015complexities, yao2022react}


The analysis layer takes the experiment workspace as its primary input. It collects information from the research topic, literature resources, hypothesis description, experiment specification, execution status, replay database, and interaction artifacts generated during execution. This design allows the analysis process to incorporate both research-level semantics and runtime evidence, instead of relying solely on isolated database tables.


For single-experiment analysis, the system first locates the target experiment directory and loads the corresponding experimental context. It then extracts the database schema, sample records, and statistical summaries from the replay database. When replay metadata is available, the analyzer uses it to interpret the semantic roles of tables and fields; otherwise, it falls back to standard SQLite schema inspection. Based on the loaded context and data summary, the analysis subagent generates initial insights and plans the subsequent analysis strategy. This process is not treated as unconstrained free-form generation. Instead, structured output contracts are used to regulate the model's behavior, requiring it to explicitly specify analysis goals, tool choices, visualization needs, and final summaries.


To support more complex analysis tasks, the system provides a controlled tool execution mechanism. The analysis agent can inspect files, search workspace content, invoke literature-search tools, or generate and execute analysis code within a restricted workspace. During code execution, the system prepares the required data files, detects dependencies, runs the generated scripts, and collects produced artifacts such as figures, tables, and intermediate results.


The resulting analysis outputs are organized into report resources. The system collects artifacts produced during execution, visualizations generated during analysis, and exploratory data analysis results. If the analysis agent does not produce sufficient exploratory evidence, the system attempts to generate fallback data summaries, ensuring that the final report remains grounded in the underlying data. Finally, the reporting component integrates the experimental context, data summaries, analysis results, visual resources, and optional literature information into researcher-facing reports. These reports can present the behavioral trajectories and measurement results of a single experiment, while also providing structured evidence for manual inspection, debugging, and paper writing.


Beyond single-experiment analysis, the module also supports batch analysis and synthesis. It can analyze multiple experiment groups under the same hypothesis, and can further synthesize results across hypotheses and experiments. This enables the system to summarize shared patterns, condition-specific differences, and comparable indicators across experimental settings.


Therefore, the analysis module serves as the research output generation layer of the overall system. It transforms the states, interactions, and measurements recorded during simulation execution into readable, inspectable, and comparable analytical results, thereby connecting experiment execution, data interpretation, and research reporting.

\subsection{Paper Generation}

\begin{figure}[!t]
    \centering
    \includegraphics[width=\textwidth]{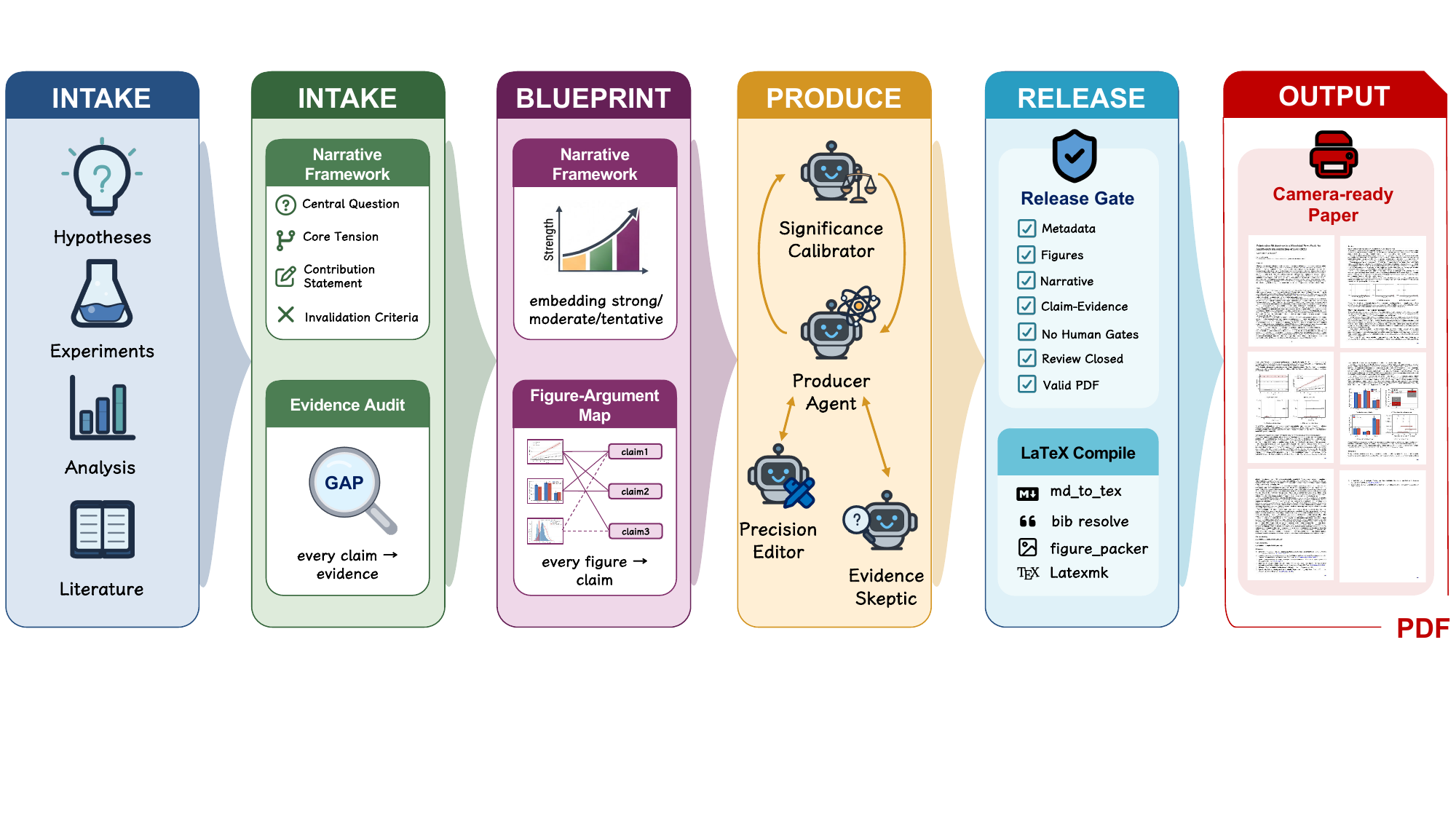}
    \vspace{-20mm}
    \caption{Architecture of the paper generation module. The workflow collects research outputs from all preceding stages, constructs a narrative framework and structured claim--evidence alignment, subjects the manuscript to multi-perspective adversarial review, and compiles the final document through a release gate.}
    \label{fig:paper-generation}
\end{figure}


The paper generation phase is the final stage of the AI social scientist workflow. Its purpose is to assist the researcher in assembling the outputs accumulated across all preceding stages (hypotheses, experimental configurations, execution traces, analysis reports, figures, and literature) into a coherent, evidence-grounded draft manuscript. The module takes on the operational work of consolidating evidence, drafting prose, and maintaining cross-references between claims and the artifacts that support them. While the analysis module produces interpretable reports and structured data summaries, these artifacts are organized around individual hypotheses and experiments rather than the integrated, argument-driven narrative that a scientific paper demands. The paper generation module therefore supports the researcher in synthesizing these diverse outputs into a unified draft with a clear storyline, logically structured claims, rigorous evidence support, and scholarly standards.


A straightforward approach would be to provide an LLM with all research materials and request a complete manuscript in a single pass. Such end-to-end generation, however, presents several critical deficiencies for scientific writing. Without explicit claim--evidence alignment, generated statements may overstate findings, invent unsupported causal assertions, or misattribute results to incorrect sources. Figures and tables, the central vehicles of evidence in empirical papers, must be logically mapped to specific claims and ordered to construct a progressive argument. This demands structured planning beyond what single-pass generation can reliably achieve. Scientific manuscripts also undergo iterative refinement through internal skepticism and external peer review, yet a single-generation pipeline provides no mechanism for systematic critique, revision routing, or quality gating. We therefore organize paper generation as a multi-phase, constraint-driven process with explicit stages for argument planning, evidence auditing, adversarial review, and readiness assessment.


The paper generation workflow begins with a standardized intake procedure that collects all research artifacts from the workspace (research topic descriptions, hypothesis statements, experiment specifications, analysis summaries, generated figures, and the curated literature corpus) and assembles them into a unified research context. This context includes provenance records that track where each piece of evidence originated and at what level of confidence, so that every claim subsequently introduced in the manuscript can be traced to a specific experiment, analysis, or literature source.


With this context in place, the system constructs a narrative framework that specifies the central question of the manuscript, core tension, contribution statement, and section-level argument logic. The framework also defines explicit conditions under which the main claims of the paper would not hold. The system then performs an evidence audit, systematically identifying gaps between the available research outputs and the evidential requirements of the intended narrative. These gaps include missing control conditions, absent robustness checks, insufficiently supported figures, and experimental configurations that require additional simulation runs. Each identified gap is classified by category, priority, and whether it can be automatically addressed through re-analysis or requires human intervention.


In the manuscript construction stage, the system produces a structured claim register in which every factual, causal, or comparative claim is explicitly recorded with its supporting evidence sources, linked figures, and an allowed wording strength derived from the confidence level of the underlying evidence. In parallel, a figure--argument alignment plan maps each figure to the specific claims it supports, the research questions it answers, and the alternative explanations it addresses. This dual structure (claim register and figure--argument map) serves as the blueprint from which the manuscript is drafted section by section. It helps ensure that the narrative remains grounded in verifiable evidence and that every quantitative display serves a defined argumentative role.


The manuscript draft then enters an adversarial review phase, where multiple reviewer agents examine the text from distinct critical perspectives. One reviewer calibrates significance claims against the strength of the underlying evidence, flagging overstatements and unsupported causal language. A second reviewer checks the precision of the writing, verifying numerical consistency, terminology alignment with the literature, and methodological clarity. A third reviewer evaluates the evidence base itself, questioning whether alternative explanations have been adequately addressed and whether the presented analyses genuinely support the conclusions drawn. Each reviewer produces a structured verdict that is routed to the appropriate revision stage. Local wording adjustments return to text editing, structural issues trigger section-level reorganization, evidence weaknesses reroute to additional analysis or experimental proposals, and framing problems escalate to narrative reconstruction. To prevent unbounded revision cycles, the system enforces per-round caps on revision actions such as figure regeneration and literature augmentation, automatically escalating to a human decision gate when these caps are exceeded.


After revisions are resolved, the manuscript enters a release gate that evaluates seven criteria: metadata completeness, narrative coherence, claim--evidence integrity, figure completeness, closure of the review round, absence of pending human decisions, and successful compilation of a valid PDF. Only when all criteria are satisfied is the manuscript marked as ready for release. The final compilation stage converts the structured manuscript into LaTeX format, resolving citations against the curated bibliography, packaging figures into publication-quality layouts, and rendering the document through a standard academic LaTeX compilation pipeline using a target journal template. The output is a camera-ready PDF with all associated source files preserved in the workspace for inspection, revision, and resubmission.


Taken together, the paper generation module serves as the capstone of the AI social scientist pipeline. It converts the cumulative outputs of literature search, hypothesis generation, experiment design, simulation execution, and result analysis into a scholarly manuscript that meets the structural and evidentiary standards of academic publication. By decomposing paper writing into structured narrative planning, evidence-constrained drafting, multi-perspective adversarial review, and tiered quality gating, the system helps ensure that generated manuscripts are not merely fluent but scientifically accountable: each claim traces back to the underlying simulation experiments and analyses. With the preceding stages, the paper generation module closes the loop from initial research question to submission-ready manuscript, moving toward an end-to-end agentic research workflow for computational social science.

\section{Agentic Data Foundation}
\label{sec:data-foundation}

\subsection{Overview}
\label{subsec:data-motivation}

Data assets provide the empirical grounding for agentic social science. Their role can be understood at three connected levels: they support social science research, enable simulation experiments, and provide the basis for constructing social generative agents.

\textbf{Data grounds social science research.}
Social science research relies on data to support empirical observation, mechanism modeling, and result interpretation. In \textit{AgentSociety$^2$} , this requirement becomes more fundamental because the AI Social Scientist coordinates a closed-loop research workflow from topic formulation, literature retrieval, hypothesis development, and experiment design to simulation execution, data analysis, and paper generation. Data assets therefore function as the infrastructure that connects real-world social evidence, agent-based simulation, and scientific conclusion generation.

\textbf{Data grounds simulation experiments.}
The key problem is how to transform heterogeneous real-world data into agentic research assets that can be discovered, interpreted, processed, and reused by the AI Social Scientist. Simulation experiments require empirical data to construct executable social scenarios, including individual attributes, population structures, social relations, environmental states, external events, and historical behavioral traces. Without such empirical grounding, simulations would rely primarily on abstract assumptions or manually written personas, making it difficult to represent heterogeneity and structural constraints in real social systems. At the same time, simulation outputs need to be compared with real-world data for analysis and interpretation. Researchers must assess whether a simulated mechanism can reproduce empirical patterns, identify agreements and deviations between simulated and observed outcomes, and use these comparisons to derive testable and interpretable social-scientific conclusions.

\textbf{Data grounds social generative agents.}
This problem is particularly important for social generative agents. A social generative agent should not be constructed solely from a simple role prompt. Instead, it should maintain stable identity attributes, dynamic internal states, experiential memory, and an iterative process of state update and behavioral decision making. Data assets enter agent construction through two complementary channels. The subjective channel initializes the inner states of agents, including preferences, values, risk perception, consumption habits, social trust, knowledge background, memory materials, and behavioral tendencies. These variables shape how agents interpret experimental situations, respond to external stimuli, form judgments, and select actions. The objective channel initializes the externally observable characteristics of agents, including age, gender, occupation, income, education, residence, household structure, spatial location, historical behavior records, and device conditions. These variables determine observable attributes, resource conditions, action boundaries, and interaction opportunities in the simulation environment. Through these two channels, data assets transform individual differences, population structures, and social environments from the real world into configurable, traceable, and analyzable experimental objects.

\subsection{From Heterogeneous Data to Agentic Data Assets}
\label{subsec:data-challenges}

Once data assets are positioned as part of the research infrastructure, \textit{AgentSociety$^2$}  must be able to incorporate datasets produced by different institutions, research communities, platforms, and measurement systems. These datasets were not originally created for LLM-based research agents, nor do they follow a shared representation standard. Before they can become usable research assets, external data integration must address three practical bottlenecks.

\textbf{Format heterogeneity.}
Available datasets appear in highly heterogeneous formats. They include structured files such as CSV, JSON, and Parquet, as well as multimodal or semi-structured materials such as text, images, audio, trajectories, survey records, and experimental logs. A rigid data standard would simplify management, but it would also force diverse data sources into a narrow representation, losing information and increasing preprocessing costs.

\textbf{Semantic opacity.}
Raw data fields are often difficult for LLMs to understand directly. Field names may contain abbreviations, business codes, internal identifiers, or implicit units, and relationships among fields may depend on contextual knowledge held by the data producer. If the AI Social Scientist directly reads such files, it may misinterpret field meanings, ignore units or constraints, or connect variables incorrectly.

\textbf{Dataset discovery and reuse.}
Dataset acquisition and discovery become difficult at scale. As the number of available datasets grows, both researchers and the AI Social Scientist need to identify datasets suitable for a given research topic, experimental objective, variable definition, and analytical requirement. Without a shared registry and searchable metadata, high-quality datasets remain scattered across local file systems, project folders, or private repositories, making reuse difficult and non-reproducible.

These challenges suggest that data management in \textit{AgentSociety$^2$}  cannot be treated as ordinary file organization. Datasets need to be represented in a form that preserves their original structure while making them accessible to an AI research agent. We therefore introduce a three-layer data foundation design that connects asset packaging, semantic documentation, and metadata-aware discovery.

Asset packaging stores datasets as standardized-yet-flexible asset packages. This design provides a common entry point, documentation structure, validation procedure, and distribution format without forcing heterogeneous data into a single schema. It allows raw data files and auxiliary access scripts to preserve their native forms, making heterogeneous datasets manageable without flattening them into an overly restrictive representation.

Semantic documentation makes dataset documentation part of the operational procedure. The AI Social Scientist follows a README-first usage pattern before reading fields, running scripts, or generating extraction code. Field meanings, units, file relations, usage constraints, and recommended access procedures are externalized into an explicit context for AI reasoning and code execution, reducing the risk of semantic misreading.

Metadata-aware discovery registers datasets through online services. Upload, registration, browsing, search, download, version management, and reuse become reproducible steps in the research workflow. Metadata further enables the AI Social Scientist to retrieve candidate datasets according to research topics, experiment types, variable requirements, and tag information. Together, these choices shift datasets from passive external files to AI-operable research assets.

\subsection{Dataset Asset Framework and Workflow}
\label{subsec:data-framework-workflow}

\begin{figure}[t]
    \centering
    \includegraphics[width=\linewidth]{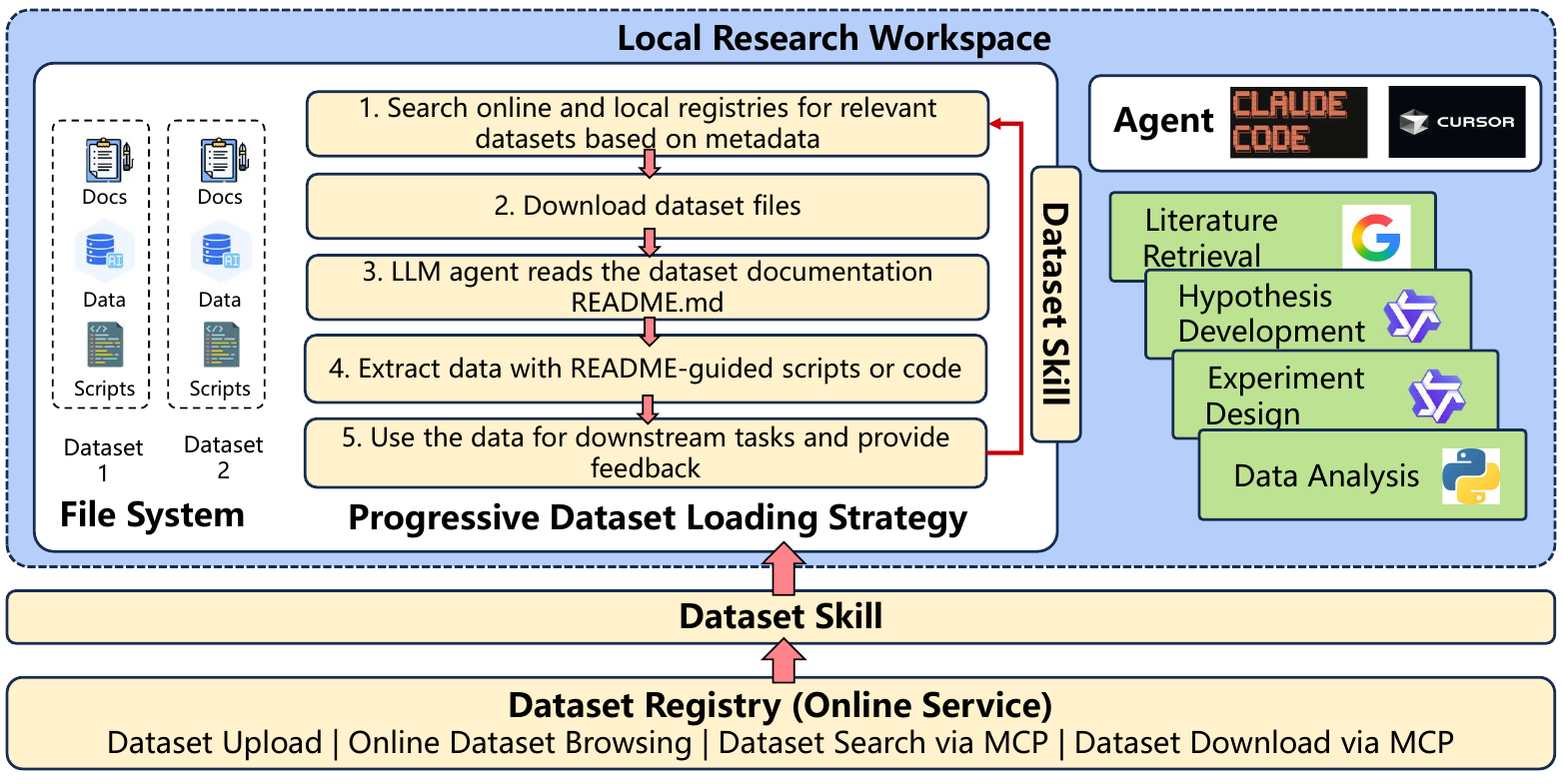}
    \caption{Dataset workflow in \textit{AgentSociety$^2$} . Datasets are uploaded, registered, searched, viewed, and downloaded through the online registry, and are progressively loaded into the local research workspace through dataset skills.}
    \label{fig:dataset-workflow}
\end{figure}

As shown in Fig.~\ref{fig:dataset-workflow}, the dataset asset framework connects the online dataset registry, the local research workspace, and the AI Social Scientist through a progressive workflow. Datasets are first uploaded, registered, searched, viewed, and downloaded through the online registry, and are then loaded into the local workspace through dataset skills. After entering the workspace, they are not exposed to the AI Social Scientist as raw files all at once. Instead, they are processed through a sequence of metadata-based discovery, package acquisition, documentation reading, README-guided extraction, and downstream task use. This workflow makes dataset use a staged research procedure rather than an ad hoc file-reading process.

\subsubsection{AgentSociety Dataset Asset Package}

The starting point of the data foundation is the dataset asset package. Each dataset is organized as a standardized-yet-flexible package containing a documentation file, a metadata file, and a data directory, instantiated as \texttt{README.md}, \texttt{dataset.json}, and \texttt{data/}. The \texttt{data/} directory may contain CSV, JSON, Parquet, multimodal files, auxiliary access scripts, or other necessary resources, preserving the native representation of data from different sources. At the same time, \texttt{README.md} and \texttt{dataset.json} provide a unified interpretive layer and retrieval layer. This package structure allows heterogeneous datasets to share a common entry point, validation procedure, and distribution format without forcing them into a single tabular schema.

The package design is intentionally minimal. Rather than imposing a universal data schema, it defines the necessary boundary conditions for an AI-operable dataset asset: the dataset must have explicit documentation, machine-readable metadata, and an organized data directory. This makes the package flexible enough to cover survey data, mobility traces, social-media networks, behavioral logs, financial records, and multimodal resources, while still ensuring that the AI Social Scientist can locate, understand, validate, and use the dataset in a reproducible way.

A central feature of this package design is the README-first usage pattern. The AI Social Scientist reads \texttt{README.md} before directly accessing fields, running scripts, or generating extraction code. The README explains what the dataset contains, what file formats and schemas are used, what each field means, what units are involved, how files are related, and how the dataset should be used in research. For complex datasets, the package may also include auxiliary access scripts, enabling the AI to perform cleaning, filtering, sampling, transformation, and feature generation after understanding the documentation. In this way, dataset documentation becomes an operational context for AI reasoning and code execution, reducing the risk of semantic misreading.

\subsubsection{Online Dataset Registry}

Once a dataset is packaged, it can enter the online registry. The registry functions as the circulation layer of the data foundation: datasets can be uploaded, registered, browsed, searched, downloaded, and versioned through the online service. Each dataset declares its unique id, name, description, category, version, tags, author, and license in \texttt{dataset.json}. The category field covers common research data types, including \texttt{agent\_profiles}, \texttt{surveys}, \texttt{experiments}, \texttt{literature}, \texttt{simulation\_results}, and \texttt{other}.

With this metadata layer, the AI Social Scientist can retrieve candidate datasets according to the research topic, experiment type, variable requirements, and tag information before deciding whether to download and inspect a dataset. The online registry therefore turns dataset discovery from an informal search over local folders or project-specific repositories into a reproducible step in the research workflow. It also supports reuse across different experiments and research projects, because datasets are no longer tied only to the original context in which they were collected or processed.

\subsubsection{Local Dataset Skills}

To make the dataset asset workflow executable by an LLM-based research agent, we design two local dataset skills that automate the two directions of the data asset lifecycle: \textbf{Dataset Creation Skill} for producing and registering data assets, and \textbf{Dataset Use Skill} for discovering, interpreting, extracting, and reusing them. These skills do not introduce a new generic skill paradigm; rather, they instantiate the skill-based harness of \textit{AgentSociety$^2$}  for data management, turning dataset production and dataset consumption into constrained, documented, and auditable procedures.

On the production side, Dataset Creation Skill initializes the dataset directory, generates the basic package structure, guides the user or the AI to add data files and complete field descriptions, checks the documentation file, metadata file, data directory, and package-size constraints, and finally packages and submits the dataset to the online registry. Through this process, scattered raw data are transformed into registered data assets with explicit semantic documentation, searchable metadata, and a distributable package.

On the consumption side, Dataset Use Skill progressively loads data assets into the local research workspace. It first searches online or local registries based on metadata, then downloads the selected dataset package when necessary. Following the README-first principle defined by the dataset package, it interprets the documentation before inspecting specific files and extracting data with README-guided scripts or code. The processed data can then be used for experiment initialization, agent profile construction, environment state configuration, behavioral calibration, result analysis, or other downstream research tasks.

\subsection{Current Dataset Assets}
\label{subsec:current-datasets}

Following the dataset asset format described above, we have organized an initial collection of reusable datasets for \textit{AgentSociety$^2$} . These datasets cover household surveys, demographic microdata, synthetic agent profiles, mobility benchmarks, disaster-response mobility, social-media networks, short-video consumption behavior, financial ownership data, corporate news events, and high-frequency stock market records. They provide empirical foundations for agent initialization, environment construction, behavioral calibration, result comparison, and downstream data analysis. Table~\ref{tab:current-dataset-assets} summarizes the current dataset assets and their basic information.

\begin{table*}[htbp]
\centering
\small
\setlength{\tabcolsep}{5pt}
\renewcommand{\arraystretch}{1.08}
\caption{Current dataset assets in \textit{AgentSociety$^2$} .}
\label{tab:current-dataset-assets}
\begin{tabular}{M{0.16\linewidth} M{0.32\linewidth} M{0.44\linewidth}}
\toprule
\textbf{Category} & \textbf{Dataset} & \textbf{Content} \\
\midrule

\multirow[c]{5}{=}[-5.0ex]{Survey and demographic data}
& ACS 2023 American Community Survey -- PUMS~\cite{census2023acs_pums}
& Housing and person microdata for U.S. population and household characteristics. \\
\cmidrule(l){2-3}
& CES 2023 Consumer Expenditure Survey -- Interview Survey~\cite{bls2023ce_pumd}
& Quarterly consumer-unit expenditure, income, and demographic records. \\
\cmidrule(l){2-3}
& PSID 2023 Family Survey~\cite{psid2023_family}
& Family-level income, employment, education, wealth, housing, and consumption variables. \\
\cmidrule(l){2-3}
& SHED 2023 Survey of Household Economics and Decisionmaking~\cite{fed2023shed}
& Household financial well-being, income stability, savings, debt, housing, and economic sentiment. \\
\cmidrule(l){2-3}
& SIPP 2023 Survey of Income and Program Participation~\cite{census2023sipp}
& Monthly income, program participation, assets, debts, employment, and insurance records. \\
\midrule

Synthetic population and spatial data
& US Census CBG Features and Synthetic Agent Profiles, ACS 2015--2019~\cite{census2019acs5year}
& CBG-level features and synthetic agent profiles for demographic and spatial initialization. \\
\midrule

\multirow[c]{2}{=}[-1.2ex]{Mobility data}
& Daily Mobility Ground Truth -- Beijing Users
& Ground-truth mobility indicators, including gyration radius, visited locations, and intention sequences. \\
\cmidrule(l){2-3}
& US Disaster Mobility -- Texas Winter Storm and Camp Fire
& De-seasonalized daily mobility outflux indices for two major U.S. hazard events. \\
\midrule

\multirow[c]{3}{=}[-2.4ex]{Social media behavior}
& Twitter Retweet Networks and Media Bias Labels 2016
& Retweet edges, media-bias labels, influencer metadata, and tweet-ID resources. \\
\cmidrule(l){2-3}
& Twitter Retweet Networks and Media Bias Labels 2020
& Retweet networks, tweet-level media-bias labels, influencer metadata, and tweet-ID resources. \\
\cmidrule(l){2-3}
& Short-Video Consumption Datasets
& Item metadata, user profiles, interaction logs, watch histories, and category mappings. \\
\midrule

\multirow[c]{3}{=}[-2.6ex]{Finance data}
& FactSet Ownership Database~\cite{factsetOwnership}
& Institutional ownership, 13F filings, fund holdings, entity profiles, and ownership analytics. \\
\cmidrule(l){2-3}
& FinnHub News Events~\cite{finnhubCompanyNews}
& Corporate news articles around major economic events, grouped by company, date, and session. \\
\cmidrule(l){2-3}
& FinnHub US Stock OHLC Data~\cite{finnhubStockCandles}
& U.S. stock OHLC records at 1-minute and 5-minute intervals across multiple years. \\
\bottomrule
\end{tabular}
\end{table*}

\section{Platform Performance}
In \textit{AgentSociety$^2$} , the LLM inference cost associated with agents' ReAct loops is unavoidable, while the main additional overhead introduced by the platform's computational framework comes from the natural language interface abstraction used to unify agent--environment interaction.
Ensuring the efficiency of this abstraction is therefore central to the platform's overall performance.
The unified interface improves the composability of environment modules, yet it also introduces two performance risks: natural-language instructions may be incorrectly mapped to environment function calls, and repeated LLM-based routing may create additional token overhead in large-scale simulations.
Thus, we conducted experiments to validate whether the agentic environments in \textit{AgentSociety$^2$}  can efficiently handle environment function calls across various module combinations.
The experiments compare the call correctness and token consumption of CodeGenRouter with baseline methods based on function calling techniques.

\subsection{Benchmark Design}

To evaluate the performance of the \textit{AgentSociety$^2$}  environment framework under realistic usage scenarios, we need to construct a benchmark that reflects these scenarios and define appropriate evaluation metrics.
In this subsection, we describe the benchmark test-case composition, metric selection, baseline methods, and LLM model selection in sequence.

\textbf{Test Case Construction.}
We constructed the benchmark from records of agent calls to the environment interface during actual simulation runs, rather than from artificially authored instructions.
The benchmark is designed to evaluate whether a routing method can recover the correct environment functions, preserve the required calling order, and ground arguments in the structured context provided by the agent.

Each agent input was labeled with ground truth, including the environment module functions to be called, function parameters, and call order.
During annotation, we further recorded four aspects of task complexity: the set of involved environment modules, temporal ordering requirements among calls, required preconditions such as observation or query before action, and argument determinism, which distinguishes explicit parameters from values inferred from context.
Based on these criteria, we categorized the inputs into six difficulty levels according to combination complexity and performed inter-level quantity balancing.
Table~\ref{tab:routing_benchmark_levels} summarizes these difficulty levels.

\begin{table}[t]
\centering
\caption{Difficulty levels of the routing benchmark.}
\label{tab:routing_benchmark_levels}
\begin{tabularx}{\linewidth}{cll}
\toprule
\textbf{Level} & \textbf{Task type} & \textbf{Complexity criterion} \\
\midrule
1 & State observation & Single-step status retrieval. \\
2 & Discovery and query & Information seeking before action. \\
3 & Atomic action & Direct execution within one environment module. \\
4 & Sequential workflow & Multi-step dependencies within a coherent workflow. \\
5 & Two-module composition & Cross-module coordination involving two modules. \\
6 & Three-module composition & Full orchestration across three modules. \\
\bottomrule
\end{tabularx}
\end{table}

\textbf{Evaluation Metrics.}
To assess call correctness, we evaluate function selection, call ordering, parameter grounding, and strict end-to-end success.
Let $E$ and $A$ denote the expected and actual call sets, respectively, and let $S_E$ and $S_A$ denote the expected and actual function sequences.
We first use function selection IoU to measure whether the correct functions are selected while ignoring order:
\[
\mathrm{IoU}(E,A)=\frac{|F(E)\cap F(A)|}{|F(E)\cup F(A)|},
\]
where $F(\cdot)$ denotes the set of function names in a call set.

We then use normalized longest common subsequence (NLCS) to evaluate whether the predicted calls follow the correct sequence:
\[
\mathrm{NLCS}(S_A,S_E)=\frac{\mathrm{LCS}(S_A,S_E)}{|S_E|},
\]
where $\mathrm{LCS}$ is the longest common subsequence length.

For parameter grounding, we compute parameter accuracy by comparing required arguments in each expected call with the best-matching actual call:
\[
\mathrm{PAcc}=\frac{1}{|E|}\sum_{e\in E}
\frac{\#\ \mathrm{matched\ parameters}(e)}{\#\ \mathrm{required\ parameters}(e)}.
\]
Finally, successful call ratio (SR) is the strictest metric.
It requires both a perfectly matched call sequence and perfectly matched parameters:
\[
\mathrm{SR} =
\mathbb{I}\left[
\mathrm{NLCS}(S_A,S_E)=1
\land
\mathrm{PAcc}=1
\right].
\]

\textbf{Baseline Methods and Experimental Models.}
To contrast generative code execution with common function-calling techniques, we compare CodeGenRouter with five baselines.
ReAct~\cite{yao2022react} represents the standard reasoning-action loop, while PlanExecute~\cite{wang-etal-2023-plan} first generates a complete action plan, executes it, and regenerates the plan when necessary.
SearchTool~\cite{braunschweiler2025toolreagt} allows the LLM to search over candidate functions before making calls.
Based on the above methods, we further construct two baselines that adopt a hierarchical selection strategy.
HierarchicalReAct first selects an environment module and then chooses the next function within that module.
HierarchicalPlanExecute similarly selects an environment module first and then generates an executable plan for the selected module.

Following BFCLv4~\cite{patilberkeley}, we selected models that perform well on function-calling tasks while covering open-source and closed-source systems as well as large and smaller model scales.
The evaluated models include GLM-4.7~\cite{glm2024chatglm}, Claude Opus 4.5, Kimi-K2-Instruct~\cite{team2025kimi}, Qwen3-235B-A22B-Instruct, Qwen3-32B~\cite{yang2025qwen3}, and Gemini 2.5 Flash~\cite{comanici2025gemini}.

\subsection{Performance Results}

\begin{table*}[thpb!]
\centering
\caption{Complete experimental results across all baseline methods and models. Bold and underlined values indicate optimal and suboptimal results under each model respectively.}
\label{tab:platform_perf_full}
\setlength{\tabcolsep}{0.5mm}
\resizebox{\textwidth}{!}{%
\begin{tabular}{llrrrrrrrr}
\hline
\multirow{2}{*}{Method} & \multirow{2}{*}{Model} & \multicolumn{5}{c}{Performance} & \multicolumn{3}{c}{Cost (Per Test)} \\
\cline{3-7} \cline{8-10}
& & IoU $\uparrow$ & NLCS $\uparrow$ & PAcc $\uparrow$ & SR $\uparrow$ & SR Imp. $\uparrow$ & Calls $\downarrow$ & InTok $\downarrow$ & OutTok $\downarrow$ \\
\hline
CodeGen & GLM-4.7 & \textbf{0.771} & \textbf{0.886} & 0.850 & \textbf{0.578} & +25.4\% & \underline{1.2} & 13775.2 & 1180.4 \\
ReAct & GLM-4.7 & 0.467 & 0.489 & 0.490 & 0.294 & -- & 2.0 & 7300.1 & \textbf{506.5} \\
PlanExecute & GLM-4.7 & 0.615 & 0.709 & \textbf{0.883} & 0.456 & -- & \textbf{1.1} & \underline{6674.3} & \underline{880.4} \\
HierarchicalReAct & GLM-4.7 & 0.497 & 0.511 & 0.531 & 0.272 & -- & 6.9 & 9913.6 & 3563.7 \\
HierarchicalPlanExecute & GLM-4.7 & 0.634 & 0.713 & \underline{0.851} & 0.400 & -- & 2.8 & \textbf{6077.8} & 1943.8 \\
SearchTool & GLM-4.7 & \underline{0.642} & \underline{0.773} & 0.778 & \underline{0.461} & -- & 6.7 & 42266.8 & 1408.2 \\
\hline
CodeGen & Claude Opus 4.5 & \textbf{0.751} & \textbf{0.845} & 0.824 & \textbf{0.533} & +11.5\% & \underline{1.4} & 16394.6 & 527.2 \\
ReAct & Claude Opus 4.5 & 0.448 & 0.446 & 0.489 & 0.267 & -- & 2.2 & 8041.8 & \underline{292.6} \\
PlanExecute & Claude Opus 4.5 & 0.575 & 0.773 & \textbf{0.913} & 0.422 & -- & \textbf{1.1} & \underline{6914.1} & 431.0 \\
HierarchicalReAct & Claude Opus 4.5 & 0.504 & 0.497 & 0.542 & 0.311 & -- & 6.8 & 9144.0 & 368.3 \\
HierarchicalPlanExecute & Claude Opus 4.5 & 0.585 & \underline{0.784} & \underline{0.867} & \underline{0.478} & -- & 2.8 & \textbf{6625.4} & 518.1 \\
SearchTool & Claude Opus 4.5 & \underline{0.628} & 0.710 & 0.687 & 0.433 & -- & 5.5 & 21389.6 & \textbf{288.4} \\
\hline
CodeGen & Kimi-K2-Instruct & \textbf{0.757} & \textbf{0.872} & \textbf{0.830} & \textbf{0.600} & +31.6\% & \underline{1.4} & 15482.6 & 435.9 \\
ReAct & Kimi-K2-Instruct & 0.431 & 0.428 & 0.427 & 0.267 & -- & 1.9 & \textbf{6201.1} & \textbf{159.5} \\
PlanExecute & Kimi-K2-Instruct & 0.465 & 0.587 & 0.608 & 0.339 & -- & \textbf{1.1} & 6997.8 & \underline{332.8} \\
HierarchicalReAct & Kimi-K2-Instruct & 0.437 & 0.433 & 0.464 & 0.283 & -- & 5.9 & 7723.3 & 363.0 \\
HierarchicalPlanExecute & Kimi-K2-Instruct & 0.541 & 0.693 & \underline{0.804} & 0.378 & -- & 2.8 & \underline{6307.8} & 379.9 \\
SearchTool & Kimi-K2-Instruct & \underline{0.641} & \underline{0.768} & 0.728 & \underline{0.456} & -- & 6.1 & 25171.5 & 558.9 \\
\hline
CodeGen & Qwen3-235B-A22B-Instruct & \textbf{0.657} & \textbf{0.729} & 0.788 & \textbf{0.478} & +2.4\% & \textbf{1.1} & 12111.9 & 476.4 \\
ReAct & Qwen3-235B-A22B-Instruct & 0.499 & 0.509 & 0.504 & 0.322 & -- & 4.8 & 12719.5 & 490.3 \\
PlanExecute & Qwen3-235B-A22B-Instruct & 0.492 & \underline{0.726} & \textbf{0.922} & \underline{0.467} & -- & \underline{1.2} & \textbf{7155.9} & 446.8 \\
HierarchicalReAct & Qwen3-235B-A22B-Instruct & 0.532 & 0.573 & 0.591 & 0.322 & -- & 9.6 & 13575.9 & \underline{393.2} \\
HierarchicalPlanExecute & Qwen3-235B-A22B-Instruct & 0.489 & 0.711 & \underline{0.864} & 0.339 & -- & 3.3 & \underline{7505.3} & 585.4 \\
SearchTool & Qwen3-235B-A22B-Instruct & \underline{0.616} & 0.691 & 0.661 & 0.406 & -- & 5.8 & 22162.5 & \textbf{305.0} \\
\hline
CodeGen & Qwen3-32B & \textbf{0.665} & \textbf{0.732} & 0.705 & \textbf{0.411} & +0.0\% & \underline{1.5} & 17350.5 & 585.2 \\
ReAct & Qwen3-32B & 0.513 & 0.506 & 0.558 & 0.339 & -- & 2.0 & \textbf{6193.3} & \textbf{83.0} \\
PlanExecute & Qwen3-32B & \underline{0.606} & \underline{0.705} & \textbf{0.884} & \underline{0.411} & -- & \textbf{1.1} & 6725.0 & 303.5 \\
HierarchicalReAct & Qwen3-32B & 0.536 & 0.533 & 0.630 & 0.356 & -- & 5.6 & 7292.8 & \underline{143.2} \\
HierarchicalPlanExecute & Qwen3-32B & 0.562 & 0.686 & \underline{0.810} & 0.367 & -- & 3.0 & \underline{6417.2} & 432.7 \\
SearchTool & Qwen3-32B & 0.603 & 0.674 & 0.700 & 0.394 & -- & 5.5 & 21543.1 & 167.4 \\
\hline
CodeGen & Gemini 2.5 Flash & \textbf{0.763} & \textbf{0.850} & 0.843 & \textbf{0.556} & +17.8\% & \underline{1.5} & 17416.8 & 575.2 \\
ReAct & Gemini 2.5 Flash & 0.453 & 0.449 & 0.479 & 0.261 & -- & 2.3 & 7808.1 & \textbf{281.2} \\
PlanExecute & Gemini 2.5 Flash & 0.562 & 0.753 & \textbf{0.923} & 0.428 & -- & \textbf{1.1} & \underline{6980.6} & 445.9 \\
HierarchicalReAct & Gemini 2.5 Flash & 0.491 & 0.488 & 0.552 & 0.311 & -- & 7.0 & 9593.1 & 365.7 \\
HierarchicalPlanExecute & Gemini 2.5 Flash & 0.586 & \underline{0.791} & \underline{0.848} & \underline{0.472} & -- & 2.8 & \textbf{6627.2} & 514.4 \\
SearchTool & Gemini 2.5 Flash & \underline{0.634} & 0.723 & 0.694 & 0.433 & -- & 5.6 & 20838.1 & \underline{294.1} \\
\hline
\end{tabular}%
}
\begin{flushleft}
\footnotesize
\textit{Abbreviations:} IoU = Intersection over Union; NLCS = Normalized Longest Common Subsequence; PAcc = Parameter Accuracy; SR = Successful call ratio; SR Imp. = CodeGen\textquotesingle s improvement percentage over the best baseline; Calls = Average LLM calls per test; InTok / OutTok = Average input/output tokens per test.
\end{flushleft}
\end{table*}

Table~\ref{tab:platform_perf_full} presents the complete experimental results across all baseline methods and models.
CodeGenRouter achieves substantial advantages over function calling methods in terms of successful call ratio.
With Kimi-K2-Instruct, it reaches 60\% successful call ratio, representing a 31.6\% improvement over the best baseline.
With GLM-4.7, it achieves 57.8\% successful call ratio, representing a 25.4\% improvement.

\begin{figure}[t]
\centering
\begin{subfigure}[t]{0.45\linewidth}
    \centering
    \includegraphics[width=\linewidth]{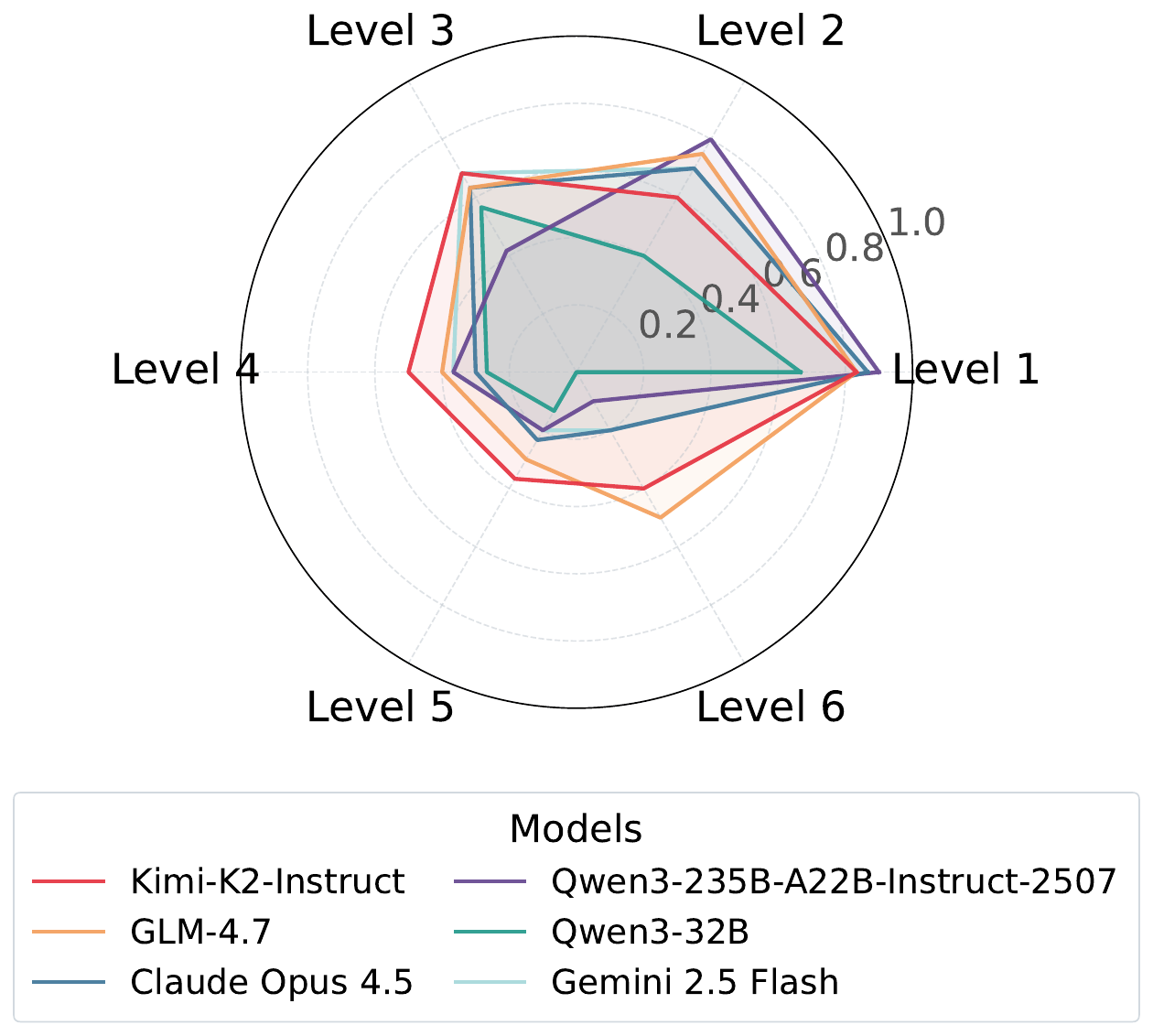}
    \caption{}
    \label{fig:platform_codegen_radar}
\end{subfigure}
\hfill
\begin{subfigure}[t]{0.45\linewidth}
    \centering
    \includegraphics[width=\linewidth]{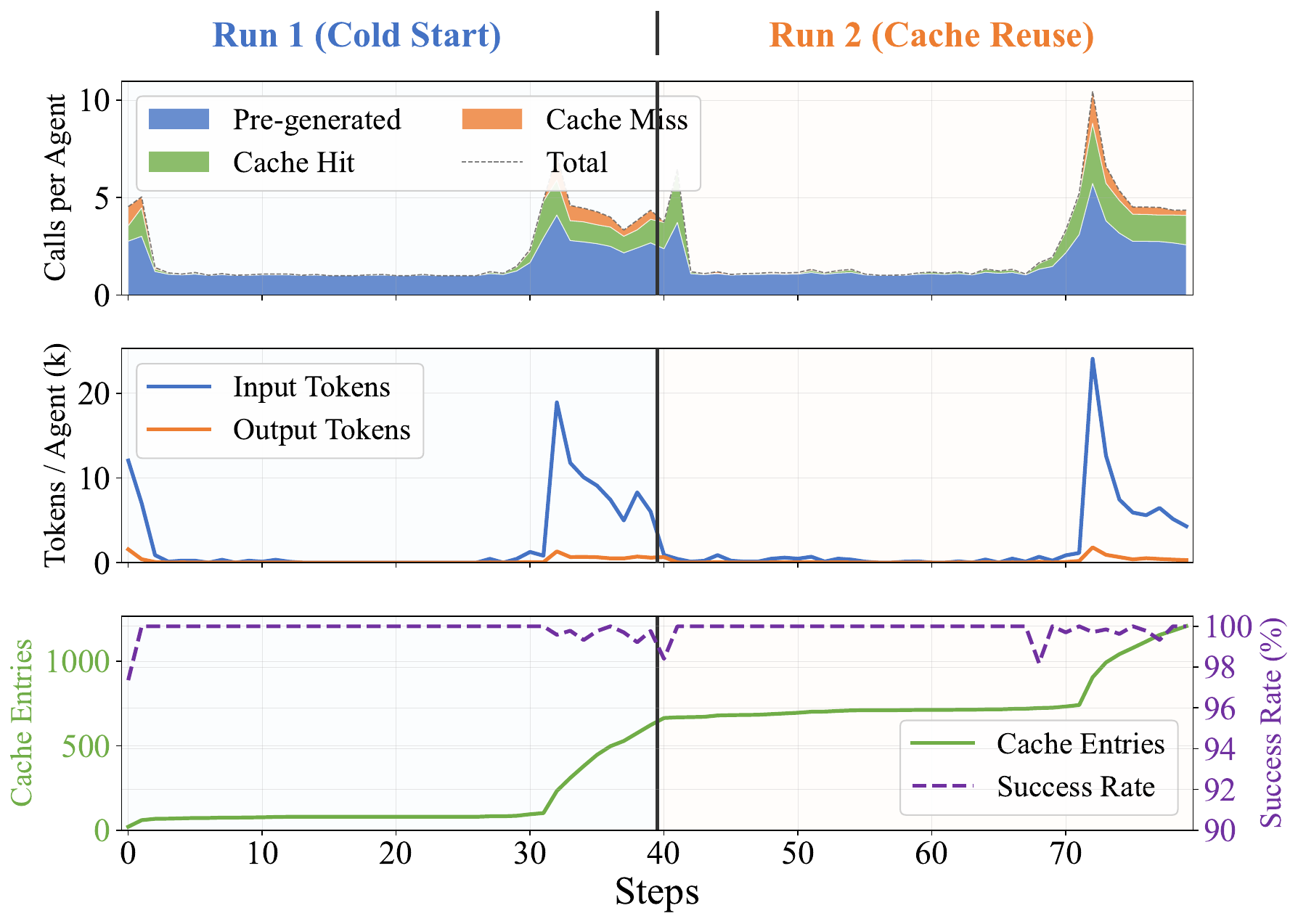}
    \caption{}
    \label{fig:platform_cache}
\end{subfigure}
\caption{(a) CodeGenRouter performance across benchmark difficulty levels; (b) Cache reuse reduces API calls and token consumption across simulation runs.}
\label{fig:platform_performance}
\end{figure}

The method brings no significant increase in the number of calls and output tokens, with increases primarily in input token usage.
In practice, this overhead can be further reduced through code caching mechanisms.
As shown in Figure~\ref{fig:platform_codegen_radar}, by comparing the performance of different models across difficulty levels, we find that all tested language models achieve similar performance on easier test cases, which avoids vendor binding risks in practical deployment.
However, harder test cases still pose challenges to model capabilities.

\subsection{Overhead Analysis}

We further evaluated the effectiveness of pre-generation and caching mechanisms in reducing framework overhead by simulating the daily behavioral activities of 100 agents.
The agent design follows \textit{AgentSociety} principles and adapts to the unified agent-environment interface.

We performed two rounds of simulation, with the first serving as a cold start for the caching mechanism and the second testing cache reuse effects.
Each round simulates 40 steps corresponding to the time period from midnight to 10 AM, covering both active and inactive periods of agent behaviors.

As shown in Figure~\ref{fig:platform_cache}, the code pre-generation mechanism handles more than 70\% of agent calls, with 74.1\% in the first run and 70.5\% in the second run.
After completing the cold start, calls at the beginning of the second run are mostly resolved by cache while maintaining high execution success rates.
Even without the cold start process, code reuse among agents in the first run reduces call requirements by 66.5\%.

The average additional token overhead per step per agent across both runs is 2,292 input tokens and 184 output tokens.
These results demonstrate that the efficiency improvement methods included in \textit{AgentSociety$^2$}  can significantly reduce introduced overhead while maintaining system efficiency.
Observations also reveal that agent active periods may involve more diverse requests, meaning cache entries from only one cold start run cannot fully satisfy all agent calls.
This highlights the importance of continuous accumulation of code cache data.

Overall, these experiments show that the unified natural-language environment interface can be made practical for large-scale agent simulations.
CodeGenRouter improves the reliability of multi-module environment invocation by generating executable call sequences rather than relying on repeated function selection.
Meanwhile, pre-generation and semantic caching amortize the additional LLM routing cost across agents and simulation steps.
The results therefore support the central design claim that \textit{AgentSociety$^2$}  can preserve flexible agent-environment composition without sacrificing the efficiency required by repeated social simulation experiments.



\section{Multi-Scale Social Science Studies}

\begin{figure}[t!]
    \centering
    \includegraphics[width=\textwidth]{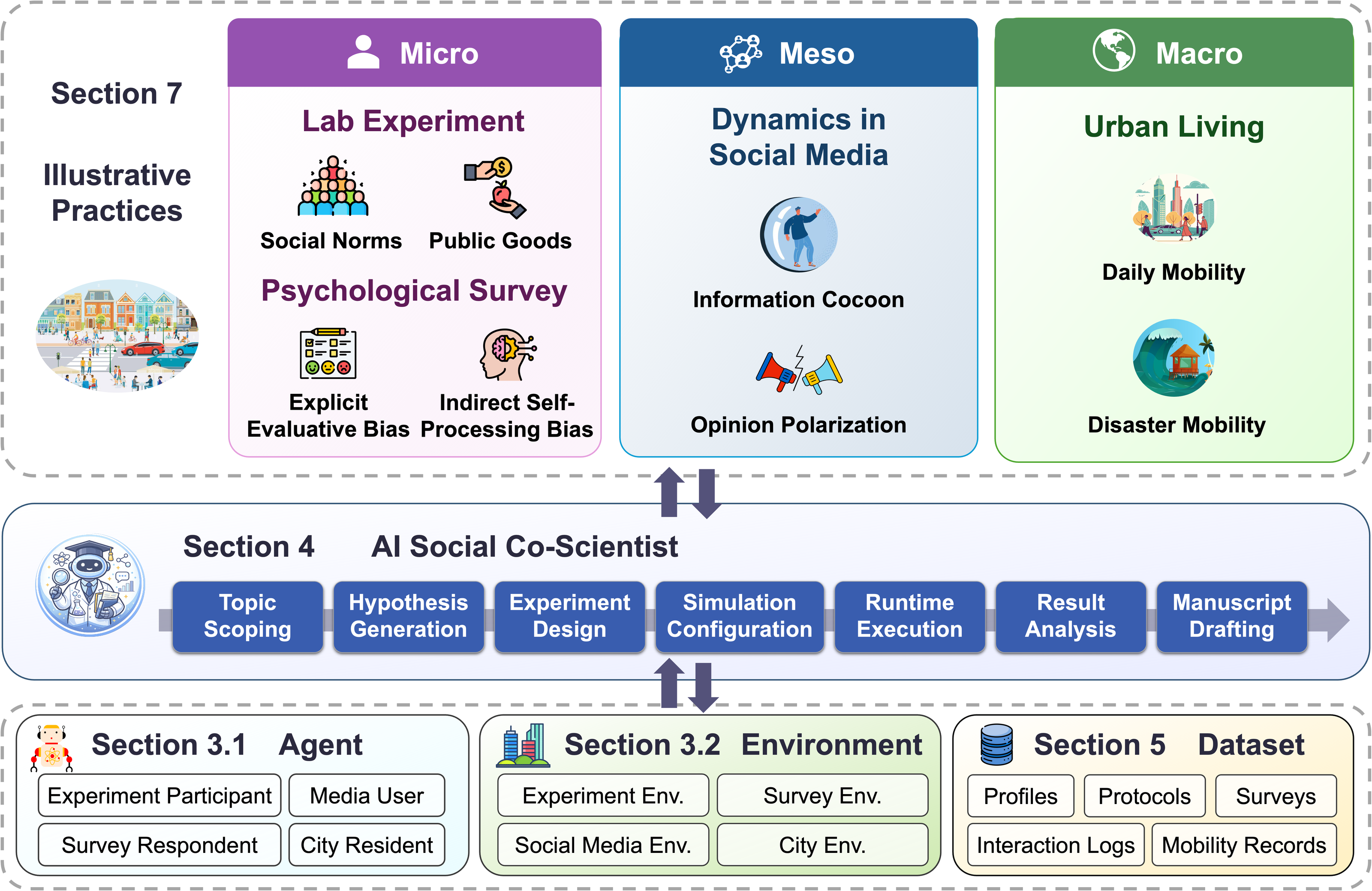}
    \caption{Overview of illustrative practices in AgentSociety$^{2}$. Env. stands for environment.}
    \label{fig:illustrative_practice_overview}
\end{figure}

To demonstrate the usability and adaptability of \textit{AgentSociety$^2$}, we present illustrative practices across micro-level behavioral studies, meso-level social network dynamics, and macro-level urban living simulations, as summarized in Figure~\ref{fig:illustrative_practice_overview}. These practices draw on classical experiments and key social phenomena to show the platform's capacity to support different forms of social science studies. At the micro-level, the platform supports two core research implementations, lab experiments and psychological surveys, that measure individual behavior and cognitive processes. Specifically, lab experiments are conducted through Axelrod’s Norms and Metanorms Games~\cite{1986An} and Fischbacher and Gächter’s public goods experiments~\cite{fischbacher2010social}, reproducing procedures to capture fundamental patterns of norm enforcement, cooperation, and free-riding. Psychological surveys replicate the comprehensive self-bias study by Qi et al.~\cite{qi2025comprehensive}, assessing agents' self-bias across both direct evaluative and indirect self-processing measures in comparison with human responses. At the meso-level, the platform models socially significant online dynamics, including information cocoon formation and opinion polarization, addressing globally relevant challenges through recommendation-driven exposure, user–content interaction, and ideology-scored consumption~\cite{shang2025large, levy2021social}. At the macro-level, \textit{AgentSociety$^2$} reproduces essential urban behaviors, including daily mobility as a fundamental human activity and disaster-response behavior as an emergent phenomenon under spatial, temporal, and event-driven constraints~\cite{uscensus2020acs, safegraph2019core, safegraph2021core}. Collectively, these illustrative cases demonstrate that \textit{AgentSociety$^2$} can support a wide range of studies and investigate social phenomena, while providing a flexible and extensible framework for large-scale research across diverse disciplines.


These practices cover the research workflow supported by \textit{AgentSociety$^2$}, from idea and hypothesis development to experiment design and execution, result analysis and interpretation, visualization, and manuscript drafting. They also illustrate how researchers can combine built-in platform components with customization for specific tasks. \textit{AgentSociety$^2$} provides \texttt{AgentBase} as the common abstraction for custom agents, \texttt{PersonAgent} as a reusable profile-based social agent, \texttt{EnvBase} as the base interface for custom environments, and router-based tool invocation for connecting agents with environment tools. Based on these abstractions, users can define and register new agents, environments, and skills under the same callable-tool interface. Table~\ref{tab:illustrative_components} summarizes the agent and environment components used across the illustrative practices, distinguishing built-in support from case specific extensions. This demonstrates that \textit{AgentSociety$^2$} can function both as an immediately usable experimental platform and as an extensible research environment for new agent-based social science studies.

\begin{sidewaystable*}[p]
\raggedright
\scriptsize
\caption{Agents, environments, and data used in the illustrative practices.}
\label{tab:illustrative_components}
\begin{tabularx}{\textheight}{
@{}
>{\raggedright\arraybackslash}p{2cm}
>{\raggedright\arraybackslash}p{3.3cm}
>{\raggedright\arraybackslash}p{5.5cm}
>{\raggedright\arraybackslash}p{5.5cm}
>{\raggedright\arraybackslash}X
@{}
}
\toprule
\textbf{Category} 
& \textbf{Practice} 
& \textbf{Agent} 
& \textbf{Environment} 
& \textbf{Data} \\
\midrule

\multirow[t]{2}{1.8cm}{Lab experiment}
& Social norms
& \textbf{Custom}: \texttt{NormsGameAgent}; stores \texttt{boldness}, \texttt{vengefulness}, payoff, and 6-bit evolutionary strategy states.
& \textbf{Custom}: \texttt{NormsGameEnv}; implements norm violation, first order punishment, metanorm punishment, selection, mutation, and population statistics.
& Experimental design and parameter settings are based on Axelrod's classic norm-evolution experiment~\cite{1986An}. \\
\cmidrule(lr){2-5}

& Public goods experiments
& \textbf{Custom}: \texttt{FGPCPublicGoodsLLMAgent}; represents participants in preference elicitation, belief updating, and repeated contribution decisions.
& \textbf{Custom}: \texttt{FGPCPublicGoodsEnv}; implements the two-stage P/C procedure, P$\rightarrow$C and C$\rightarrow$P order conditions, and records preferences, beliefs, contributions, and payoffs.
& Experimental protocol, task settings, and human response patterns are based on Fischbacher and G\"achter's public-goods experiment~\cite{fischbacher2010social}. \\

\midrule

\multirow[t]{2}{1.8cm}{Psychological survey}
& Evaluative biases
& \textbf{Built-in}: \texttt{PersonAgent}; injects real participant profiles without modifying the core agent class.
& \textbf{Custom}: \texttt{EndowmentEffectPaperEnv} for WTA/WTP pricing and \texttt{SelfEnhancementPaperEnv} for reverse coded percentile self-ratings.
& Participant profiles, experiment materials, and human responses are from the self-bias study by Qi et al.~\cite{qi2025comprehensive}. \\
\cmidrule(lr){2-5}

& Self-processing biases
& \textbf{Built-in}: \texttt{PersonAgent}; uses the same profile based respondent population as the evaluative bias tasks.
& \textbf{Custom}: \texttt{IATPaperEnv} for key press, reaction time, and accuracy trials; \texttt{SelfReferenceEffectPaperEnv} for trait encoding and recognition judgments with Remember/Know labels.
& Participant profiles, experiment materials, and human responses are from the self-bias study by Qi et al.~\cite{qi2025comprehensive}. \\

\midrule

\multirow[t]{2}{1.8cm}{Dynamics in social media}
& Information cocoon
& \textbf{Custom}: \texttt{FinalCustomUserAgent}; uses user profiles, interaction histories, and category specific viewing rules.
& \textbf{Custom}: \texttt{VideoRecommendationEnv}; implements recommendation, user--video interaction, feedback updates, logging, and cocoon measurement.
& User profiles, item information, category labels, and interaction histories are from a large-scale mobile short-video dataset~\cite{shang2025large}. \\
\cmidrule(lr){2-5}

& Opinion polarization
& \textbf{Custom}: \texttt{NewsConsumerAgent}; models ideological users with subscription, browsing, reading/sharing, and affective thermometer reporting behavior.
& \textbf{Custom}: \texttt{NewsPolarizationEnv}; implements ideology scored news sources, algorithmic filtering, subscription--feed--reading--sharing chains, and attitude exposure conditions.
& News materials, treatment settings, and human attitude responses are based on Levy's field experiment on social media polarization~\cite{levy2021social}. \\

\midrule

\multirow[t]{2}{1.8cm}{Urban living}
& Daily mobility
& \textbf{Built-in}: \texttt{PersonAgent}; represents daily residents with profile derived home, work, and activity context, without introducing a custom mobility specific agent.
& \textbf{Built-in}: \texttt{CodeGenRouter} coordinates tool use, and \texttt{MobilitySpace} supports city scale daily travel through POI search, route planning, and movement execution on the Beijing map.
& Map information, POI data, demographic profiles, activity contexts, and human mobility references are from the DailyMobility benchmark built on the processed dataset released by Shao et al.~\cite{shao2024chain}. \\
\cmidrule(lr){2-5}

& Disaster mobility
& \textbf{Built-in}: \texttt{PersonAgent}; represents residents under disaster conditions, with intention questioning disabled to focus on observed mobility responses.
& \textbf{Built-in}: \texttt{CodeGenRouter} coordinates tool use; \texttt{MobilitySpace} simulates travel, \texttt{EventSpace} records disaster related events, and \texttt{GlobalInformationEnv} broadcasts daily emergency information.
& Agent profiles are based on American Community Survey estimates~\cite{uscensus2020acs}, while empirical disaster-mobility responses for the Texas Winter Storm and California Camp Fire scenarios are based on SafeGraph mobility records~\cite{safegraph2019core,safegraph2021core}. \\

\bottomrule
\end{tabularx}
\end{sidewaystable*}

\subsection{Emergence of Social Norms}
To evaluate the ability of AgentSociety\textsuperscript{2} to construct mechanism based social simulation experiments, we reproduce Axelrod's classic model of norm evolution by comparing the Norms Game and the Metanorms Game\cite{1986An}. The experiment initializes a population of computational agents with boldness and vengefulness strategies and places them in a custom environment that implements norm violation, punishment, metanorm punishment, payoff updating, selection, and mutation. The main outputs are population level mean boldness, mean vengefulness, mean payoff, and strategy distribution, which measure the evolution of norm violation and punishment willingness. Since the original platform does not provide a ready made Axelrod norm evolution module, the experiment demonstrates how AgentSociety\textsuperscript{2} supports this task through custom agent and environment development, executable experiment configuration, simulation logging, and result analysis.
\subsubsection{Idea \& Hypothesis Development}


This experiment aims to use AgentSociety\textsuperscript{2} to reproduce Axelrod's (1986) classic model of norm evolution. The hypotheses are grounded in Axelrod's theory of norm enforcement. In the Norms Game, agents are expected to evolve toward higher boldness and lower vengefulness, indicating the collapse of norms. In the Metanorms Game, agents are expected to evolve toward lower boldness and higher vengefulness, indicating stable norm enforcement.

\begin{tcolorbox}[
    colback=black!3,     
    colframe=black!40,   
    fontupper=\normalfont\scriptsize, 
    fonttitle=\bfseries\scriptsize,
    width=\textwidth,
    sharp corners,
    boxrule=0.5pt,
    enhanced,
    breakable,
    title=Key Prompt - Literature Understanding and Experiment Generation,
    coltitle=white,
    colbacktitle=green!40!black!60, 
    attach boxed title to top left={yshift=-2mm, xshift=2mm},
    top=3mm,
    before upper={\linespread{1.21}\selectfont\parskip=0.3em\parindent=1.5em} 
]

You are an AI social scientist. Your task is to reproduce, using the platform’s capabilities, the experimental conclusions reported in Robert Axelrod’s (1986) paper "An Evolutionary Approach to Norms." Please start from scratch:

1. Independently read the paper and understand the objectives and conclusions of the experiments.

2. Independently design the experiment, including but not limited to agent types, strategy dimensions, environment settings, interaction rules, and evolutionary rules. Use the platform’s existing agents, environments, and skills; create new ones if necessary.

3. Output a complete description of the experimental design and implement it directly on the platform.

\end{tcolorbox}

\subsubsection{Experiment Design \& Execution}

\noindent\textbf{Simulation Setup.}
This experiment adopts a two group comparative design. The control condition is the Norms Game, in which agents are only allowed to impose first order punishment on norm violators. The treatment condition is the Metanorms Game, in which second order metanorm punishment against agents who fail to punish violators is added to first order punishment. Both conditions include 20 computational agents and run for 100 generations. Each generation consists of 4 repeated interaction rounds, so each full simulation contains 400 interaction steps in total. The initial boldness and vengefulness of each agent are randomly sampled from 0 to 7. Boldness represents the tendency to violate norms, while vengefulness represents the willingness to punish.

\noindent\textbf{Agent Development.}
This experiment constructs a custom agent type named \texttt{NormsGameAgent}. Each agent mainly contains two strategic variables, namely boldness and vengefulness. The former determines whether an agent tends to violate norms, while the latter determines whether an agent is willing to punish norm violators. Unlike agents that rely on free natural language decision making, the agents in this experiment mainly serve as carriers of evolutionary strategies. Their behavior is controlled by explicit rules and parameters, which ensures the interpretability and reproducibility of the experimental mechanism.

\noindent\textbf{Environment Development.}
The experimental environment is implemented through the custom module \texttt{NormsGameEnv}. This environment manages norm violation, observer punishment, metanorm punishment, payoff updating, intergenerational selection, and mutation. In the Norms Game, a violator receives temptation $T=3$, social members receive hurt $H=-1$, punishment imposes $P=-9$ on the violator, and the punisher bears an enforcement cost $E=-2$. In the Metanorms Game, agents who fail to punish violators may further receive meta punishment $MP=-9$, and agents who carry out metanorm punishment bear a meta enforcement cost $ME=-2$.

\noindent\textbf{Experimental Protocol.}
In each generation, agents decide whether to violate norms according to their own boldness, and observers decide whether to punish violators according to their vengefulness. Under the metanorms condition, the system further checks which agents fail to punish violators and allows other agents to impose metanorm punishment on them. At the end of each generation, the system performs strategy selection and mutation based on accumulated payoffs. The experiments are run under \texttt{game\_mode="norms"} and \texttt{game\_mode="metanorms"}, respectively. Finally, population average boldness, population average vengefulness, and strategy distribution are recorded to compare the differences in norm evolution between the two mechanisms.

\subsubsection{Result Analysis \& Interpretation}
The simulation results show different patterns across the two experimental conditions. In the Norms Game, average boldness decreases from 1.90 to 0.10, while average vengefulness slightly decreases from 2.90 to 2.55. This indicates that norm-violating behavior declines even without metanorms, but the willingness to punish violators is not reinforced. Therefore, the condition without metanorms provides a relatively weak foundation for maintaining the norm, since violations are suppressed without a corresponding increase in punishment motivation.

\begin{figure}[t]
    \centering
    \begin{subfigure}[t]{0.48\linewidth}
        \centering
        \includegraphics[width=\linewidth]{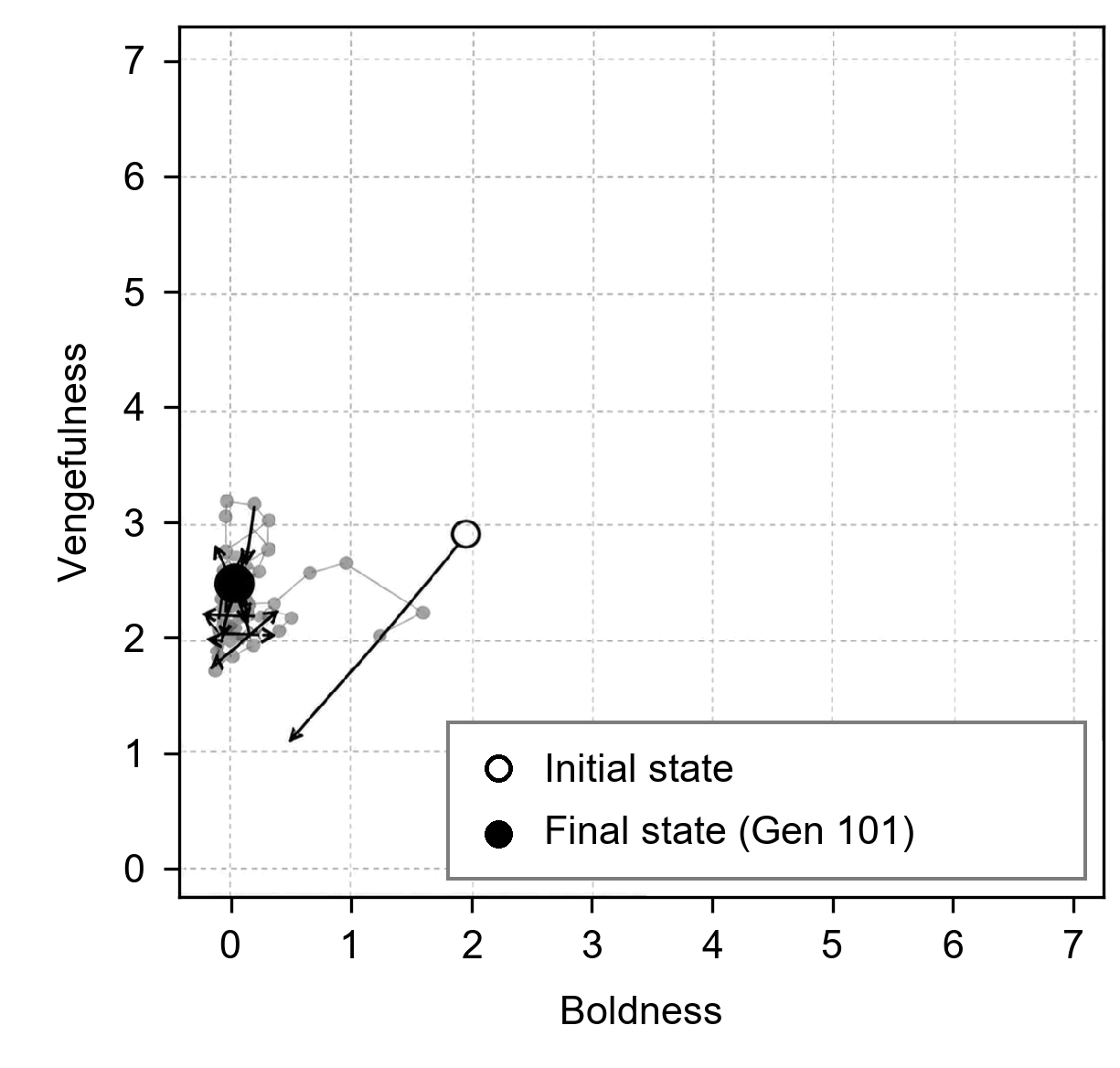}
        \caption{Norms Game}
        \label{fig:norms_game_result}
    \end{subfigure}
    \hfill
    \begin{subfigure}[t]{0.48\linewidth}
        \centering
        \includegraphics[width=\linewidth]{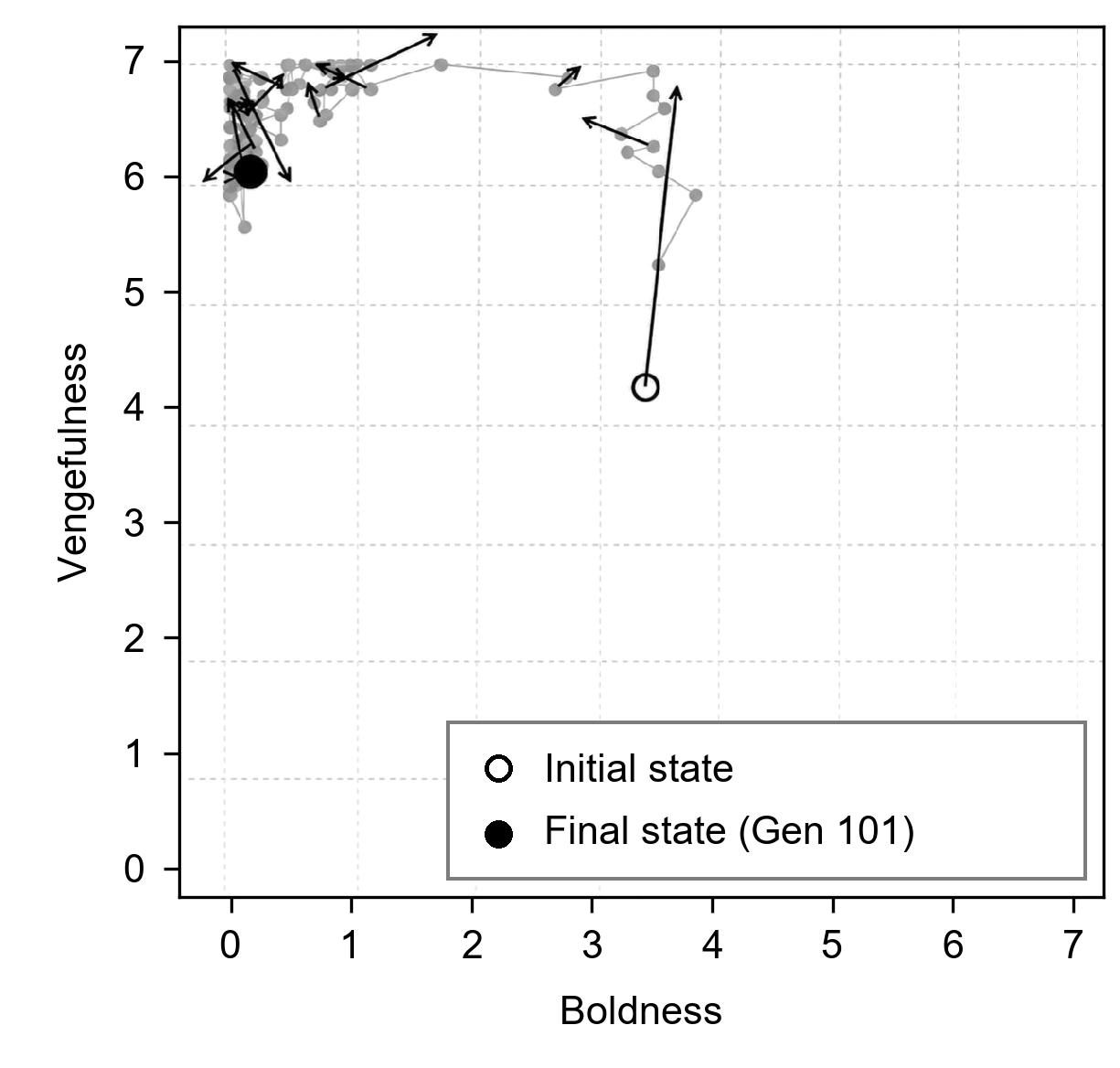}
        \caption{Metanorms Game}
        \label{fig:metanorms_game_result}
    \end{subfigure}
    \caption{Population-level dynamics in Norms and Metanorms Games. Norms Game: boldness decreases, vengefulness stable; Metanorms Game: boldness decreases, vengefulness increases. Open circles are initial states, black circles are final states.}
    \label{fig:norms_metanorms_game_result}
\end{figure}

As shown in Figure~\ref{fig:norms_game_result}, the population moves from a state with relatively high boldness and moderate vengefulness toward a state with very low boldness and slightly lower vengefulness. This trajectory suggests that first-order punishment can reduce norm violations under the current implementation. However, because vengefulness does not increase, the norm is not supported by a strengthened enforcement motivation.

In the Metanorms Game, average boldness decreases from 3.35 to 0.15, while average vengefulness increases from 4.40 to 6.15. This result supports the central prediction of the metanorm model. When agents can punish those who fail to punish violators, they have stronger incentives to maintain or increase their willingness to enforce norms.

As shown in Figure~\ref{fig:metanorms_game_result}, the population moves toward a region with very low boldness and high vengefulness. This pattern indicates that metanorms not only suppress norm-violating behavior, but also strengthen punishment willingness. Therefore, with metanorms, norm enforcement is more likely to become stable.

\begin{figure}[t]
    \centering
    \begin{subfigure}[t]{0.48\linewidth}
        \centering
        \includegraphics[width=\linewidth]{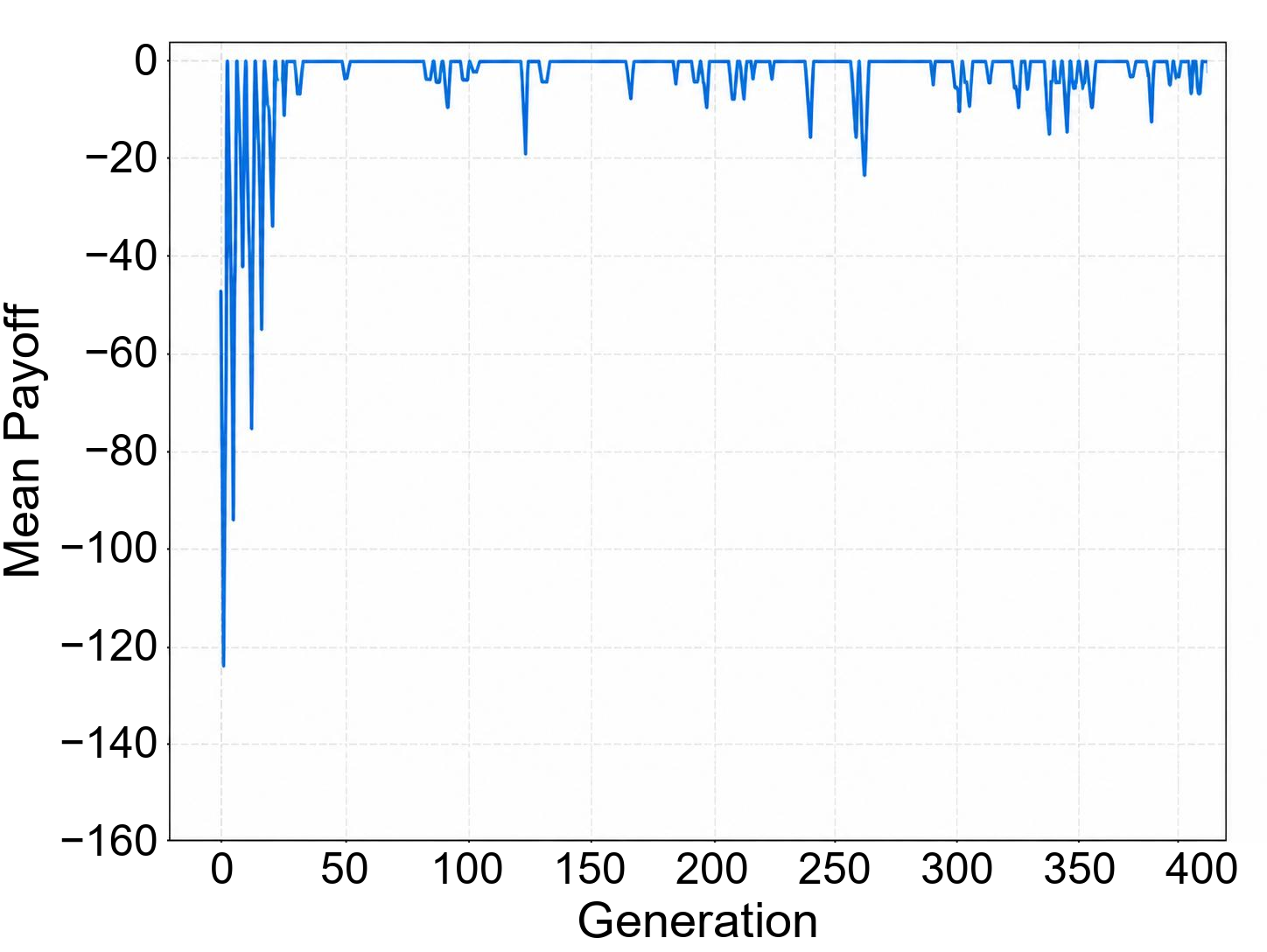}
        \caption{Norms Game}
        \label{fig:norms_payoff_result}
    \end{subfigure}
    \hfill
    \begin{subfigure}[t]{0.48\linewidth}
        \centering
        \includegraphics[width=\linewidth]{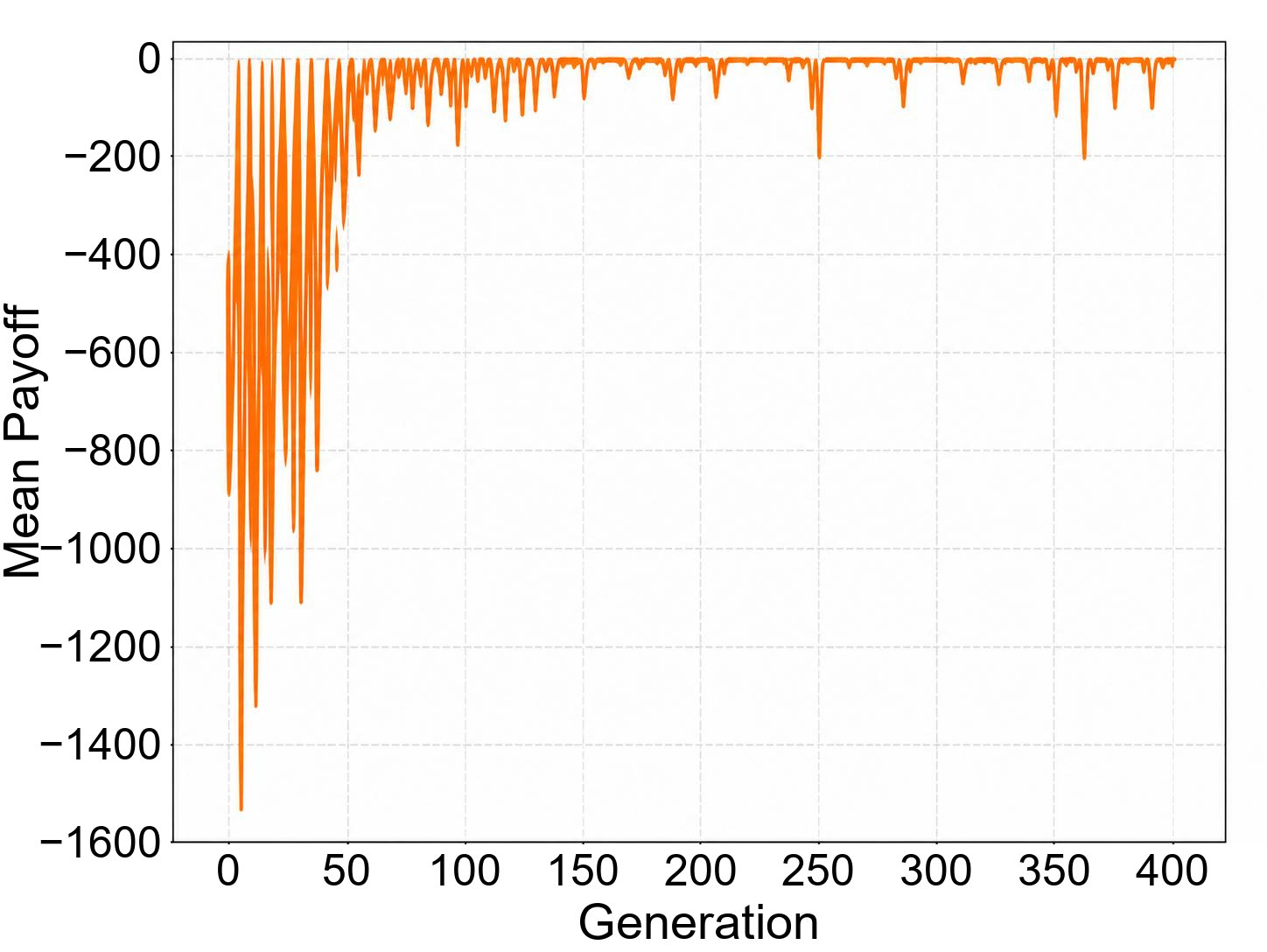}
        \caption{Metanorms Game}
        \label{fig:metanorms}
    \end{subfigure}
    \caption{Mean payoff over 400 simulation steps in the Norms Game and Metanorms Game. Norms Game without metanorms and Metanorms Game with metanorms are shown.}
    \label{fig:payoff_result}
\end{figure}

The payoff dynamics provide an additional perspective on the cost of norm enforcement. As shown in Figure~\ref{fig:payoff_result}, the left panel shows that the Norms Game produces large negative payoff fluctuations in the early stage, but the mean payoff gradually stabilizes near zero as boldness declines. This suggests that once norm-violating behavior is suppressed, fewer punishment events occur and the average payoff loss becomes smaller. 

The right panel of Figure~\ref{fig:payoff_result} shows that the Metanorms Game has a similar long-run tendency toward payoff stabilization, but its early-stage payoff losses are much larger than those in the Norms Game. This difference is consistent with the structure of the metanorm mechanism. In addition to first-order punishment against norm violators, agents may also impose second-order punishment on those who fail to punish violators. As a result, the early stage of the Metanorms Game involves higher enforcement and meta-enforcement costs. However, as vengefulness increases and boldness decreases, the frequency of norm violations and punishment events declines, and mean payoff also approaches zero.

Therefore, the payoff results suggest a trade-off between enforcement cost and norm stability. Metanorms help sustain a higher willingness to punish and make norm enforcement more stable, but they may also generate larger short-term payoff losses during the transition process. In contrast, the Norms Game has lower early enforcement costs, but it does not strengthen punishment willingness to the same extent.

Compared with the Norms Game, the most important difference is not the final level of boldness, since both conditions converge to very low boldness. Instead, the key difference is the level and direction of change in vengefulness. Without metanorms, norm-violating behavior declines, but punishment willingness does not increase, which means that the norm is maintained on a relatively weak enforcement basis. With metanorms, norm-violating behavior declines while punishment willingness rises, so the norm is more likely to become stably established.

This pattern is partially consistent with Axelrod's original theory. The Metanorms Game reproduces the effect that second-order punishment sustains high vengefulness and stabilizes norm enforcement. However, the Norms Game does not show full norm collapse in this run, because boldness remains very low at the end of the simulation. According to the explanations provided by the AgentSociety\textsuperscript{2}, one plausible reason is that first-order punishment in the current implementation may be strong enough to suppress violations, for example because multiple observers may independently punish the same violator. The platform also suggests other possible factors, including payoff parameter sensitivity, the evolutionary selection rule, and random seed effects. Building on this comprehensive analysis, Figure~\ref{fig:manuscript-drafting}(a) illustrates how \textit{AgentSociety$^2$} supports manuscript drafting by organizing the Axelrod-inspired hypothesis, custom norm-game implementation, evolutionary dynamics, payoff analysis, and interpretation into a complete academic manuscript. In sum, this study demonstrates how \textit{AgentSociety$^2$} can translate a formal theory of norm enforcement into an executable simulation, while exploring deviations from the original prediction traceable to explicit modeling choices such as punishment strength, selection rules, and parameter settings.

\subsection{Public Goods Experiments}



To evaluate the end-to-end replication capability of \textit{AgentSociety$^2$} for social science experiments, we select the public goods experiment by Fischbacher and Gächter as a case study~\cite{fischbacher2010social}. This study examines the relationship among individual social preferences, belief updating, and the dynamics of free riding in public goods games, making it a representative setting for assessing whether simulated agents can reproduce key mechanisms in social science experiments.

We provide the original paper PDF to the AI social scientist and ask it to extract the experimental procedure from the paper, construct the corresponding agent and environment modules, and generate executable experiment configurations. Based on the generated configuration, the system runs a public goods replication experiment with 24 agents, which are randomly assigned to six four-person groups. The experiment preserves the original P/C two-stage design: in the P stage, agents’ conditional contribution preferences are elicited; in the C stage, agents make public goods contribution decisions over ten rounds. We record agents’ actual contributions in each round, their beliefs about the contributions of other group members, and the conditional contribution schedules elicited in the P stage. Based on these records, we analyze contribution trajectories, belief trajectories, and the discrepancy between actual contributions and preference-predicted contributions.

\subsubsection{Idea \& Hypothesis Development}



In the idea development stage, AI social scientist transformed the original paper from an empirical replication target into a testable simulation hypothesis. Rather than treating declining contributions in repeated public goods games as a simple consequence of selfish learning, AI social scientist identified the paper’s central mechanism as the interaction between conditional cooperative preferences and dynamically updated beliefs about others’ contributions. It therefore formulated the hypothesis that agents with stable conditional cooperation preferences may still exhibit declining contributions when their beliefs about others’ cooperation deteriorate over repeated interactions. This hypothesis serves as the conceptual bridge between the original behavioral experiment and the executable simulation design in AgentSociety$^2$.

\subsubsection{Experiment Design \& Execution}

\begin{tcolorbox}[
    colback=black!3,
    colframe=black!40,
    fontupper=\normalfont\scriptsize, 
    fonttitle=\bfseries\scriptsize,
    width=\textwidth,
    sharp corners,
    boxrule=0.5pt,
    enhanced,
    breakable,
    title=Paper to Experiment Instruction,
    coltitle=white,
    colbacktitle=green!40!black!60,
    attach boxed title to top left={yshift=-2mm, xshift=2mm},
    top=3mm,
    before upper={\linespread{1.21}\selectfont\parskip=0.3em\parindent=1.5em}
]

Use the AgentSociety2 experimental platform to reproduce the social science experiment described in Social Preferences, Beliefs, and the Dynamics of Free Riding in Public Goods Experiments.

Read the original paper carefully and extract its research question, theoretical mechanism, experimental design, interaction rules, variables, data records, and evaluation metrics. Apart from the paper title and the general capabilities of AgentSociety2, do not assume any predefined experimental structure. All stages, parameters, variables, and metrics should be derived from the paper or reasonably inferred from its content.

Based on the paper, formulate a testable hypothesis suitable for AgentSociety2 and translate it into an executable agent-based simulation. Specify the required agent modules, environment modules, state variables, interaction procedures, runtime parameters, data to be recorded, and evaluation methods. If existing modules are insufficient, design new ones as needed.

If the original paper contains treatment groups, control groups, order comparisons, or other experimental conditions, convert them into executable settings in AgentSociety2. Output a structured experimental plan that can be further converted into configuration files and runtime steps.

\end{tcolorbox}


After understanding the paper and assessing the capabilities of \textit{AgentSociety$^2$}, AI social scientist identified that the existing modules were insufficient to fully implement the original experimental protocol. With human approval, it extended the platform by creating \texttt{FGPCPublicGoodsLLMAgent} and \texttt{FGPCPublicGoodsEnv}, where the latter supports the two-stage P/C procedure required by the study. ASS then specified a P→C main condition and a C→P comparison condition, yielding an executable, comparable, and evaluable experiment design within \textit{AgentSociety$^2$}.


AI social scientist then created an initialization script and generated the required \textit{AgentSociety$^2$} configuration files. Using the platform’s built-in execution and validation capabilities, it  iteratively checked whether the agent modules, environment modules, experimental procedure, and runtime parameters could be correctly loaded and executed. After the configuration passed validation, ASS launched the experiment in the background. Once the first-round results were produced, AI social scientist inspected the generated data files, execution logs, and intermediate states, identified several minor implementation issues, fixed them, and restarted the experiment. Through periodic monitoring, the experiment was completed successfully. The main experiment produced complete data records, including 24 preference-elicitation records in the P stage and 240 round-level contribution records in the C stage.

\subsubsection{Result Analysis \& Interpretation}

\begin{figure}[ht]
    \centering
    \includegraphics[width=\textwidth]{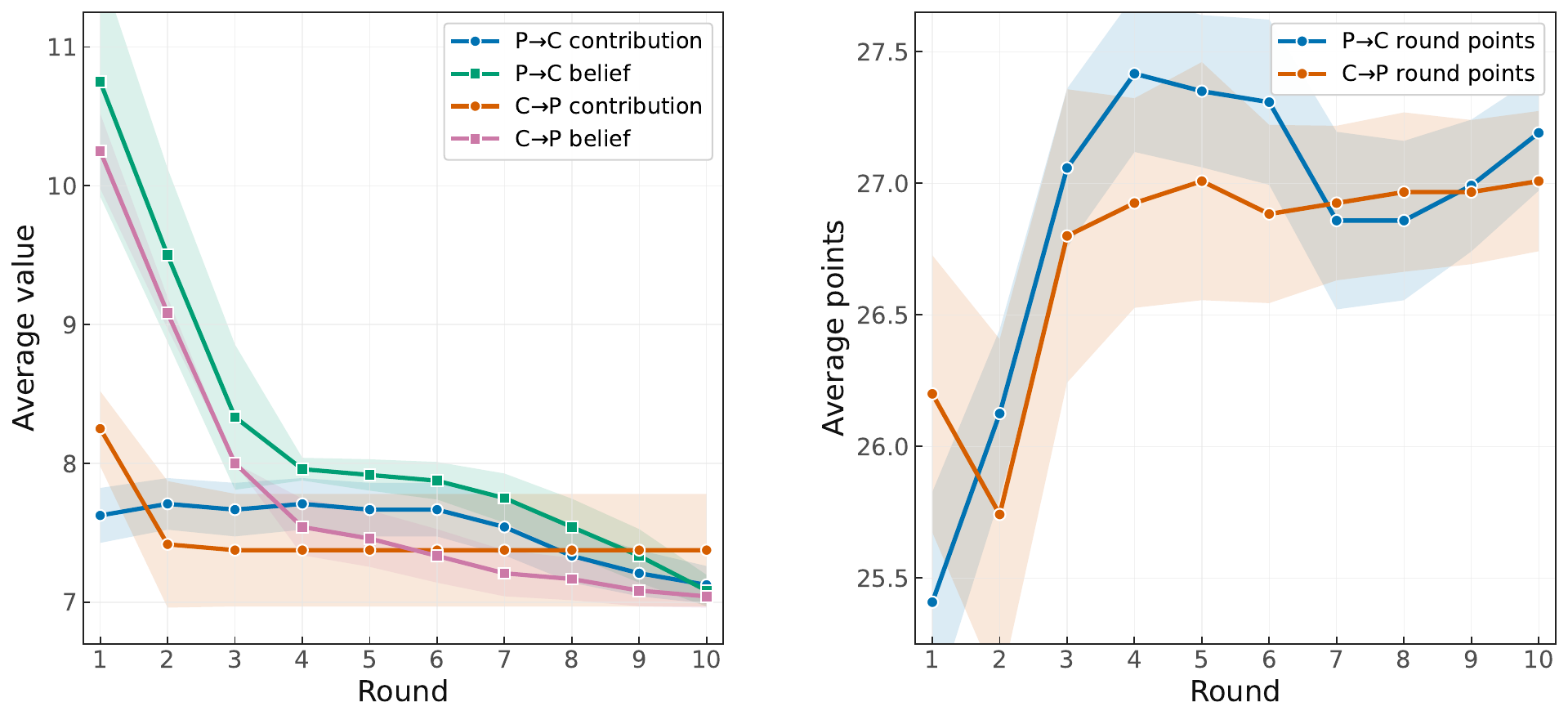}
    \caption{Round-level dynamics under different experimental orders. Contributions, beliefs, and payoffs are tracked across ten rounds for the P->C and C->P orders. Beliefs adjust over time, while contributions decline only slightly before stabilizing, indicating no sustained shift toward free riding.}
\end{figure}

\begin{figure}[ht]
    \centering
    \includegraphics[width=\textwidth]{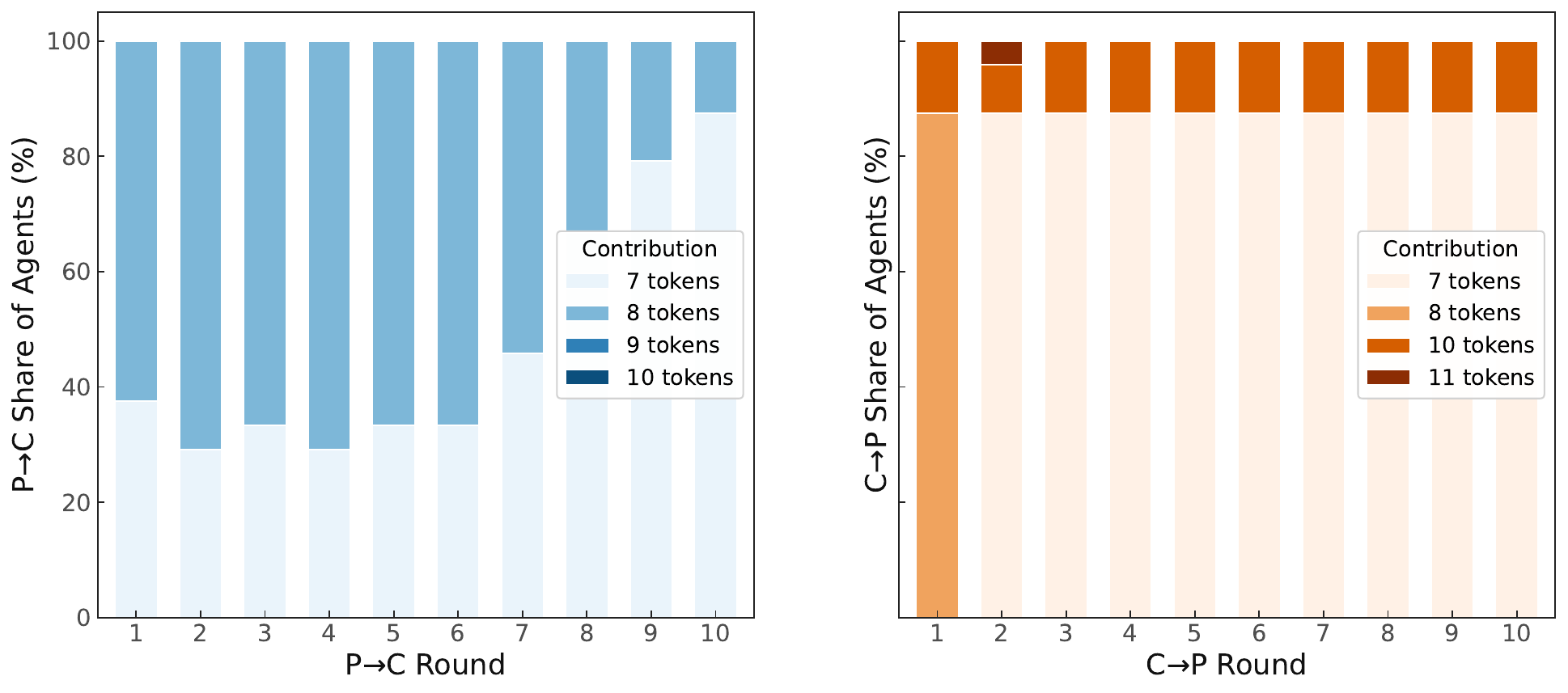}
    \caption{Shift in contribution distributions across rounds. Stacked bars show the percentage share of each contribution level (tokens) per round. Under the P→C order (left), contributions gradually shift from 8 toward 7 tokens; under the C→P order (right), the distribution locks into a narrow 7–8 token band after Round 1 with virtually no further change.}
\end{figure}

\begin{figure}[ht]
    \centering
    \includegraphics[width=\textwidth]{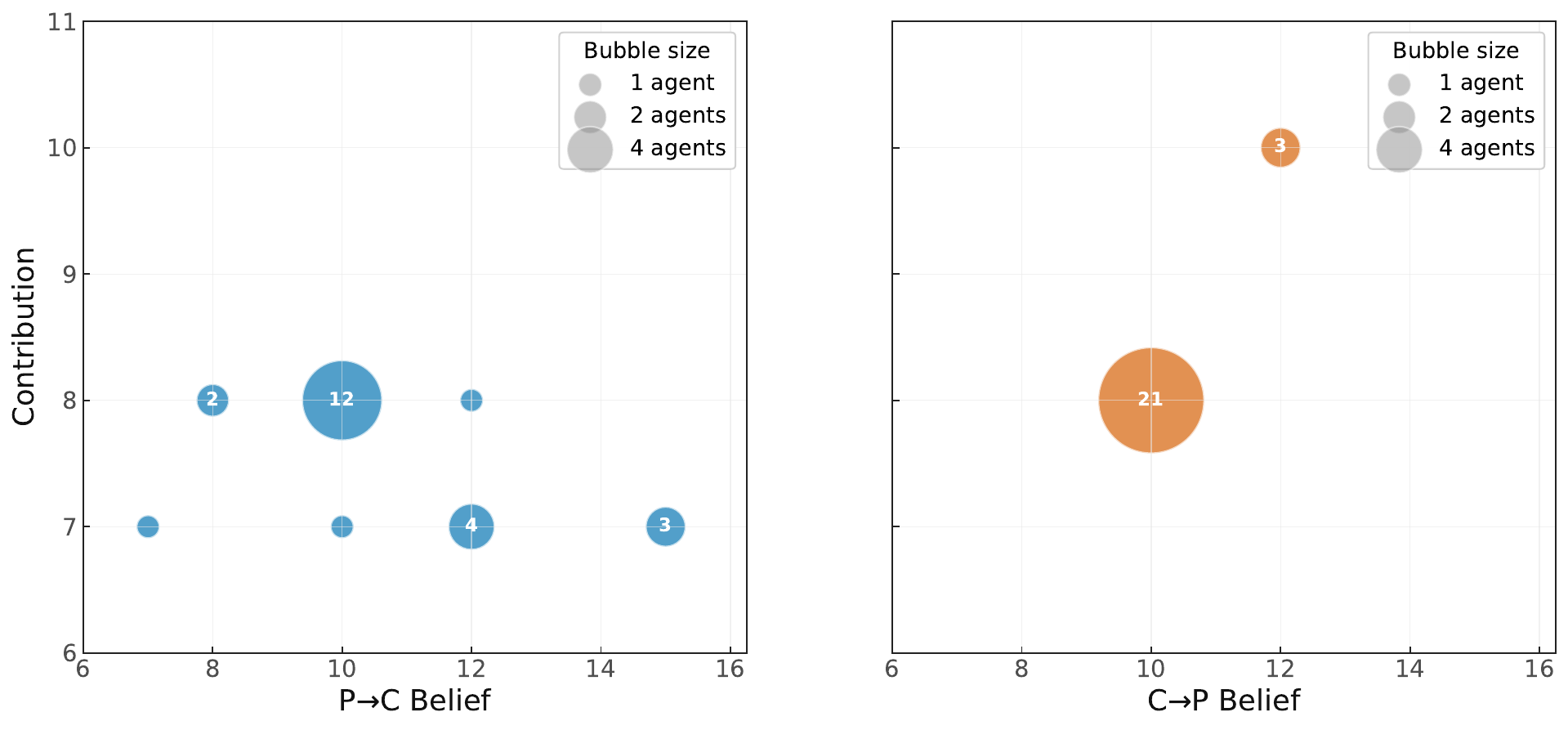}
    \caption{First-round belief-contribution distributions under different experimental orders. Bubble size indicates the number of agents at each belief-contribution pair. Contributions cluster at moderate-to-high levels despite varying beliefs, suggesting that initial contributions are only weakly tied to belief differences.}
\end{figure}


The results show that the simulation partially reproduces the key mechanisms identified in the original paper, but does not fully reproduce the sustained dynamics of free riding. In the preference-elicitation stage, 17 out of 24 agents were classified as incomplete conditional cooperators: they increased their intended contributions as others’ contributions increased, but at a lower rate than perfect conditional cooperation. This finding is consistent with the original mechanism that conditional cooperation is common among participants.


However, the contribution trajectory in the C stage does not show sustained cooperation decay. The average contribution decreased from 8.25 in the first round to an early minimum of approximately 7.79, but then recovered and stabilized around 8. By the tenth round, the average contribution was 8.04, only 0.21 lower than in the first round. This suggests that agents adjusted their behavior in the early rounds, but the decline did not accumulate into the persistent free-riding dynamics observed in the original human experiment.


A more notable discrepancy appears in the comparison between actual contributions and preference-predicted contributions. In later rounds, agents’ actual contributions were often higher than the values predicted by their P-stage preference schedules. For example, in the tenth round, 70.83\% of agents contributed more than their preference-predicted levels. This contrasts with the mechanism in the original paper, where actual contributions tend to fall below preference-based predictions, reflecting an increasing tendency toward free riding. These findings indicate that although LLM agents can display conditional cooperation and early behavioral adjustment, they do not continuously shift toward lower contribution levels in repeated interactions. Instead, they exhibit a relatively strong tendency toward cooperative stability. Following the analysis stage, \textit{AgentSociety$^2$} further develops this completed public-goods study into a research manuscript shown in Figure~\ref{fig:manuscript-drafting}(b), where the P/C experimental design, agent and environment construction, contribution and belief dynamics, and comparison with the original free-riding mechanism are organized into a structured academic manuscript. Overall, this study showcases that \textit{AgentSociety$^2$} can support the full workflow of behavioral-game replication, from extracting a paper-based protocol and implementing custom components to analyzing both reproduced mechanisms and deviations from human findings.

\subsection{Psychological Survey}

Survey study is a core social science design for assessing both self-reported attitudes and judgments and less directly observable cognitive processes through structured responses and task-based measures. To demonstrate \textit{AgentSociety$^2$}'s capacity in carrying out psychological survey studies, we reproduce the human self-bias study by Qi et al. (2025)~\cite{qi2025comprehensive} within \textit{AgentSociety$^2$} and compare simulated agent responses with the original human data. We select four self-bias tasks from the original study to cover both \textit{explicit evaluative measures} and \textit{indirect self-processing measures}. The explicit evaluative measures include Endowment Effect (EE), which tests whether people value owned items more than unowned items, and Self-Enhancement (SE), which tests whether people evaluate themselves more favorably than an average peer. The indirect self-processing measures include Implicit Association Test of Self-Esteem (IAT), which measures automatic self-positive associations through reaction time, and Self-Reference Effect (SRE), which measures whether self-referenced information is remembered better. EE and SE rely on direct numerical judgments, while IAT and SRE assess self-bias through reaction time categorization and memory recognition performance.

Our case study demonstrates that the platform can generate simulated respondents from participant profiles, construct task-specific survey environments, execute structured procedures, and produce evaluation metrics comparable to human baselines. \textit{AgentSociety$^2$} constructs a simulated respondent pool from the 134 participant profiles provided by Qi et al. (2025)~\cite{qi2025comprehensive}, injects their demographic and psychometric attributes into the built-in \texttt{PersonAgent}, and keeps the core agent class unchanged across tasks. Four task-specific \texttt{EnvBase} environments reproduce the original protocols and record structured outputs, including valuation judgments, self-ratings, reaction-time traces, accuracy records, and memory-recognition judgments. The analysis workflow then computes task-specific indicators, compares them with human baselines, and generates visualizations and interpretations.

\subsubsection{Idea \& Hypothesis Development}

The hypotheses are grounded in classical psychological findings on ownership valuation~\cite{kahneman1990experimental}, self-enhancement~\cite{taylor1988illusion,heine1999there}, implicit self-esteem~\cite{greenwald2000using}, and self-reference memory~\cite{rogers1977self}. For EE, simulated agents are expected to report higher WTA than WTP, although the price gap may be smaller than that of humans because agents do not have embodied ownership. For SE, agents may differ from the human pattern because the scale is reverse coded, with 0 as best and 100 as worst, which may interact with conservative LLM self-assessment. For IAT, agents are expected to respond faster in self-positive blocks than self-negative blocks. For SRE, agents are expected to recognize self-encoded traits better than friend- or other-encoded traits. These hypotheses allow us to evaluate whether profile-based agents can reproduce self-bias patterns across direct evaluative judgments and indirect self-processing measures.

\begin{tcolorbox}[
    colback=black!3,
    colframe=black!40,
    fontupper=\normalfont\scriptsize, 
    fonttitle=\bfseries\scriptsize,
    width=\textwidth,
    sharp corners,
    boxrule=0.5pt,
    enhanced,
    breakable,
    title=Key Prompt --- Literature Investigation,
    coltitle=white,
    colbacktitle=green!40!black!60,
    attach boxed title to top left={yshift=-2mm, xshift=2mm},
    top=3mm,
    before upper={\linespread{1.21}\selectfont\parskip=0.3em\parindent=1.5em}
]
Please investigate the source paper Qi et al. (2025) at \{SOURCE\_PAPER\_PATH\} and its supplementary materials at \{SUPPLEMENTARY\_MATERIALS\_DIR\}. Carefully read the paper and extract the research motivation, participant information, task designs, experimental procedures, stimuli, response formats, and human baseline results for EE, SE, IAT, and SRE.

Please also search for related literature connected to the source paper and the four self-bias tasks, especially ownership valuation, self-enhancement, implicit self-esteem, and self-reference effects. Use the source paper and related literature to summarize what each task measures and what details should be preserved when reproducing the study in \textit{AgentSociety$^2$}.
\end{tcolorbox}

\begin{tcolorbox}[
    colback=black!3,
    colframe=black!40,
    fontupper=\normalfont\scriptsize, 
    fonttitle=\bfseries\scriptsize,
    width=\textwidth,
    sharp corners,
    boxrule=0.5pt,
    enhanced,
    breakable,
    title=Key Prompt --- Hypothesis Formation,
    coltitle=white,
    colbacktitle=green!40!black!60,
    attach boxed title to top left={yshift=-2mm, xshift=2mm},
    top=3mm,
    before upper={\linespread{1.21}\selectfont\parskip=0.3em\parindent=1.5em}
]
Based on the source paper and the related literature, formulate hypotheses for reproducing EE, SE, IAT, and SRE with profile-based LLM agents in \textit{AgentSociety$^2$}.

For each task, explain the expected human pattern, the possible agent pattern, the metric for comparison, and why agents may align with or diverge from human participants.
\end{tcolorbox}

\subsubsection{Experiment Design \& Execution}

\noindent\textbf{Data Preparation.}
We instruct the platform to extract all 134 participant profiles from the supplementary materials of Qi et al. (2025)~\cite{qi2025comprehensive}. These profiles initialize the simulated respondent population, so each \texttt{PersonAgent} corresponds to one real participant profile rather than a generic persona. Each profile contains participant ID, age, gender, Big Five personality traits, Rosenberg self-esteem, self-construal ratings, individualism collectivism measures, subjective well-being scores, and dark triad traits. A profile loader assigns participant profiles to agent IDs and formats each profile into a neutral prompt fragment, which is injected into \texttt{PersonAgent} during initialization.

\noindent\textbf{Agent Development.}
We use the platform-provided \texttt{PersonAgent} without modifying the agent class. The only agent-side adaptation is profile injection through initialization arguments. This design keeps the simulated respondent model stable across EE, SE, IAT, and SRE, so differences across tasks come from task-specific environments and response procedures rather than changes to the agent architecture. Each experiment also uses a task completion skill that guides the agent to call the relevant environment tools, submit all required responses, and stop only after the full procedure is completed.

\noindent\textbf{Environment Development.}
\textit{AgentSociety$^2$} helps generate four custom \texttt{EnvBase} modules, one for each psychological task. Each environment specifies participant-facing instructions, stimuli, task rules, response format, measurement interface, and logging schema. The environment modules expose Pydantic-typed \texttt{@tool} methods for structured interaction and store responses in the experiment database for later analysis. \texttt{EndowmentEffectPaperEnv} records WTA/WTP prices for EE. \texttt{SelfEnhancementPaperEnv} records percentile self-ratings for SE. \texttt{IATPaperEnv} records key presses, reaction times, and accuracies for IAT. \texttt{SelfReferenceEffectPaperEnv} records encoding ratings and recognition judgments with Remember/Know labels for SRE.

\noindent\textbf{Experimental Protocol.}
Each task runs independently with the same profile-based \texttt{PersonAgent} population. Agents are initialized, connected to the corresponding custom environment, and guided by task skills to complete all required tool calls. The EE task presents four items, pen, plate, glass, and doll, and collects WTA and WTP prices from 0 to 100 CNY under ownership and non-ownership conditions. The SE task obtains self-ratings across eight dimensions, including intelligence, cooperation, appearance, morality, sociability, health, honesty, and generosity, using a reverse-coded percentile scale where 0 means best, 100 means worst, and 50 means median. The IAT task records key presses, response times, and accuracy across identity practice, valence practice, congruent trials, identity switch, and incongruent trials, with 140 trials in total. The SRE task contains an encoding phase and a recognition phase, where agents first rate 120 traits assigned to self, friend, or familiar-other identities and then judge 240 traits as old or new with Remember/Know labels. For evaluation, each task is mapped to the metric used in the original human study. EE is evaluated by the WTA--WTP gap and the proportion of agents with WTA higher than WTP. SE is evaluated by the overall and dimension-level percentile ratings relative to the neutral midpoint of 50. IAT is evaluated by reaction-time differences and accuracy differences between congruent and incongruent blocks. SRE is evaluated by old/new recognition accuracy and identity-conditioned old-item accuracy across self, friend, and familiar-other encoding conditions. These responses are extracted from per-experiment SQLite logs and used for statistical testing, visualization, and comparison with human baselines.

\begin{tcolorbox}[
    colback=black!3,
    colframe=black!40,
    fontupper=\normalfont\scriptsize, 
    fonttitle=\bfseries\scriptsize,
    width=\textwidth,
    sharp corners,
    boxrule=0.5pt,
    enhanced,
    breakable,
    title=Key Prompt --- Experiment Protocol Design,
    coltitle=white,
    colbacktitle=green!40!black!60,
    attach boxed title to top left={yshift=-2mm, xshift=2mm},
    top=3mm,
    before upper={\linespread{1.21}\selectfont\parskip=0.3em\parindent=1.5em}
]
Please design a full \textit{AgentSociety$^2$} experiment protocol for reproducing the four self-bias tasks from Qi et al. (2025). Use the paper's methodology and supplementary materials as the main reference.

The protocol should cover participant initialization, task procedure, stimuli, response collection, evaluation metrics, and comparison with human baselines. Please keep the four tasks faithful to the original study while making them executable within \textit{AgentSociety$^2$}.
\end{tcolorbox}

\begin{tcolorbox}[
    colback=black!3,
    colframe=black!40,
    fontupper=\normalfont\scriptsize, 
    fonttitle=\bfseries\scriptsize,
    width=\textwidth,
    sharp corners,
    boxrule=0.5pt,
    enhanced,
    breakable,
    title=Key Prompt --- Agent Initialization,
    coltitle=white,
    colbacktitle=green!40!black!60,
    attach boxed title to top left={yshift=-2mm, xshift=2mm},
    top=3mm,
    before upper={\linespread{1.21}\selectfont\parskip=0.3em\parindent=1.5em}
]
Please review the supplementary materials of Qi et al. (2025), especially the participant profiles under \texttt{/user\_data/Self\_bias\_dataset/0.Self\_reported\_scales/Participant\_profiles}. Use these real participant profiles to initialize simulated respondents in \textit{AgentSociety$^2$}.

Each simulated respondent should correspond to one real participant profile. Inject the available demographic and psychometric information into the built-in \texttt{PersonAgent} without modifying the core agent class, so that the simulated population remains aligned with the original human sample.
\end{tcolorbox}

\begin{tcolorbox}[
    colback=black!3,
    colframe=black!40,
    fontupper=\normalfont\scriptsize, 
    fonttitle=\bfseries\scriptsize,
    width=\textwidth,
    sharp corners,
    boxrule=0.5pt,
    enhanced,
    breakable,
    title=Key Prompt --- Experiment Construction and Execution,
    coltitle=white,
    colbacktitle=green!40!black!60,
    attach boxed title to top left={yshift=-2mm, xshift=2mm},
    top=3mm,
    before upper={\linespread{1.21}\selectfont\parskip=0.3em\parindent=1.5em}
]
Please create and execute \textit{AgentSociety$^2$} experiments for EE, SE, IAT, and SRE based on Qi et al. (2025) and the extracted protocols.

Follow the original task logic, stimuli, response formats, and evaluation metrics as closely as possible. Please decide the appropriate platform implementation, create any necessary task-specific components, and make sure each simulated respondent can complete the full procedure through structured interactions.

After execution, please record all agent responses in a structured form suitable for statistical analysis, visualization, and comparison with the human baselines from the source paper.
\end{tcolorbox}

\subsubsection{Result Analysis \& Interpretation}

\begin{figure}[!t]
    \centering
    \begin{subfigure}[t]{0.47\textwidth}
        \centering
        \includegraphics[width=\linewidth]{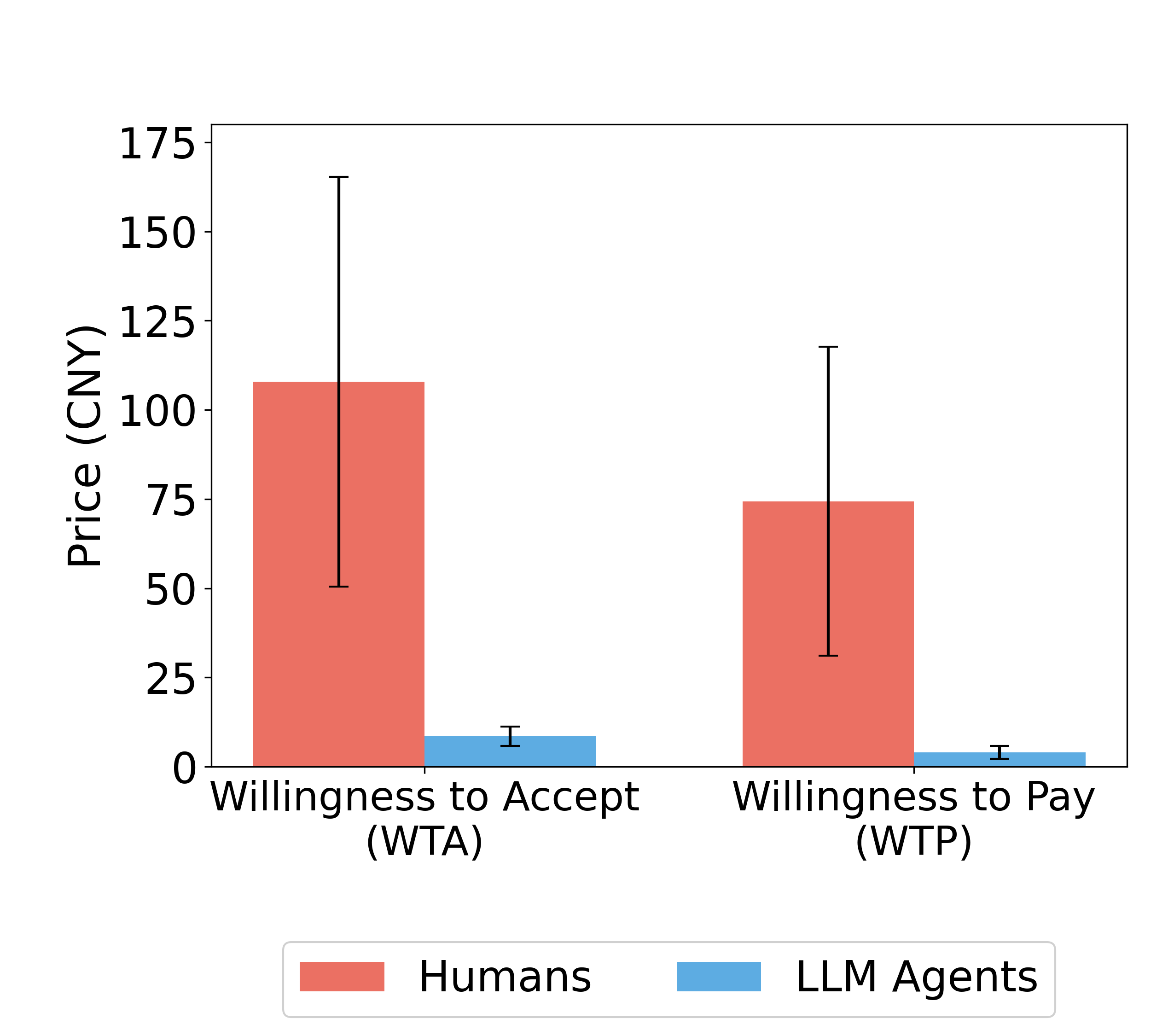}
        \caption{Endowment Effect}
        \label{fig:ee-sub}
    \end{subfigure}
    \hfill
    \begin{subfigure}[t]{0.47\textwidth}
        \centering
        \includegraphics[width=\linewidth]{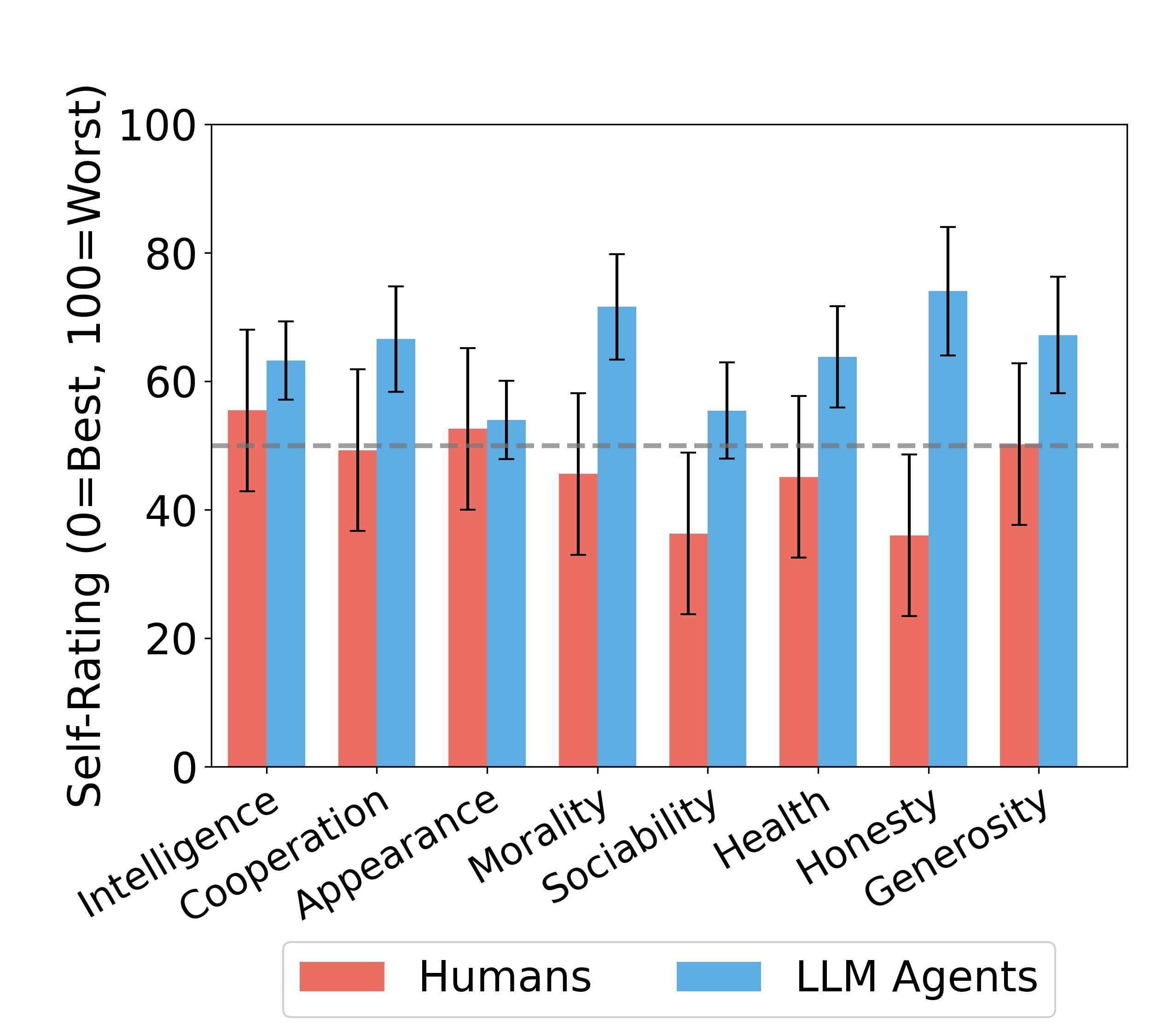}
        \caption{Self-Enhancement}
        \label{fig:se-sub}
    \end{subfigure}

    \begin{subfigure}[t]{0.47\textwidth}
        \centering
        \includegraphics[width=\linewidth]{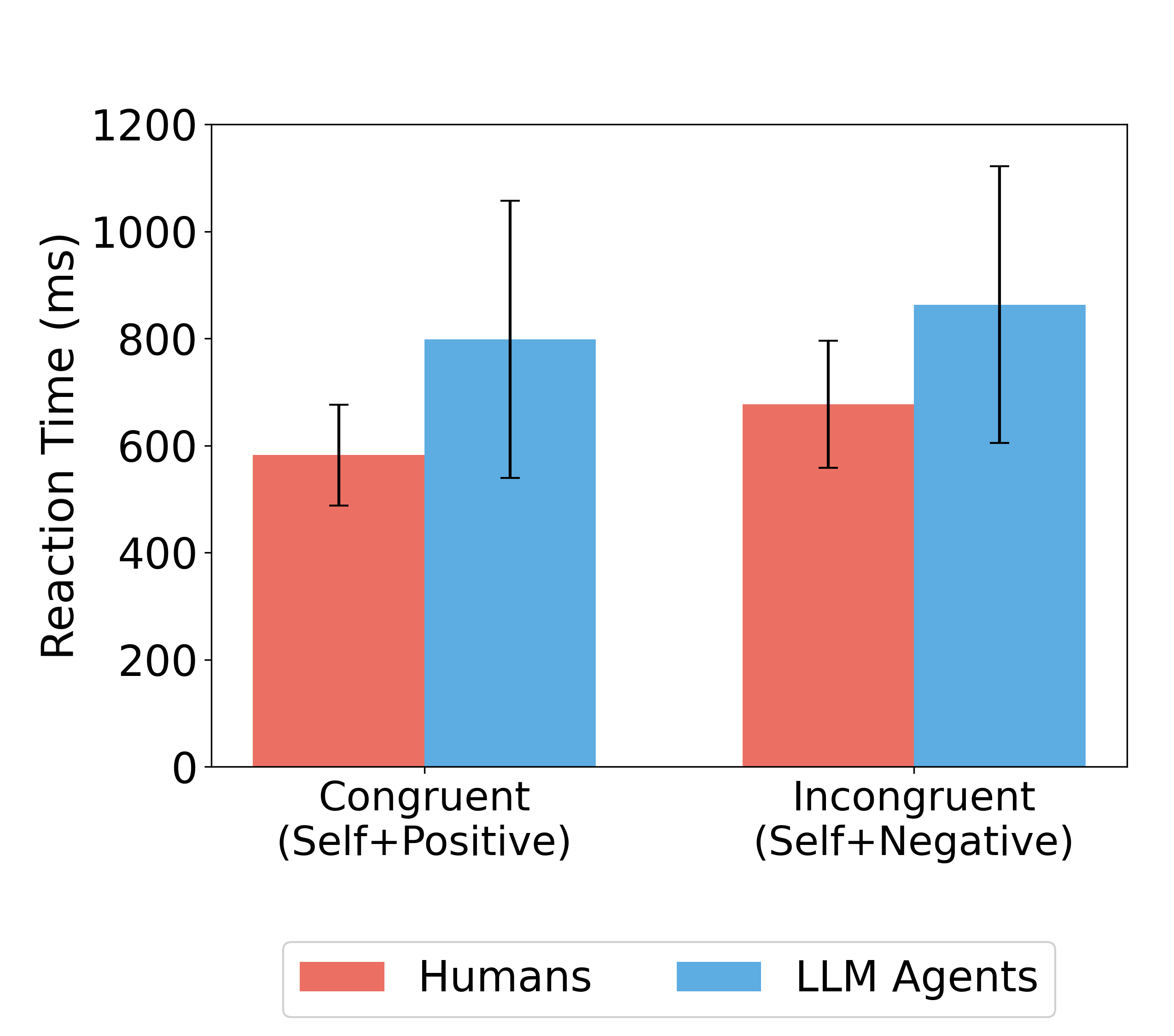}
        \caption{Implicit Association Test}
        \label{fig:iat-sub}
    \end{subfigure}
    \hfill
    \begin{subfigure}[t]{0.47\textwidth}
        \centering
        \includegraphics[width=\linewidth]{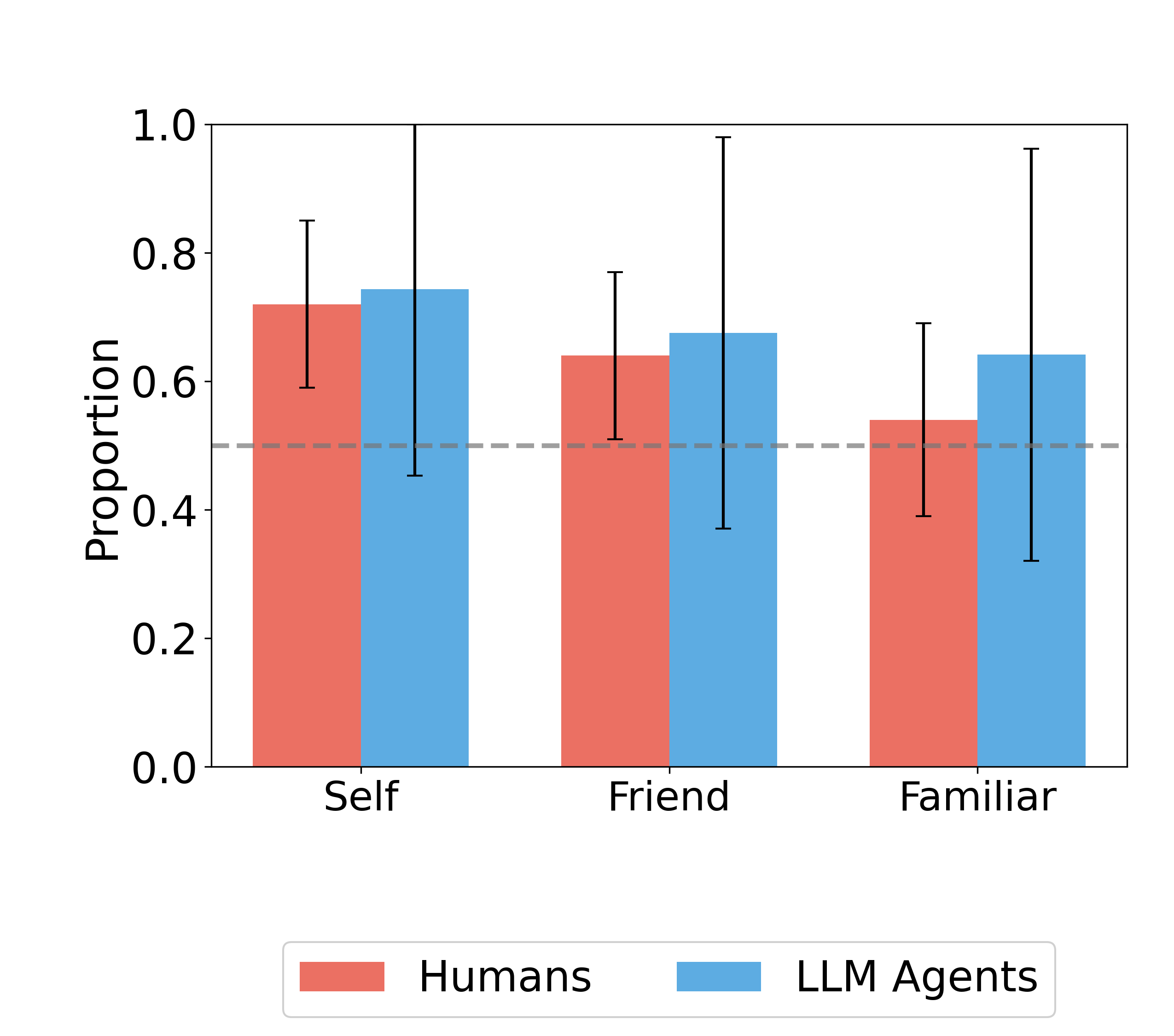}
        \caption{Self-Reference Effect}
        \label{fig:sre-sub}
    \end{subfigure}

    \caption{
        Comparative results across four self-bias tasks between
        agents and humans:
        (a) Endowment Effect,
        (b) Self-Enhancement,
        (c) Implicit Association Test, and
        (d) Self-Reference Effect.
    }
    \label{fig:psychology-survey-figure}
\end{figure}

Figure~\ref{fig:psychology-survey-figure} summarizes the comparison between simulated agents and human baselines across the four self-bias tasks. We interpret the results by comparing two types of measures within the unified psychological survey suite: explicit evaluative measures, including EE and SE, and indirect self-processing measures, including IAT and SRE. Agents complete all four tasks and produce valid structured outputs, but their alignment with human data differs substantially across these two types of measures.

For the explicit evaluative measures, EE and SE are operationalized as direct numerical judgment tasks over the same profile-based \texttt{PersonAgent} population. EE elicits object valuation under ownership and non-ownership conditions, while SE elicits self-evaluation relative to an average peer. These tasks test whether agents can express self-bias through direct valuation and self-rating responses.

In the EE task, agents reproduce the core direction of the endowment effect. Their WTA is higher than WTP, with WTA = 63.32 CNY and WTP = 48.37 CNY. The pattern appears in 97.8\% of participants, with paired $t(133)=19.94$, $p<.001$, and $d=1.72$. Compared with human data from Qi et al. (2025)~\cite{qi2025comprehensive}, where WTA = 107.84 and WTP = 74.36, the direction is preserved while the magnitude is smaller. This compression suggests that agents can represent the semantic structure of ownership and loss aversion, but their valuation scale remains more conservative than that of human participants.

The SE task shows a more mixed pattern. Human participants show slight self-enhancement, with an overall mean of 43.61 under the reverse coded scale. Agents instead show mild self-deprecation, with an overall mean of 53.72, one-sample $t(133)=5.20$ against 50, $p<.001$, and $d=0.45$. Dimension-level results are mixed: agents rate themselves better than average on sociability and honesty, but worse than average on intelligence and cooperation. This pattern suggests that explicit self-evaluation in agents is sensitive to scale direction, reference group framing, and calibration pressure from the underlying LLM. Taken across EE and SE, the explicit evaluative tasks indicate that direct numerical judgments are feasible for profile-based agents, although valuation intensity and self-evaluation calibration remain important sources of divergence from human baselines.

The indirect self-processing measures pose a different challenge. IAT and SRE are implemented as structured performance tasks that infer self-related cognition from behavioral traces rather than direct self-report. IAT records response latency and accuracy for identity-valence categorization, while SRE records recognition performance after identity-conditioned trait encoding. These tasks therefore require agents not only to understand the task procedure, but also to generate process-sensitive signals related to reaction time, identity-conditioned encoding, and memory retrieval.

In the IAT task, agents show near-identical reaction times in congruent and incongruent conditions, with 807.7 ms versus 809.3 ms, paired $t(133)=0.25$, $p=0.81$. Accuracy is 88.5\% in congruent trials and 86.9\% in incongruent trials. This result indicates that agents can follow the categorical structure of the IAT task and produce valid key presses and trial-level outputs, while the latency signal does not encode the expected difference between self-positive and self-negative associations. Symbolic categorization is therefore easier for current agents than latency-based association modeling.

In the SRE task, agents also complete the required procedure, but they do not show the expected self-reference memory advantage. Their old-item accuracy is 58.4\%, and their new-item accuracy is 63.4\%. By encoding identity, old-item accuracy is 57.2\% for self, 59.2\% for friend, and 58.7\% for other. The expected Self $>$ Friend $>$ Other gradient is not observed. This result suggests that current agents do not automatically bind trait encoding, identity labels, and later recognition judgments in the way required by SRE-style memory tasks. Although agents can complete both the encoding and recognition phases, their later memory judgments do not reliably reflect the identity condition attached to each trait during encoding.

These findings show that \textit{AgentSociety$^2$} supports the full psychological survey workflow, covering literature search, hypothesis generation, participant profile injection, custom environment construction, experiment execution, statistical testing, visualization generation, result interpretation, and report generation. The results also reveal where simulated respondents align with human baselines and where they diverge. For explicit evaluative measures, agents reproduce the direction of the endowment effect with compressed magnitude and show dimension-specific self-enhancement patterns under the reverse-coded scale. For indirect self-processing measures, agents complete the IAT and SRE procedures, but do not show stable latency differences for implicit self-association or the expected self-reference memory advantage. This contrast suggests that explicit evaluative measures are easier for profile-based agents to approximate, whereas indirect self-processing measures require stronger mechanisms for latency modeling, identity-conditioned encoding, and memory retrieval. The findings further point to concrete design directions for improving simulated agents, including scale checking and reference group calibration for explicit self-evaluation, association-based latency modeling for IAT, and memory components that bind traits, identities, and retrieval judgments for SRE. Building on this analysis, Figure~\ref{fig:manuscript-drafting}(c) shows the academic manuscript draft for this study as produced with support from \textit{AgentSociety$^2$}, where the literature-grounded hypotheses, EE, SE, IAT, and SRE protocols, human-agent comparisons, statistical tests, visualizations, and cross-task interpretation are organized into a unified academic paper. This study therefore presents \textit{AgentSociety$^2$} not only as a practical infrastructure for survey study replication, but also as a research system for comparing simulated agents with human respondents and guiding future agent mechanisms for social science simulation.



\subsection{Emergence of Information Cocoons}
Social media and its impact on users are a central topic in sociological research. To validate AgentSociety$^{2}$'s capability to model classic sociological phenomena and to demonstrate its capacity for large-scale experiments, this illustrative practice simulates user behavior under a short-video recommendation system with 10,000 users for an equivalent of six months. We aim to reproduce the information cocoon effect in social media and to examine its age-related inequality. The experiment uploads user profiles and an item dataset as the basis for recommendation-system pretraining and agent decision-making. Building on the platform-generated custom agent framework (FinalCustomUserAgent) and a custom environment module (VideoRecommendationEnv), we simulate users' multi-round viewing behaviors in response to recommended short videos. The simulation outputs complete interaction logs, which are then used to compute viewing entropy and other information-cocoon-related indicators for subsequent analysis.

\subsubsection{Idea \& Hypothesis Development} 

To reproduce the classic sociological phenomenon of age-related unfairness in information cocoons, we first clarified the research theme within AgentSociety$^{2}$ and conducted a comprehensive review of related literature using the platform's built-in literature retrieval capability. Leveraging the platform's idea generation and hypothesis management functions, and further refining the ideas with additional human prompts, we finally formulated two core hypotheses: \textbf{H1} (universality)---after interacting with a social-media recommendation system for a certain period, user groups of all ages will exhibit significant information cocoon effects (i.e., a decrease in overall viewing entropy); and \textbf{H2} (age gradient)---older user groups will fall into deeper information cocoons, reflected by a larger decline in viewing entropy.

\subsubsection{Experiment Design \& Execution}

\noindent\textbf{Data preparation.} The experiment is constructed based on a large-scale real-world short-video dataset. We randomly sample 10,000 users as the agent population, and use 86,201 short videos as the item candidate pool, covering 35 categories such as gaming and news. User interaction logs are partitioned along the time dimension by usage duration: the first 25 hours of interactions are used as the pretraining set for the recommendation system. We also use the same first-25-hour data to model user profiles and to solidify user decision rules. Finally, the full item set, user profiles, personalized decision code, and users' historical interaction data are packaged as custom User Data and managed within the platform.

\begin{tcolorbox}[
    colback=black!3,
    colframe=black!40,
    fontupper=\normalfont\scriptsize, 
    fonttitle=\bfseries\scriptsize,
    width=\textwidth,
    sharp corners,
    boxrule=0.5pt,
    enhanced,
    breakable,
    title=Key Prompt --- Agent Development,
    coltitle=white,
    colbacktitle=green!40!black!60,
    attach boxed title to top left={yshift=-2mm, xshift=2mm},
    top=3mm,
    before upper={\linespread{1.21}\selectfont\parskip=0.3em\parindent=1.5em}
]
I want to design a custom agent framework for my experiment. The agent should represent a simple user who can read my User Data (e.g., profile attributes, dataset metadata, and a set of user-specific decision rules).

For each step, the environment will provide a list of recommended short videos. The agent must decide whether to watch each video using only rule-based decision-making (no free-form generation): it should first identify the video category, then call the corresponding user-specific decision code from User Data/rules for that category, and finally output a watch/not-watch decision.

Please implement the agent interface, clearly specifying the required inputs (from User Data and from the environment), the outputs (decision format and any logs to store), and how the agent loads and executes the user-specific decision code.
\end{tcolorbox}

\begin{tcolorbox}[
    colback=black!3,
    colframe=black!40,
    fontupper=\normalfont\scriptsize, 
    fonttitle=\bfseries\scriptsize,
    width=\textwidth,
    sharp corners,
    boxrule=0.5pt,
    enhanced,
    breakable,
    title=Key Prompt --- Environment Development,
    coltitle=white,
    colbacktitle=green!40!black!60,
    attach boxed title to top left={yshift=-2mm, xshift=2mm},
    top=3mm,
    before upper={\linespread{1.21}\selectfont\parskip=0.3em\parindent=1.5em}
]
I need to create a custom recommendation-system environment module and design the interaction logic between users and the recommender.

1) Backbone recommendation model: DIN.

2) Recommendation workflow: based on real interaction data. Cold-start pretraining: build training samples from each user's first 5 sequences and use them to pretrain the model before the simulation starts. Online interaction loop: (i) generate recommendations, (ii) apply category-diversity constraints and deduplication, (iii) simulate user interactions via agents, (iv) optionally perform supplementary/extra recommendations, and (v) update the model.

3) Experiment protocol: specify how the environment coordinates batched recommendation, logging, periodic incremental training, and any round-based scheduling.

Please define the module's APIs (inputs/outputs), internal state, and how it integrates with the agent framework.
\end{tcolorbox}

\noindent\textbf{Agent development.} Using AgentSociety$^{2}$'s agent-development utilities, we built a custom agent framework, FinalCustomUserAgent, which makes viewing decisions based on user profiles and personalized rule code. The agent directly inherits from the platform's base class (AgentBase), reads User Data (e.g., profiles and decision code), interacts with the environment module, and decides whether to watch recommended videos. Beyond core behavioral logic, we further designed a state-management structure to record each user's simulation state, interaction history, memory, and other signals, enabling richer interactions with the environment.

\noindent\textbf{Experimental protocol development.} The experiment includes all 10,000 users and runs for ten simulation rounds, corresponding to roughly six months of short-video consumption in the real world. In the initialization stage, the recommender in the environment module is pretrained; the formal simulation then iterates the following loop: the recommender generates a recommendation list and delivers it to users, users make batch viewing decisions, interaction logs are stored, and decisions are fed back to the recommender for updates and optimization. After the simulation ends, we obtain complete user interaction records for subsequent analysis.

\subsubsection{Result Analysis \& Interpretation}

\begin{figure}[!t]
    \centering
    \begin{subfigure}[b]{0.485\textwidth}
        \centering
        \includegraphics[width=\textwidth]{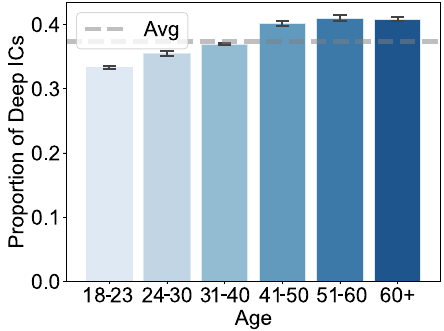}
        \caption{Real Dataset}
        \label{fig:ics_age_proportion_real}
    \end{subfigure}
    \hfill
    \begin{subfigure}[b]{0.485\textwidth}
        \centering
        \includegraphics[width=\textwidth]{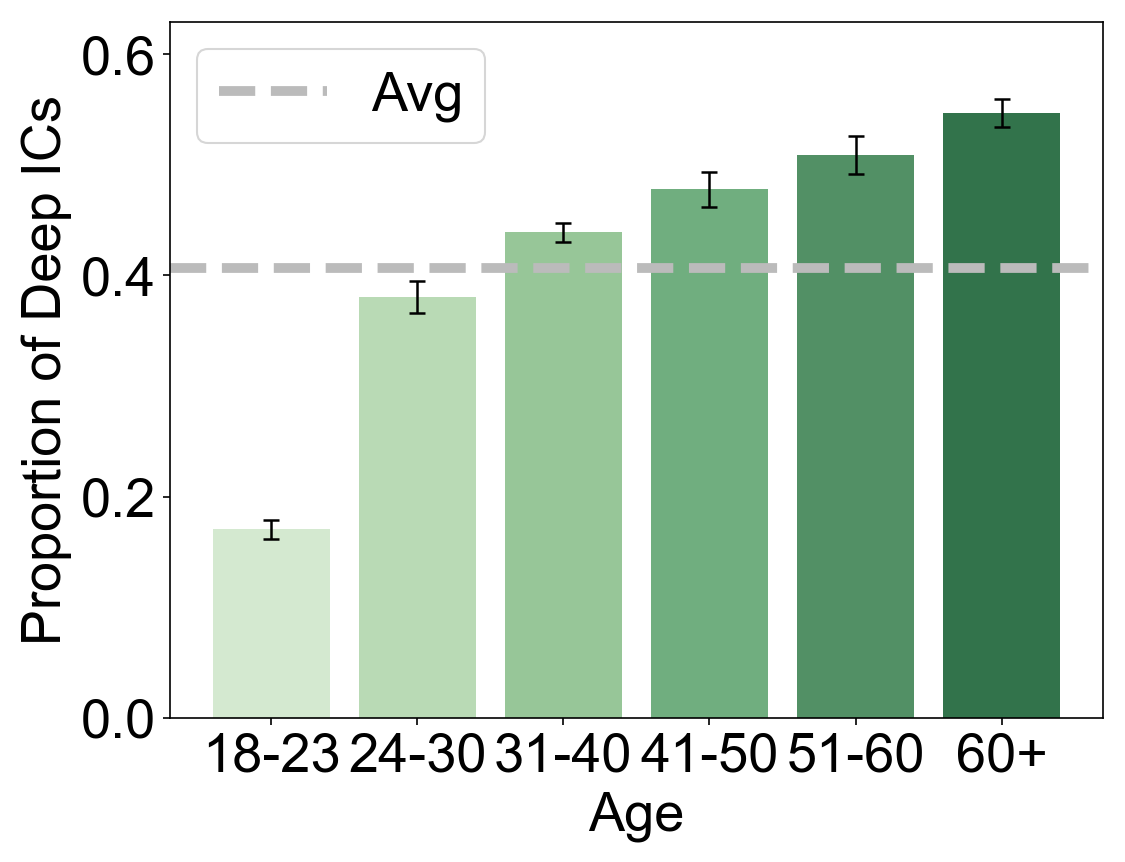}
        \caption{Experiment Result}
        \label{fig:ics_age_proportion_exp}
    \end{subfigure}
    \caption{Proportion of users falling into deep information cocoons across age groups: (a) real dataset and (b) experimental (simulated) result.}
    \label{fig:ics_age_proportion}
\end{figure}

After the simulation, we used the AI social scientist to analyze the results. It loaded the experimental data, computed information-cocoon-related metrics in an automated and multi-perspective manner, and conducted age-grouped analyses aligned with our hypotheses. Figure~\ref{fig:ics_age_proportion} shows the proportion of users trapped in deep information cocoons across different age groups, comparing the real dataset with our experimental (simulated) results.

\begin{figure}[!t]
    \centering
    \begin{subfigure}[b]{0.32\textwidth}
        \centering
        \includegraphics[width=\textwidth]{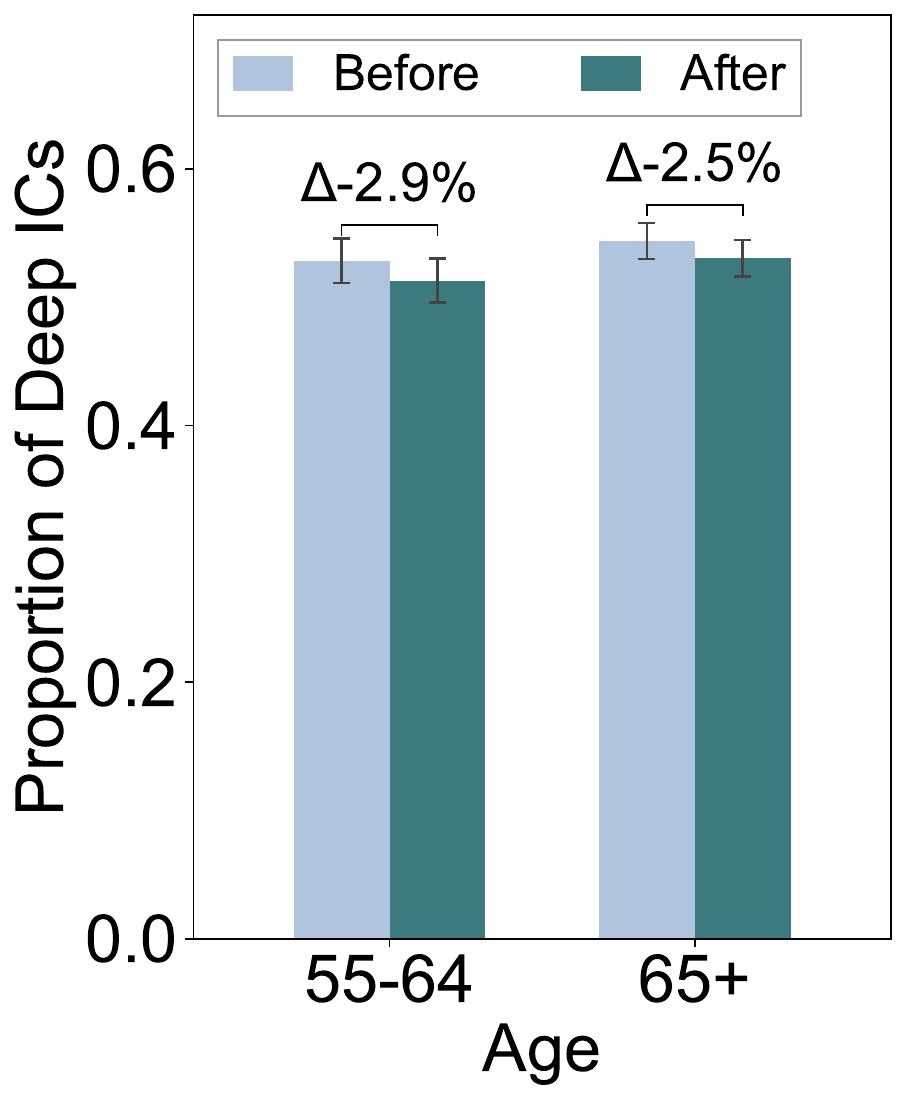}
        \caption{Strategy 1}
        \label{fig:ics_intervention_strategy1}
    \end{subfigure}
    \hfill
    \begin{subfigure}[b]{0.32\textwidth}
        \centering
        \includegraphics[width=\textwidth]{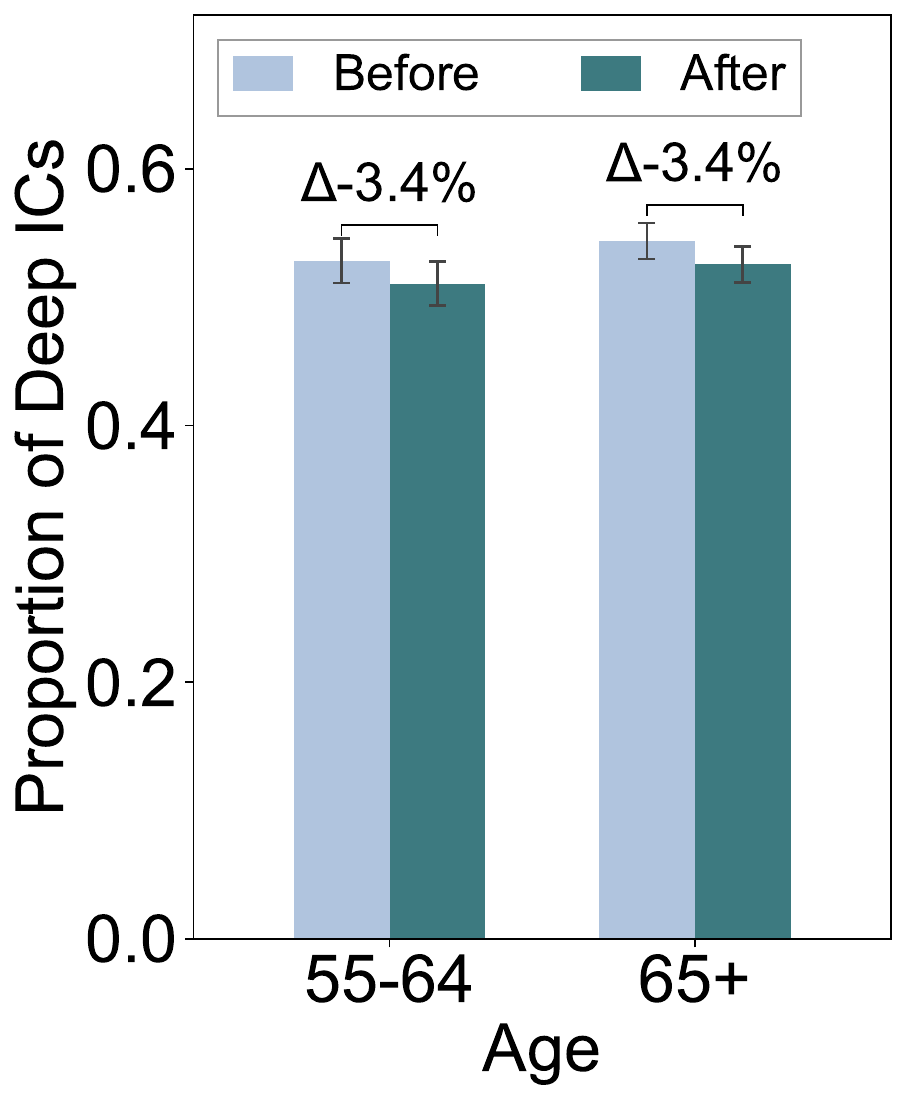}
        \caption{Strategy 2}
        \label{fig:ics_intervention_strategy2}
    \end{subfigure}
    \hfill
    \begin{subfigure}[b]{0.32\textwidth}
        \centering
        \includegraphics[width=\textwidth]{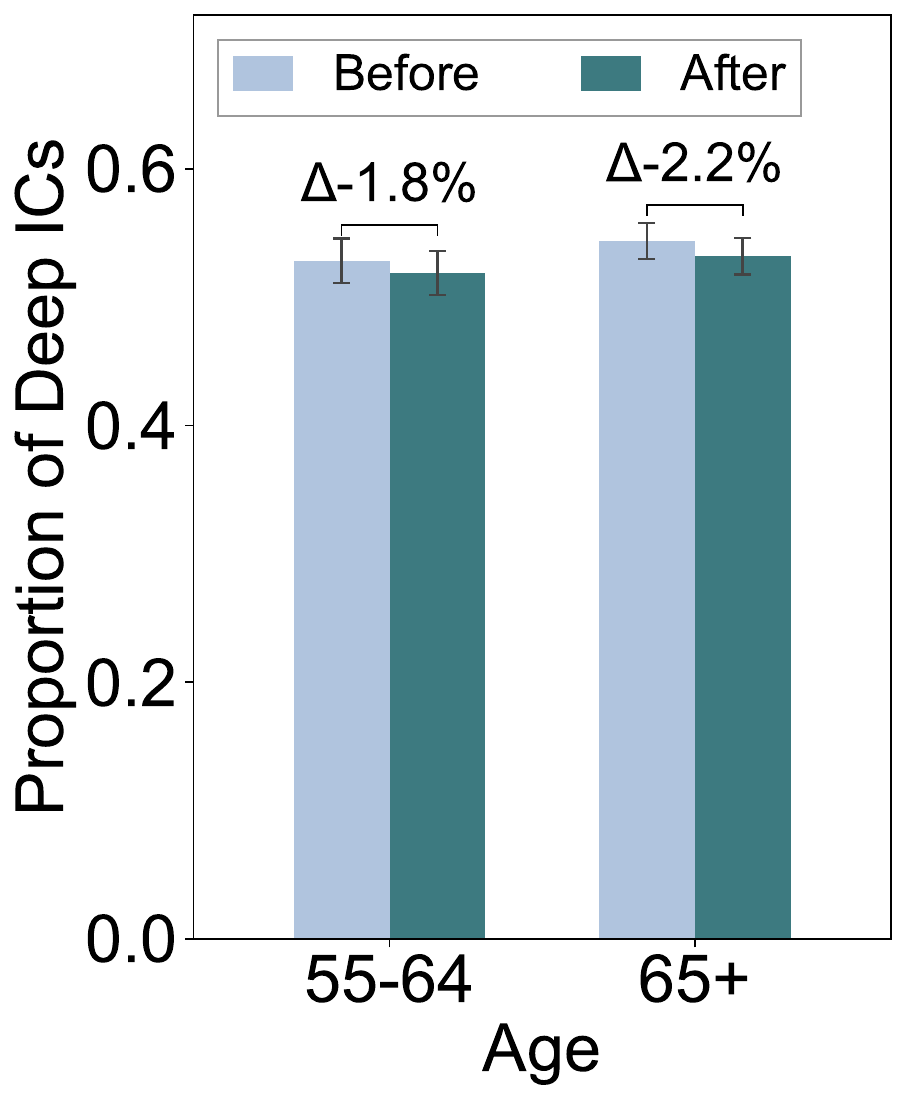}
        \caption{Strategy 3}
        \label{fig:ics_intervention_strategy3}
    \end{subfigure}
    \caption{Effects of intervention strategies on deep IC prevalence across age groups. (a)--(c) correspond to Strategy 1--Strategy 3, respectively, comparing the proportion of deep-IC users before and after intervention.}
    \label{fig:ics_intervention}
\end{figure}

Synthesizing our initial hypotheses, the relevant literature, and the simulation evidence, the AI social scientist---after several rounds of interaction with us---consolidated prior theoretical and empirical findings into three actionable, quota-aware reranking strategies tailored to older users (aged 50+), a cohort empirically found to be more susceptible to deep information cocoons. Specifically, it proposed: (1) \textit{Tail QuotaTransfer}, which transfers 30\% of the exposure quota from each top-10 category to the bottom-10 categories and then reranks under the updated quotas to increase tail exposure and diversity; (2) \textit{Efficiency Boost}, which reallocates 30\% of the top-10 quota to the top-10 non-head categories ranked by engagement efficiency $\rho=\text{watch}/\text{exposure}$ to better align exposure with user consumption; and (3) \textit{UnderServedBoost}, which, among non-top-10 categories, ranks content by a potential score that combines high efficiency with low recommendation volume and shifts 30\% of the top-10 quota to these targets to correct exposure inequality. Through tail-traffic migration, efficiency-weighted promotion, and underserved-content prioritization, these strategies reallocate exposure from mainstream (head) categories to a more diverse set of alternative domains. Empirically, all three strategies consistently reduce the probability that older users become trapped in deep information cocoons, showing measurable intervention effects across the 50+ age subgroups (see Figure~\ref{fig:ics_intervention}). Furthermore, the reporting output generated with support from \textit{AgentSociety$^2$} is shown in Figure~\ref{fig:manuscript-drafting}(d), where the manuscript brings the recommendation environment, profile-based user construction, age-group susceptibility analysis, reranking interventions, empirical effects, and interpretation into a complete research paper. 


Overall, this experiment faithfully reproduces the general emergence of information cocoons and the differentiated imbalance across age groups observed in real-world social media. It also comprehensively demonstrates AgentSociety$^{2}$'s strengths and potential in large-scale sociological research, spanning idea generation, environment construction, and end-to-end experimental support.

\subsection{Mechanisms behind Opinion Polarization}

Political attitudes on social media are jointly shaped by what users choose to follow and what platforms choose to rank. We reconstruct the field experiment of Levy~\cite{levy2021social} in a Facebook-style news feed and separate two simulated mechanisms, voluntary subscription and algorithmic down-ranking, to examine how each contributes to news reading, affective polarization, and issue-specific opinions over an eight-week treatment window.


\subsubsection{Idea \& Hypothesis Development}

The idea development stage treats the polarization experiment as a causal chain that begins with subscription offers, propagates through feed exposure and reading behavior, and ends in endline attitudes. The chain follows Levy's~\cite{levy2021social} field design, where counter-attitudinal offers were found to raise cross-cutting exposure and lower affective polarization at endline while issue opinions barely moved. Reproducing this chain in simulation requires keeping the voluntary subscription decision and the platform's ranking algorithm as separate operations, since the cross-cutting share of any user's feed depends on both how many counter-attitudinal sources that user actually accepts and how the platform reorders posts from those sources. Recording outcomes at every step then ensures that the effect of each operation can be read off the data rather than inferred indirectly.


From this design, we derive four hypotheses that the simulation should be able to evaluate against Levy~\cite{levy2021social}:
\begin{itemize}[leftmargin=15pt]
    \item \textbf{H1 (Subscription to Exposure).} A counter-attitudinal subscription offer raises the share of cross-cutting posts that an agent sees in the feed at endline relative to the control condition, while a pro-attitudinal offer lowers it.
    \item \textbf{H2 (Exposure to Reading).} The cross-cutting share that is realized in the feed accumulates into cumulative reading volume over the eight-week window, so the treatment difference is clearer in cumulative reads than in any single weekly exposure snapshot.
    \item \textbf{H3 (Reading to Affective Polarization).} Higher cross-cutting reading reduces the change in affective polarization at endline, while the pro-attitudinal condition raises it.
    \item \textbf{H4 (Issue Opinions Comparatively Stable).} The eight-week treatment leaves the 20-item political-opinion profile comparatively unchanged across conditions, so any measurable polarization shift comes mainly from affective evaluations rather than from issue positions.
\end{itemize}


\begin{tcolorbox}[
    colback=black!3,
    colframe=black!40,
    fontupper=\normalfont\scriptsize, 
    fonttitle=\bfseries\scriptsize,
    width=\textwidth,
    sharp corners,
    boxrule=0.5pt,
    enhanced,
    breakable,
    title=Key Prompt --- Literature Understanding and Hypothesis Formation,
    coltitle=white,
    colbacktitle=green!40!black!60,
    attach boxed title to top left={yshift=-2mm, xshift=2mm},
    top=3mm,
    before upper={\linespread{1.21}\selectfont\parskip=0.3em\parindent=1.5em}
]
Please read Levy (2021), \textit{Social Media, News Consumption, and Polarization: Evidence from a Field Experiment}, and identify the causal mechanism that should be preserved in an \textit{AgentSociety$^2$} simulation.

The hypothesis should distinguish between pro-attitudinal and counter-attitudinal subscription offers, and should explain how each condition changes subscriptions, feed exposure, reading and sharing behavior, affective polarization, and political opinions. Please also specify why algorithmic ranking should be modeled as a separate mechanism from user subscription choices, and define the main outcomes needed to test this mechanism in a simulated news-feed environment.
\end{tcolorbox}

\subsubsection{Experiment Design \& Execution}

\noindent\textbf{Data Preparation.}
The experiment generator builds three executable configurations from the same population template. Each configuration contains 200 agents, with 100 liberal agents and 100 conservative agents sampled from ideology scores on the two sides of the political spectrum. For each agent, the configuration records demographic attributes, baseline news habits, baseline subscriptions, baseline feeling-thermometer values, affective-index components, endline completion status, and a 20-item political-opinion profile. Treatment assignments then determine which outlets are offered. The control condition provides no new subscription offer; the pro-attitudinal condition offers liberal outlets to liberal agents and conservative outlets to conservative agents; and the counter-attitudinal condition reverses this mapping. Offered outlets are capped at four and use replacement outlets when an agent already follows a primary outlet at baseline.


\noindent\textbf{Agent Development.}
The simulation uses a custom \texttt{NewsConsumerAgent} to represent a social media news user. At initialization, each agent receives an ideology score, treatment label, baseline attitudes, baseline political opinions, news habits, and demographic attributes. During each simulation step, the agent first records the response to the subscription offer, then retrieves a personalized feed, reads a small number of posts according to rank and ideological relevance, and shares selected posts with a calibrated probability. The agent then reports affective measures derived from accumulated exposure and reading balances. At the final step, agents marked as endline respondents submit the 20-item political-opinion battery.


\begin{tcolorbox}[
    colback=black!3,
    colframe=black!40,
    fontupper=\normalfont\scriptsize, 
    fonttitle=\bfseries\scriptsize,
    width=\textwidth,
    sharp corners,
    boxrule=0.5pt,
    enhanced,
    breakable,
    title=Key Prompt --- Agent Development,
    coltitle=white,
    colbacktitle=green!40!black!60,
    attach boxed title to top left={yshift=-2mm, xshift=2mm},
    top=3mm,
    before upper={\linespread{1.21}\selectfont\parskip=0.3em\parindent=1.5em}
]
Please implement a custom \texttt{NewsConsumerAgent} for the polarization experiment. The agent should store ideology, treatment assignment, baseline attitudes, baseline political opinions, news habits, and demographic attributes.

During each step, the agent should respond to the subscription offer, request the personalized feed, select posts to read according to rank and ideological relevance, share selected posts with a calibrated probability, and report affective measures based on accumulated exposure and reading behavior. At the final step, agents marked as endline respondents should submit the political-opinion battery.
\end{tcolorbox}

\noindent\textbf{Environment Development.}
The custom \texttt{NewsPolarizationEnv} represents the platform and the news supply. It defines ideology-scored liberal and conservative outlets, generates topic-specific posts for the 2018 study period, records baseline and treatment subscriptions, and exposes structured tools for offer response, feed retrieval, reading, sharing, attitude reporting, political-opinion reporting, consumption statistics, and exposure-gap decomposition. The feed is built from outlets followed by each agent and ranked with a fixed penalty for counter-attitudinal posts. In the current configuration, the feed contains up to 20 posts, each outlet generates three posts per step, and \texttt{algorithm\_filter\_strength} is set to 0.4. This setting keeps the subscription intervention and the platform ranking mechanism observable as separate parts of the experiment.


\begin{tcolorbox}[
    colback=black!3,
    colframe=black!40,
    fontupper=\normalfont\scriptsize, 
    fonttitle=\bfseries\scriptsize,
    width=\textwidth,
    sharp corners,
    boxrule=0.5pt,
    enhanced,
    breakable,
    title=Key Prompt --- Environment Development,
    coltitle=white,
    colbacktitle=green!40!black!60,
    attach boxed title to top left={yshift=-2mm, xshift=2mm},
    top=3mm,
    before upper={\linespread{1.21}\selectfont\parskip=0.3em\parindent=1.5em}
]
Please implement a custom \texttt{NewsPolarizationEnv} for the simulated news-feed platform. The environment should define ideology-scored liberal and conservative outlets, generate topic-specific posts for the study period, record baseline and treatment subscriptions, and expose structured tools for offer response, feed retrieval, reading, sharing, attitude reporting, political-opinion reporting, consumption statistics, and exposure-gap decomposition.

The feed should be built from outlets followed by each agent. Apply a fixed ranking penalty to counter-attitudinal posts, keep the feed size configurable, and record pro-attitudinal and counter-attitudinal exposure separately.
\end{tcolorbox}

\noindent\textbf{Experimental Protocol.}
Each condition follows the same eight-week schedule. The simulation starts on February 28, 2018 with a baseline survey that records party feeling thermometers and, in treatment conditions, the subscription-offer notice. The system then runs eight weekly interaction steps. In each step, agents respond to the current subscription state, browse the ranked feed, read and share posts, and update affective reports through environment tools. After the weekly steps, the endline survey collects feeling thermometers, perspective-taking items, party-idea ratings, the child-marriage affective item, and the 20 political-opinion items. The resulting records support comparisons across control, pro-attitudinal, and counter-attitudinal conditions at the levels of subscription compliance, feed exposure, reading and sharing behavior, affective polarization, political opinions, and exposure-gap decomposition.


\begin{tcolorbox}[
    colback=black!3,
    colframe=black!40,
    fontupper=\normalfont\scriptsize, 
    fonttitle=\bfseries\scriptsize,
    width=\textwidth,
    sharp corners,
    boxrule=0.5pt,
    enhanced,
    breakable,
    title=Key Prompt --- Configuration Generation,
    coltitle=white,
    colbacktitle=green!40!black!60,
    attach boxed title to top left={yshift=-2mm, xshift=2mm},
    top=3mm,
    before upper={\linespread{1.21}\selectfont\parskip=0.3em\parindent=1.5em}
]
Please generate three \textit{AgentSociety$^2$} experiment configurations based on Levy (2021): a control condition, a pro-attitudinal subscription condition, and a counter-attitudinal subscription condition.

Each configuration should contain 200 ideological news users, with balanced liberal and conservative groups. For each user, include ideology score, demographic attributes, baseline news habits, baseline subscriptions, baseline affective measures, endline completion flag, and a 20-item political-opinion profile. Assign outlet offers according to the condition and use replacement outlets when a primary outlet is already followed at baseline.
\end{tcolorbox}

\begin{tcolorbox}[
    colback=black!3,
    colframe=black!40,
    fontupper=\normalfont\scriptsize, 
    fonttitle=\bfseries\scriptsize,
    width=\textwidth,
    sharp corners,
    boxrule=0.5pt,
    enhanced,
    breakable,
    title=Key Prompt --- Execution Schedule,
    coltitle=white,
    colbacktitle=green!40!black!60,
    attach boxed title to top left={yshift=-2mm, xshift=2mm},
    top=3mm,
    before upper={\linespread{1.21}\selectfont\parskip=0.3em\parindent=1.5em}
]
Please create the step schedule for the eight-week treatment process. The schedule should begin with a baseline survey, run eight weekly interaction steps, and end with an endline survey.

The baseline survey should record feeling thermometers and treatment notices when applicable. The endline survey should collect feeling thermometers, perspective-taking items, party-idea ratings, the child-marriage affective item, and the 20 political-opinion items. Store all records in a format suitable for comparing subscription compliance, feed exposure, reading and sharing behavior, affective polarization, political opinions, and exposure-gap decomposition across the three conditions.
\end{tcolorbox}

\subsubsection{Result Analysis \& Interpretation}

\begin{figure}[!t]
    \centering
    \begin{subfigure}[t]{0.32\textwidth}
        \centering
        \includegraphics[width=\linewidth]{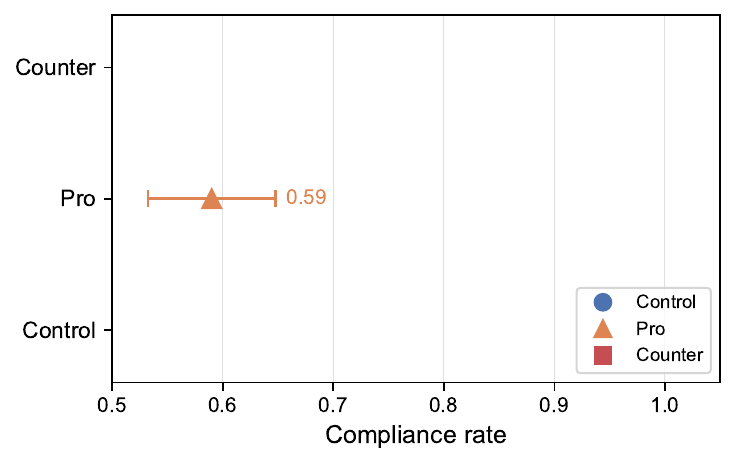}
        \caption{Subscription compliance}
        \label{fig:polarization-mechanism-compliance}
    \end{subfigure}
    \hfill
    \begin{subfigure}[t]{0.32\textwidth}
        \centering
        \includegraphics[width=\linewidth]{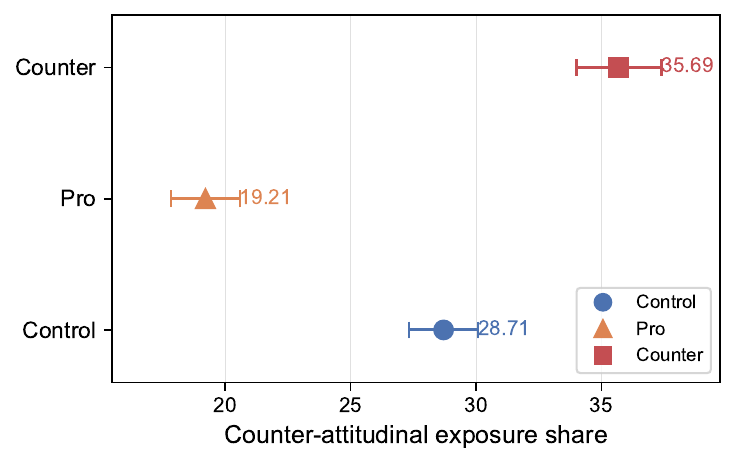}
        \caption{Counter-attitudinal exposure}
        \label{fig:polarization-mechanism-exposure}
    \end{subfigure}
    \hfill
    \begin{subfigure}[t]{0.32\textwidth}
        \centering
        \includegraphics[width=\linewidth]{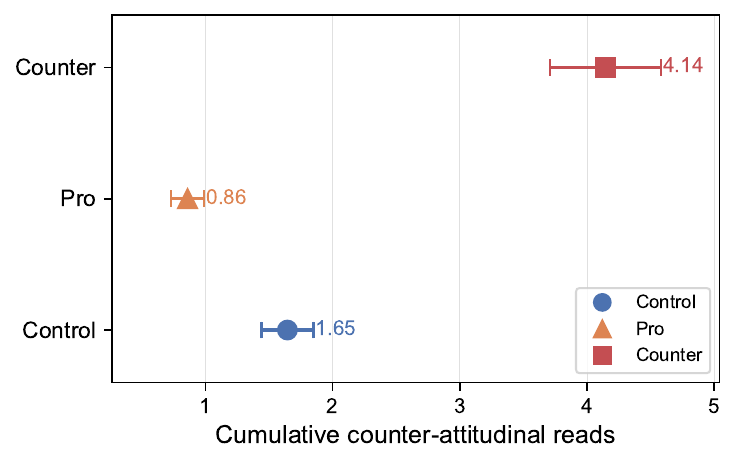}
        \caption{Counter-attitudinal reading}
        \label{fig:polarization-mechanism-reading}
    \end{subfigure}
    \caption{Behavioral cascade from subscription offers to cross-cutting consumption. Pro and Counter denote pro-attitudinal and counter-attitudinal treatment conditions. Points indicate group means and horizontal intervals indicate 90\% confidence intervals.}
    \label{fig:polarization_mechanism}
\end{figure}

\noindent\textbf{Behavioral Cascade from Subscriptions to Reading.} After the simulation, \textit{AgentSociety$^2$} analyzes the structured records stored in \texttt{news\_polarization\_agent\_state} and \texttt{news\_polarization\_event}. The analysis follows the causal chain of the experiment from subscription uptake through feed exposure to reading. Figure~\ref{fig:polarization_mechanism} shows that the three steps order the conditions differently. Subscription compliance is highest in the pro-attitudinal condition at 0.59, with 1.36 accepted treatment outlets on average, and lower in the counter-attitudinal condition at 0.43 with 0.87 accepted outlets. The downstream feed exposure reverses this ordering: counter-attitudinal exposure reaches 35.69\% of the feed in the counter-attitudinal condition, against 28.71\% in the control condition and 19.21\% in the pro-attitudinal condition. Cumulative counter-attitudinal reading widens the gap further: counter-attitudinal agents read 4.15 cross-cutting posts on average, control agents read 1.65, and pro-attitudinal agents read 0.86. Together these three steps show that exposure outcomes can move in the opposite direction from raw subscription uptake once feed ranking and reading selection enter the chain.

\begin{figure}[!t]
    \centering
    \begin{subfigure}[t]{0.42\textwidth}
        \centering
        \includegraphics[width=\linewidth]{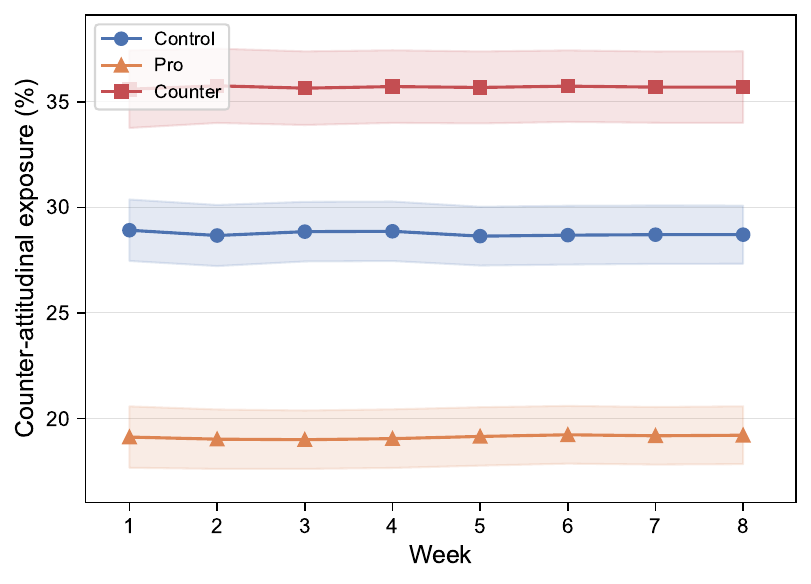}
        \caption{Exposure over time}
        \label{fig:polarization-dynamics-exposure}
    \end{subfigure}%
    \hspace{0.05\textwidth} 
    \begin{subfigure}[t]{0.42\textwidth}
        \centering
        \includegraphics[width=\linewidth]{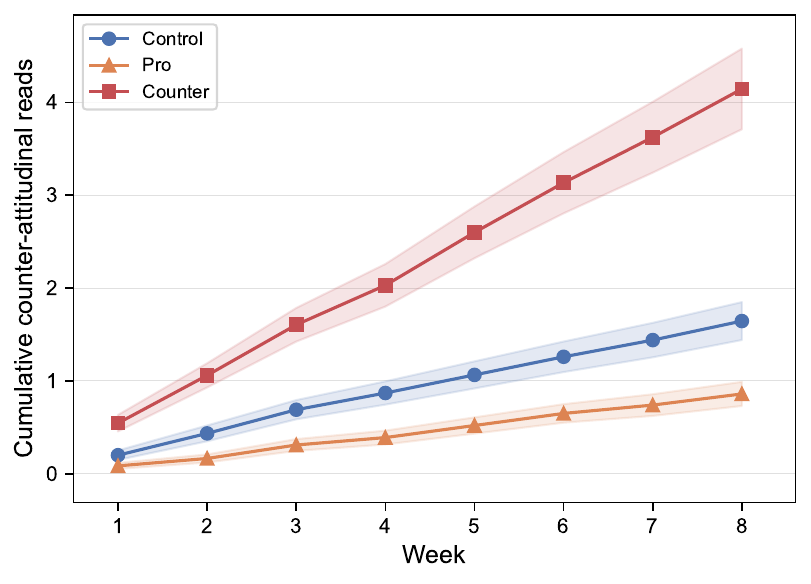}
        \caption{Reading over time}
        \label{fig:polarization-dynamics-reading}
    \end{subfigure}

    \vspace{-0.2em}
    
    \begin{subfigure}[t]{0.42\textwidth}
        \centering
        \includegraphics[width=\linewidth]{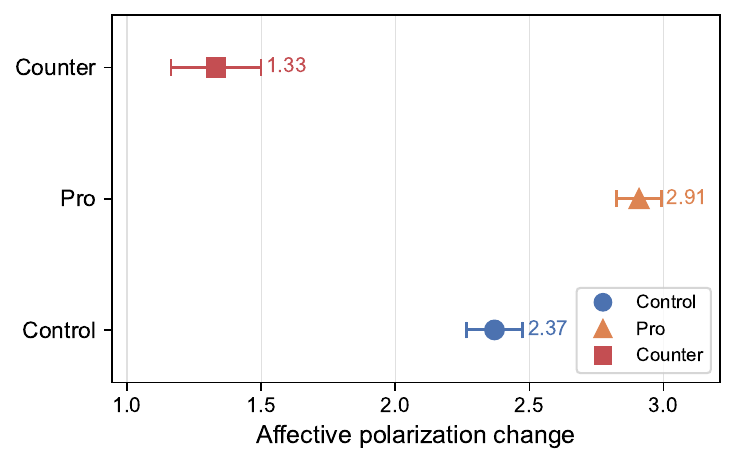}
        \caption{Affective polarization (endline change)}
        \label{fig:polarization-dynamics-affective}
    \end{subfigure}%
    \hspace{0.05\textwidth} 
    \begin{subfigure}[t]{0.42\textwidth}
        \centering
        \includegraphics[width=\linewidth]{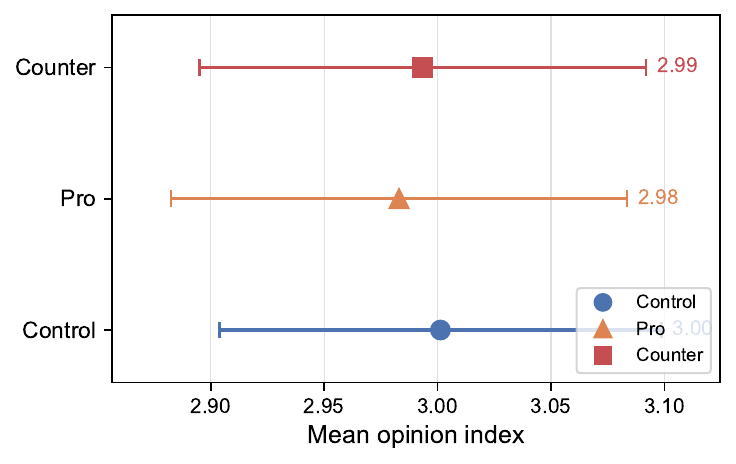}
        \caption{Political opinions (endline index)}
        \label{fig:polarization-dynamics-opinions}
    \end{subfigure}
    
    \vspace{0.5em} 
    \caption{Temporal dynamics (a, b) and endline outcomes (c, d) in the polarization experiment. Pro and Counter denote pro-attitudinal and counter-attitudinal treatment conditions. Shaded regions and horizontal intervals indicate 90\% confidence intervals.}
    \label{fig:polarization_dynamics_outcomes}
\end{figure}

\noindent\textbf{Reading Accumulates the Exposure Gap.} Figure~\ref{fig:polarization_dynamics_outcomes} examines whether the mechanism persists over the eight-week treatment window. The temporal panels show that counter-attitudinal exposure is stable, while reading differences accumulate over repeated interactions. From week 1 to week 8, counter-attitudinal exposure changes only slightly in all three conditions: from 28.92\% to 28.71\% in the control condition, from 19.13\% to 19.21\% in the pro-attitudinal condition, and from 35.59\% to 35.69\% in the counter-attitudinal condition. In contrast, cumulative counter-attitudinal reading grows from 0.20 to 1.65 posts in the control condition, from 0.09 to 0.86 posts in the pro-attitudinal condition, and from 0.55 to 4.15 posts in the counter-attitudinal condition. The main treatment difference therefore appears as accumulated reading opportunities under a stable exposure regime, with minimal week-by-week drift in feed composition.

\noindent\textbf{Temporal Dynamics: Stable Exposure, Accumulating Reads.} Figure~\ref{fig:polarization_dynamics_outcomes}(a, b) examines whether the cascade persists over the eight-week treatment window. Counter-attitudinal exposure is essentially stable from week 1 to week 8: from 28.92\% to 28.71\% in the control condition, from 19.13\% to 19.21\% in the pro-attitudinal condition, and from 35.59\% to 35.69\% in the counter-attitudinal condition. Cumulative counter-attitudinal reading instead grows monotonically: from 0.20 to 1.65 posts in the control condition, from 0.09 to 0.86 in the pro-attitudinal condition, and from 0.55 to 4.15 in the counter-attitudinal condition. Once subscriptions stabilize, feed composition behaves as a constant supply regime, and the treatment differences accumulate through repeated reading rather than through any growing weekly gap in exposure.


\noindent\textbf{Affective Polarization Shifts More than Issue Opinions.} Figure~\ref{fig:polarization_dynamics_outcomes}(c, d) compares the two endline outcomes. The change in affective polarization between baseline and endline separates clearly across conditions: the increase is 2.36 in the control condition, 2.80 in the pro-attitudinal condition, and 1.28 in the counter-attitudinal condition. The 20-item political-opinion index is, in contrast, almost identical across conditions, with values of 3.01 in the control condition, 3.04 in the pro-attitudinal condition, and 2.99 in the counter-attitudinal condition. The simulated counter-attitudinal offer therefore reduces the change in affective polarization by 1.08 points relative to the control condition and raises it by 0.44 points under the pro-attitudinal offer, while issue opinions move by at most 0.06 points on the same index. Within the eight-week window, the simulated treatment shifts attitudes toward political groups much more than it shifts positions on specific issues.


\begin{figure}[!t]
    \centering
    \begin{subfigure}[t]{0.48\textwidth}
        \centering
        \includegraphics[width=\linewidth]{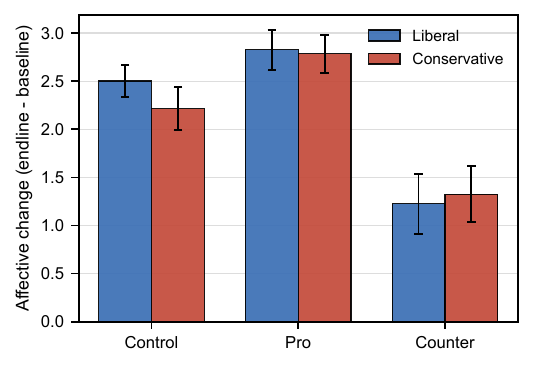}
        \caption{Affective change by ideology}
        \label{fig:polarization-heterogeneity-affective}
    \end{subfigure}
    \hfill
    \begin{subfigure}[t]{0.48\textwidth}
        \centering
        \includegraphics[width=\linewidth]{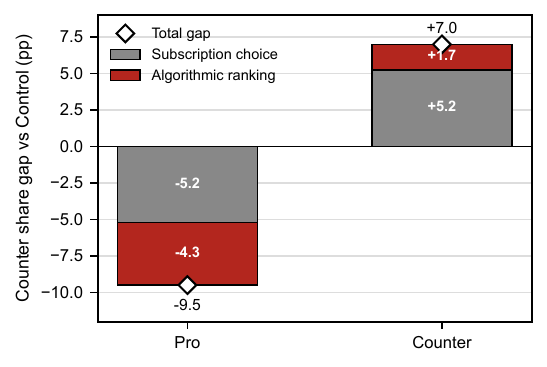}
        \caption{Exposure-gap decomposition}
        \label{fig:polarization-heterogeneity-decomposition}
    \end{subfigure}

    \vspace{0.8em}

    \begin{subfigure}[t]{0.48\textwidth}
        \centering
        \includegraphics[width=\linewidth]{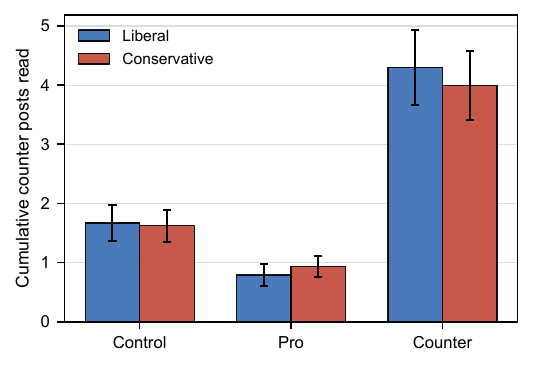}
        \caption{Cumulative cross-cutting reads by ideology}
        \label{fig:polarization-heterogeneity-reads}
    \end{subfigure}
    \hfill
    \begin{subfigure}[t]{0.48\textwidth}
        \centering
        \includegraphics[width=\linewidth]{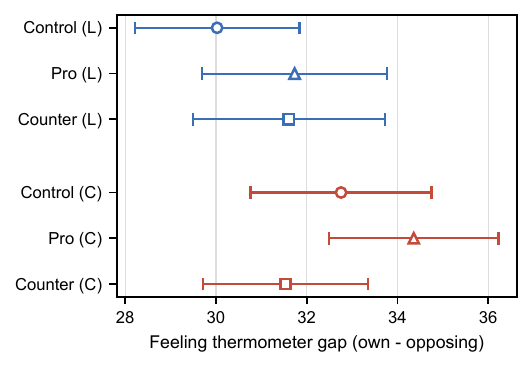}
        \caption{Endline feeling-thermometer gap by ideology}
        \label{fig:polarization-heterogeneity-feeling}
    \end{subfigure}
    \caption{Heterogeneity by ideology (a, c, d) and the algorithm-versus-choice decomposition of the cross-cutting exposure gap (b). Lib and Con denote liberal and conservative agent subgroups. Error bars in (a), (c), (d) indicate 90\% confidence intervals around the subgroup mean.}
    \label{fig:polarization_heterogeneity}
\end{figure}

\noindent\textbf{Heterogeneity and the Algorithm-versus-Choice Decomposition.} Figure~\ref{fig:polarization_heterogeneity} examines whether the cascade is symmetric across ideological subgroups and decomposes the exposure gap into two parts: the subscription mix that an agent voluntarily accepts, and the additional ranking penalty applied by the platform. Panel (a) shows that the affective-change pattern is symmetric across liberal and conservative agents in each condition, with the counter group reducing the change in both subgroups (1.23 for liberal agents, 1.32 for conservative agents) relative to the control group (2.50 and 2.21 respectively). Panel (c) shows the same symmetry for cumulative cross-cutting reads, with the counter group reading roughly four cross-cutting posts in both subgroups. Panel (d) reports a small additional asymmetry at endline: conservative agents start from a larger feeling-thermometer gap between own and opposing parties (32.76 in control versus 30.03 for liberal agents) and respond more strongly to the pro-attitudinal offer (34.36 versus 31.73), while both subgroups end at a similar 31.5 to 31.6 gap under the counter-attitudinal offer. Panel (b) reports the exposure-gap decomposition relative to the control condition: in the pro-attitudinal condition, subscription choice contributes $-5.23$ pp and the ranking penalty contributes $-4.27$ pp, for a total gap of $-9.50$ pp; in the counter-attitudinal condition, subscription choice contributes $+5.24$ pp and the ranking penalty contributes $+1.75$ pp, for a total of $+6.99$ pp. In this simulation, the larger share of the exposure gap is therefore generated by which outlets each agent decides to follow, with the ranking algorithm amplifying that choice asymmetrically across conditions.


\begin{figure}[!t]
    \centering
    \includegraphics[width=\textwidth]{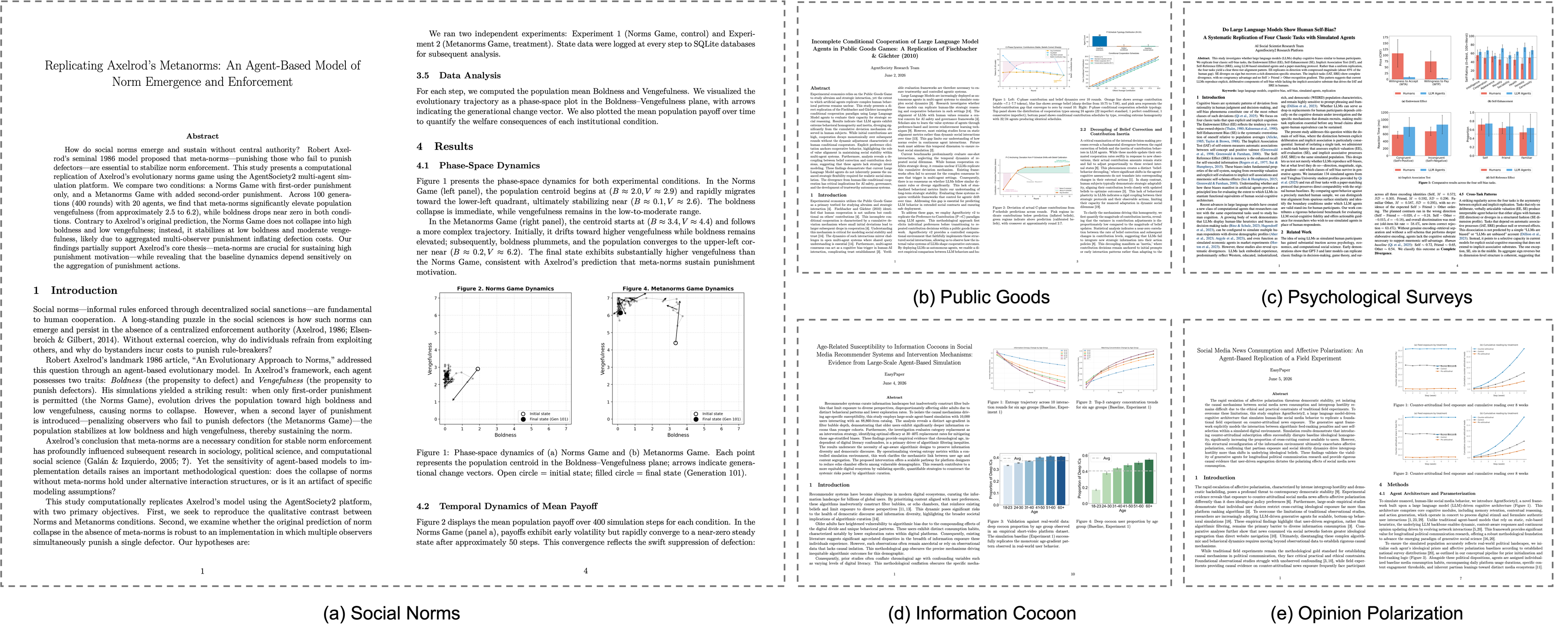}
    \caption{
Academic manuscript drafting across social science studies using \textit{AgentSociety$^2$}: 
(a) Social Norms, 
(b) Public Goods, 
(c) Psychological Surveys, 
(d) Information Cocoon, and 
(e) Opinion Polarization.
}
    \label{fig:manuscript-drafting}
\end{figure}

\noindent\textbf{Alignment with Levy (2021).} The simulated cascade reproduces the direction of every outcome that Levy~\cite{levy2021social} reports: counter-attitudinal offers raise cross-cutting exposure and reading, reduce affective polarization at endline, and leave issue opinions comparatively stable, while subscription compliance is somewhat lower for counter-attitudinal offers than for pro-attitudinal offers (Table~\ref{tab:polarization_levy_alignment}). The pro-attitudinal condition mirrors these patterns in the opposite direction. The simulation diverges from Levy's field study on two points of magnitude rather than direction. The compliance gap between conditions reaches 16 pp in the simulation (0.59 versus 0.43) against 11 pp in the field study (0.59 versus 0.48), which we attribute to more uniform agent behavior under a fixed prompt than to the heterogeneity of human survey respondents. Levy also concludes that the bulk of the cross-cutting exposure gap is generated by algorithmic ranking rather than by subscription choice (his Section~VI and Figure~10), whereas the simulated decomposition assigns the larger share to subscription choice in both treatment conditions. The latter discrepancy is consistent with our calibration of \texttt{algorithm\_filter\_strength} = 0.4 in \texttt{NewsPolarizationEnv}, which keeps the ranking penalty mild relative to the change in subscriptions, so the relative weight of the two channels remains a calibration target rather than a fixed outcome. As part of the final reporting stage, Figure~\ref{fig:manuscript-drafting}(e) illustrates the manuscript produced with \textit{AgentSociety$^2$}, presenting the study from research motivation and social-media environment design to the subscription--exposure--reading cascade, affective-polarization outcomes, heterogeneity analysis, and alignment with Levy's field experiment. In summary, this practice shows that \textit{AgentSociety$^2$} can host social-media experiments in which platform structure and user agency are both observable and separately controllable.



\begin{table}[!t]
\centering
\small
\caption{Alignment of simulated polarization outcomes with Levy~\cite{levy2021social}.}
\label{tab:polarization_levy_alignment}
\begin{tabular}{p{0.30\textwidth} p{0.24\textwidth} p{0.26\textwidth} c}
\toprule
\textbf{Outcome (Counter vs Control unless noted)} & \textbf{Levy 2021 (field)} & \textbf{Simulation} & \textbf{Match} \\
\midrule
Subscription compliance (Pro / Counter)            & 0.59 / 0.48           & 0.59 / 0.43                         & Yes \\
Cross-cutting exposure                              & Positive              & $+6.99$ pp                         & Yes \\
Cumulative cross-cutting reads                      & Higher in Counter     & 4.15 vs 1.65 posts                 & Yes \\
Affective polarization change                       & $-0.03$ SD (ITT), $-0.06$ SD (TOT) & $-1.08$ on raw-change scale & Yes \\
Issue-opinion index                                 & Near zero             & $\leq 0.06$ index points            & Yes \\
Main driver of exposure gap                         & Algorithmic ranking   & Subscription choice                 & Partial \\
\bottomrule
\end{tabular}
\begin{flushleft}
\small \textit{Note: ITT and TOT denote intent-to-treat and treatment-on-the-treated estimates as reported in Levy~\cite{levy2021social}. The ``Main driver'' row compares the decomposition in Figure~\ref{fig:polarization_heterogeneity}(b) with Levy's Section~VI and Figure~10.}
\end{flushleft}
\end{table}

\subsection{Daily Mobility Simulation}

The development of LLMs provides a feasible approach to simulating complex behavioral patterns of individuals, enabling the reconstruction of microscopic and realistic human societal dynamics. As demonstrated by the \textit{AgentSociety} framework, deploying large-scale LLM agents to simulate human societies not only assists practitioners in social and management sciences in obtaining new scientific discoveries but also improves real-world planning and decision-making. To verify the human-like capabilities of LLM agents under complex urban constraints, this study uses AI social scientist to reproduce and evaluate the DailyMobility benchmark.

Daily mobility simulation evaluates whether profile-based agents can generate realistic daily activity and movement patterns in a city-scale environment. In this experiment, we construct a Beijing daily mobility simulation with 100 agents initialized from on-platform datasets. Each profile provides demographic information as well as home and work AOI (Area of Interest) identifiers, which are used for agent initialization and routine mobility generation. The simulation follows the DailyMobility benchmark setting and evaluates both activity intentions and spatial movement outcomes.

\subsubsection{Idea \& Hypothesis Development}

To develop the simulation design for DailyMobility, AI social scientist shifted the focus from static distribution matching to the generative capabilities of LLM agents. The primary research question centers on whether agents can synthesize demographic profiles and spatial constraints into authentic daily routines. By initializing agents with specific home and work AOIs, the experiment tests if autonomous decision-making can effectively mirror the spatial-temporal patterns found in ground-truth data. This approach establishes a functional connection between the original benchmark and the AgentSociety2 platform.

It therefore formulated the experiment as a benchmark-oriented simulation hypothesis: if agents are initialized with demographic information, home AOI, and work AOI, then their autonomous daily planning and mobility decisions should reproduce the temporal intention patterns and spatial movement statistics observed in the DailyMobility ground-truth data. This hypothesis serves as the conceptual bridge between the original benchmark dataset and the executable simulation design in AgentSociety2.

\subsubsection{Experiment Design \& Execution}

\noindent\textbf{Data Preparation.}
The experiment is designed as a benchmark reproduction task based on the DailyMobility dataset. The data preparation stage constructs a Beijing daily mobility population with 100 profile-based agents. Each agent profile contains demographic attributes as well as home and work AOI (Area of Interest) identifiers, which provide the basic spatial anchors for daily routine generation. These profiles are used to initialize agents in the city-scale mobility environment. The benchmark ground-truth data provides reference distributions for activity intentions and spatial mobility outcomes, including intention sequences, aggregate intention proportions, gyration radius, and daily visited-location numbers.

\noindent\textbf{Agent Development.}
The simulation uses profile-based \texttt{PersonAgent}s to represent urban residents. At initialization, each agent receives demographic information, a home AOI, and a work AOI. These attributes define the agent's basic daily constraints and are used to support activity planning, destination selection, and movement behavior. During the simulation, agents autonomously generate daily activity intentions and corresponding mobility decisions under a 15-minute temporal grid. Their behaviors are recorded as both intention-level outputs and trajectory-level outputs, allowing the benchmark to evaluate whether LLM agents can transform individual profile information into realistic daily mobility patterns.

\begin{tcolorbox}[
    colback=black!3,
    colframe=black!40,
    fontupper=\normalfont\scriptsize, 
    fonttitle=\bfseries\scriptsize,
    width=\textwidth,
    sharp corners,
    boxrule=0.5pt,
    enhanced,
    breakable,
    title=Key Prompt --- Agent Development,
    coltitle=white,
    colbacktitle=green!40!black!60,
    attach boxed title to top left={yshift=-2mm, xshift=2mm},
    top=3mm,
    before upper={\linespread{1.21}\selectfont\parskip=0.3em\parindent=1.5em}
]
Please initialize profile-based \texttt{PersonAgent}s for the Beijing DailyMobility benchmark. Each agent should receive demographic information, a home AOI, and a work AOI.

During the simulation, each agent should generate full-day activity intentions and corresponding mobility decisions under a 15-minute temporal grid. The generated behaviors should be recorded as intention sequences and spatial trajectories so that they can be compared with the DailyMobility ground-truth data.
\end{tcolorbox}

\noindent\textbf{Environment Development.}
The mobility environment represents the Beijing city-scale spatial context and provides the AOI-based structure required for daily movement simulation. It supports the initialization of home and work locations, records agent trajectories, and exports visited-location statistics for post-experiment evaluation. In the AI social scientist workflow, the environment is configured to preserve both microscopic trajectory traces and aggregate mobility indicators. This design allows the experiment to evaluate not only whether agents choose plausible activity intentions, but also whether their spatial movements match the empirical distribution observed in the benchmark.

\begin{tcolorbox}[
    colback=black!3,
    colframe=black!40,
    fontupper=\normalfont\scriptsize, 
    fonttitle=\bfseries\scriptsize,
    width=\textwidth,
    sharp corners,
    boxrule=0.5pt,
    enhanced,
    breakable,
    title=Key Prompt --- Environment Development,
    coltitle=white,
    colbacktitle=green!40!black!60,
    attach boxed title to top left={yshift=-2mm, xshift=2mm},
    top=3mm,
    before upper={\linespread{1.21}\selectfont\parskip=0.3em\parindent=1.5em}
]
Please configure a Beijing mobility environment for the DailyMobility benchmark. The environment should support AOI-based home and work initialization, city-scale movement simulation, trajectory recording, and visited-location statistics.

The environment should export each agent's mobility traces and aggregate spatial indicators, including gyration radius and daily visited-location numbers, for comparison with the DailyMobility ground-truth data.
\end{tcolorbox}

\noindent\textbf{Experimental Protocol.}
The experiment follows the DailyMobility benchmark setting. The simulation runs a full-day schedule with a 15-minute temporal grid. At each time step, agents make activity and mobility decisions based on their profiles, initialized spatial anchors, and the current simulation context. The system records activity intention sequences, aggregate intention proportions, spatial trajectories, gyration radius, and daily visited-location numbers. After execution, these outputs are compared with the DailyMobility ground-truth data using Jensen-Shannon divergence. The final evaluation therefore measures both temporal activity realism and spatial mobility realism, providing a quantitative benchmark for the human-like daily mobility generation capability of LLM agents.

\begin{tcolorbox}[
    colback=black!3,
    colframe=black!40,
    fontupper=\normalfont\scriptsize, 
    fonttitle=\bfseries\scriptsize,
    width=\textwidth,
    sharp corners,
    boxrule=0.5pt,
    enhanced,
    breakable,
    title=Key Prompt --- Configuration Generation,
    coltitle=white,
    colbacktitle=green!40!black!60,
    attach boxed title to top left={yshift=-2mm, xshift=2mm},
    top=3mm,
    before upper={\linespread{1.21}\selectfont\parskip=0.3em\parindent=1.5em}
]
Please generate an AI social scientist experiment configuration for the Beijing DailyMobility benchmark. The configuration should contain 100 profile-based agents initialized from the benchmark population data.

For each agent, include demographic attributes, home AOI, and work AOI. Use these attributes to initialize the agent's daily mobility context and prepare the simulation for full-day activity intention generation and trajectory recording.
\end{tcolorbox}

\begin{tcolorbox}[
    colback=black!3,
    colframe=black!40,
    fontupper=\normalfont\scriptsize, 
    fonttitle=\bfseries\scriptsize,
    width=\textwidth,
    sharp corners,
    boxrule=0.5pt,
    enhanced,
    breakable,
    title=Key Prompt --- Execution Schedule,
    coltitle=white,
    colbacktitle=green!40!black!60,
    attach boxed title to top left={yshift=-2mm, xshift=2mm},
    top=3mm,
    before upper={\linespread{1.21}\selectfont\parskip=0.3em\parindent=1.5em}
]
Please create the execution schedule for a full-day Beijing DailyMobility benchmark simulation. The schedule should use a 15-minute temporal grid and run through the complete daily mobility process.

During execution, collect each agent's activity intention sequence, aggregate intention proportions, spatial trajectory, gyration radius, and daily visited-location number. After the simulation, compare these outputs with the DailyMobility ground-truth data using Jensen-Shannon divergence.
\end{tcolorbox}

\subsubsection{Result Analysis \& Interpretation}

To quantify the realism of the generated daily mobility behavior, we calculate the Jensen-Shannon Divergence (JSD) between the simulated distributions and real-world ground truth across four key dimensions: (a) \textbf{Intention Sequences} (\textit{Seq.}), measuring the logical ordering and transition patterns of daily activities; (b) \textbf{Intention Proportions} (\textit{Prop.}), measuring the aggregate balance of different activity types across the population; (c) \textbf{Gyration Radius} (\textit{Radius}), measuring the spatial dispersion and characteristic travel range of each agent; and (d) \textbf{Daily Location Numbers} (\textit{Locs.}), measuring the frequency and variety of distinct locations visited per day. A lower JSD value indicates a closer approximation of empirical human mobility patterns. The results and comparisons with classical generative baselines are shown in Table~\ref{tab:mobility_jsd}.

\begin{table}[htbp]
\centering
\caption{JSD Results for Daily Mobility Simulation Dimensions Compared with Baseline Models}
\label{tab:mobility_jsd}
\begin{tabular}{lcccc}
\toprule
\textbf{Method} & \textbf{Intention Seq.} & \textbf{Intention Prop.} & \textbf{Gyration Radius} & \textbf{Daily Locs.} \\
\midrule
TimeGeo~\cite{jiang2016timegeo}      & 0.536  & 0.297  & 0.254  & 0.258  \\
Movesim~\cite{feng2018predicting}     & 0.904  & 0.154  & 0.233  & 0.051  \\
Volunteer~\cite{volunteer_baseline}   & 0.804  & 0.318  & 0.455  & 0.049  \\
DiffTraj~\cite{zhu2023difftraj}       & 0.597  & 0.695  & 0.027  & 0.647  \\
Act2Loc~\cite{zheng2024act2loc}       & 0.391  & \textbf{0.131} & \textbf{0.024} & \textbf{0.042} \\
\midrule
Ours (\textit{qwen-next-80b-a3b})    & \textbf{0.1379} & 0.3147 & 0.5978 & 0.7358 \\
\bottomrule
\end{tabular}
\begin{flushleft}
\small \textit{Note: A lower JSD value indicates a closer approximation of empirical benchmarks. Intention Seq., Intention Prop., and Daily Locs. denote Intention Sequences, Intention Proportions, and Daily Location Numbers, respectively.}
\end{flushleft}
\end{table}

The results show a clear trade-off between cognitive intent alignment and fine-grained spatial movement accuracy. Notably, for \textbf{Intention Sequences}, our proposed agent configuration achieves the lowest JSD value of \textbf{0.1379} , outperforming all established baseline models, including the advanced machine learning-based Act2Loc (0.391) and traditional mechanistic models like Movesim (0.904). This significant improvement demonstrates that large language models excels at capturing the macro-level behavioral logic, causal ordering, and scheduling routines inherent in human daily activities. For \textbf{Intention Proportions}, our model yields a moderate JSD of 0.3147 , which is highly competitive with the Volunteer baseline (0.318).

Conversely, in the spatial evaluation dimensions, specialized trajectory generation models demonstrate superior performance. Act2Loc establishes the state-of-the-art across \textbf{Gyration Radius} (0.024) and \textbf{Daily Location Numbers} (0.042), whereas our model exhibits larger divergences (0.5978 and 0.7358, respectively). This discrepancy indicates that while the cognitive architecture effectively guides what agents plan to do, grounding these high-level intentions into accurate, fine-grained urban coordinates remains a critical challenge. Without specialized trajectory-matching algorithms or real-world distance constraints, the purely LLM-driven generation tends to scatter spatially, highlighting an important area for future integration with geographical priors.

\subsection{Disaster Mobility Simulation}

Disaster mobility simulation evaluates whether profile-based agents can reproduce aggregate mobility changes under external shocks, thereby assessing the dynamic adaptability of LLM agents in response to evolving environmental conditions. We test two historical scenarios: the 2021 Texas Winter Storm and the 2018 California Camp Fire. The empirical ground truth is derived from SafeGraph-style mobility outflux indices, where mobility is measured as departures from residential census block groups and adjusted for weekly seasonality. Since the released values represent relative mobility intensity rather than absolute trip counts, we compare simulated and empirical sequences after total-sum normalization.

\subsubsection{Idea \& Hypothesis Development}

During the initial idea development, AI social scientist re-envisioned disaster mobility data as a foundation for testable hypotheses rather than a simple curve-fitting exercise. The core of the simulation lies in the dynamic interplay between an agent's profile-driven needs and the evolving flow of disaster-related information. Consequently, the model posits that agents within a city-scale environment will naturally demonstrate mobility suppression and recovery as they process disaster narratives. This theoretical framework links empirical SafeGraph benchmarks directly to the AgentSociety2 execution environment.

Specifically, the Texas Winter Storm and Camp Fire are used as benchmark scenarios. AI social scientist converts each event into a structured simulation setting with profile-based agents, a mobility environment, an event-tracking module, and a global information channel that broadcasts the evolving disaster state. The empirical mobility indices are not used as absolute trip-count targets; instead, they provide normalized reference sequences for evaluating whether the simulated population reproduces the temporal shape of disaster-induced mobility change.

\subsubsection{Experiment Design \& Execution}

\noindent\textbf{Data Preparation.}
The experiment generator builds two executable disaster-mobility configurations from the same population and map template. Each configuration uses profile-based agents initialized from demographic records with gender, race, education, transport mode, commuting-time statistics, income and age statistics, household-size statistics, home AOI identifiers, and work AOI identifiers.The empirical benchmark is constructed from SafeGraph-style mobility outflux indices for the 2021 Texas Winter Storm and the 2018 California Camp Fire. Because these indices represent relative mobility intensity rather than absolute trip counts, the simulated movement sequence and the empirical reference sequence are compared after total-sum normalization.

\begin{tcolorbox}[
    colback=black!3,
    colframe=black!40,
    fontupper=\normalfont\scriptsize, 
    fonttitle=\bfseries\scriptsize,
    width=\textwidth,
    sharp corners,
    boxrule=0.5pt,
    enhanced,
    breakable,
    title=Key Prompt --- Data Preparation,
    coltitle=white,
    colbacktitle=green!40!black!60,
    attach boxed title to top left={yshift=-2mm, xshift=2mm},
    top=3mm,
    before upper={\linespread{1.21}\selectfont\parskip=0.3em\parindent=1.5em}
]
Please prepare two disaster-mobility benchmark configurations using the same profile-based population and Houston mobility map. Each agent should include demographic attributes, transport mode, commuting-time statistics, income and age statistics, household-size statistics, a home AOI identifier, and a work AOI identifier.

Use SafeGraph-style mobility outflux indices for the 2021 Texas Winter Storm and the 2018 California Camp Fire as empirical reference sequences. Since the reference values represent relative mobility intensity, store both simulated and empirical mobility sequences in a format suitable for total-sum normalization and temporal-pattern comparison.
\end{tcolorbox}

\noindent\textbf{Agent Development.}
The simulation uses the built-in \texttt{PersonAgent} to represent profile-based urban residents. At initialization, each agent receives a compact profile describing demographic attributes, travel mode, commuting characteristics, household context, and the AOI identifiers of home and workplace. During each simulation step, agents observe the current global disaster information, reason about whether and how to move under the prevailing constraints, and interact with the mobility environment through structured tools. This design keeps individual movement decisions dependent on both stable profile attributes and time-varying disaster conditions, while leaving the final evaluation at the aggregate mobility level.

\begin{tcolorbox}[
    colback=black!3,
    colframe=black!40,
    fontupper=\normalfont\scriptsize, 
    fonttitle=\bfseries\scriptsize,
    width=\textwidth,
    sharp corners,
    boxrule=0.5pt,
    enhanced,
    breakable,
    title=Key Prompt --- Agent Development,
    coltitle=white,
    colbacktitle=green!40!black!60,
    attach boxed title to top left={yshift=-2mm, xshift=2mm},
    top=3mm,
    before upper={\linespread{1.21}\selectfont\parskip=0.3em\parindent=1.5em}
]
Please configure profile-based \texttt{PersonAgent}s for the disaster mobility experiment. Each agent should store demographic attributes, transport mode, commuting-time information, household context, home AOI, and work AOI.

During each simulation step, the agent should observe the current global disaster information, reason about travel needs and safety constraints, and use the available mobility tools to decide whether to remain in place or complete a movement. The resulting behavior should support aggregate comparison with empirical disaster-period mobility patterns.
\end{tcolorbox}

\noindent\textbf{Environment Development.}
The environment combines \texttt{MobilitySpace}, \texttt{EventSpace}, and \texttt{Global\allowbreak Information\allowbreak Env}. \texttt{MobilitySpace} provides the city-scale spatial substrate, initialized from the Houston map and each agent's home AOI. \texttt{EventSpace} records each agent's current behavioral event, allowing ongoing activities and movement-related states to be tracked during the simulation. \texttt{GlobalInformationEnv} serves as the disaster information channel. For each scenario, it is updated according to a structured event timeline: the Texas Winter Storm timeline progresses from cold-wave arrival and freezing conditions to infrastructure disruption and recovery, while the Camp Fire timeline progresses from high fire risk to ignition, evacuation, smoke and traffic control, and early containment.

\begin{tcolorbox}[
    colback=black!3,
    colframe=black!40,
    fontupper=\normalfont\scriptsize, 
    fonttitle=\bfseries\scriptsize,
    width=\textwidth,
    sharp corners,
    boxrule=0.5pt,
    enhanced,
    breakable,
    title=Key Prompt --- Environment Development,
    coltitle=white,
    colbacktitle=green!40!black!60,
    attach boxed title to top left={yshift=-2mm, xshift=2mm},
    top=3mm,
    before upper={\linespread{1.21}\selectfont\parskip=0.3em\parindent=1.5em}
]
Please configure the disaster mobility environment with \texttt{MobilitySpace}, \texttt{EventSpace}, and \texttt{GlobalInformationEnv}. \texttt{MobilitySpace} should load the Houston map and initialize agents at their home AOIs. \texttt{EventSpace} should track agents' current activities and movement-related states. \texttt{GlobalInformationEnv} should broadcast the evolving disaster condition.

Create two disaster narratives. The Texas Winter Storm narrative should cover cold-wave arrival, freezing precipitation, power and water disruptions, restricted travel, and gradual recovery. The Camp Fire narrative should cover high fire risk, ignition and evacuation, peak destruction, smoke and traffic control, and early containment.
\end{tcolorbox}

\noindent\textbf{Experimental Protocol.}
Each scenario follows the same eleven-day simulation protocol with a one-hour temporal grid. The Texas Winter Storm simulation starts on February 11, 2021, and the Camp Fire simulation starts on November 6, 2018. At the beginning of each disaster phase, the global information channel is updated to reflect the corresponding event state. In each hourly step, agents reason under the current disaster condition, execute mobility actions through the environment tools, and update their behavioral states. The system records completed movements and aggregates them over the evaluation window. The resulting simulated sequence is normalized by its total sum and compared with the corresponding empirical reference sequence using temporal-shape metrics such as correlation and mean absolute error.

\begin{tcolorbox}[
    colback=black!3,
    colframe=black!40,
    fontupper=\normalfont\scriptsize, 
    fonttitle=\bfseries\scriptsize,
    width=\textwidth,
    sharp corners,
    boxrule=0.5pt,
    enhanced,
    breakable,
    title=Key Prompt --- Configuration Generation,
    coltitle=white,
    colbacktitle=green!40!black!60,
    attach boxed title to top left={yshift=-2mm, xshift=2mm},
    top=3mm,
    before upper={\linespread{1.21}\selectfont\parskip=0.3em\parindent=1.5em}
]
Please generate two \textit{AgentSociety$^2$} disaster mobility experiment configurations. Both configurations should use the same profile-based agent population, the Houston mobility map, \texttt{MobilitySpace}, \texttt{EventSpace}, and \texttt{GlobalInformationEnv}. The first configuration should implement the 2021 Texas Winter Storm scenario, and the second should implement the 2018 California Camp Fire scenario.

Each configuration should contain profile-based agents with demographic attributes, transport mode, commuting-time statistics, home AOI, and work AOI. The generated configuration should support an eleven-day simulation window with a one-hour temporal grid and should store records suitable for aggregating completed movements across the evaluation window.
\end{tcolorbox}

\begin{tcolorbox}[
    colback=black!3,
    colframe=black!40,
    fontupper=\normalfont\scriptsize, 
    fonttitle=\bfseries\scriptsize,
    width=\textwidth,
    sharp corners,
    boxrule=0.5pt,
    enhanced,
    breakable,
    title=Key Prompt --- Execution Schedule,
    coltitle=white,
    colbacktitle=green!40!black!60,
    attach boxed title to top left={yshift=-2mm, xshift=2mm},
    top=3mm,
    before upper={\linespread{1.21}\selectfont\parskip=0.3em\parindent=1.5em}
]
Please create the execution schedule for two eleven-day disaster mobility simulations with a one-hour temporal grid. The Texas Winter Storm schedule should start on February 11, 2021, and the Camp Fire schedule should start on November 6, 2018.

During the simulation, update the global disaster information according to the corresponding event timeline, let agents make mobility decisions under the current condition, record completed movements, aggregate the movement sequence over the evaluation window, normalize simulated and empirical sequences by their total sums, and compare their temporal patterns.
\end{tcolorbox}

\subsubsection{Result Analysis \& Interpretation}
To evaluate the temporal dynamics of human mobility under extreme events, we aggregate the total number of completed movements across all agents on a daily basis. To ensure a consistent comparison between the simulated mobility counts and the empirical SafeGraph mobility intensity, we apply a sum-normalization to both sequences independently:
\begin{equation}
    \tilde{x}t = \frac{x_t}{\sum{s=1}^{T} x_s},
\end{equation}
where $x_t$ represents the raw scalar value on day $t$ (for simulation, the total daily move counts; for empirical data, the disaster-period mobility intensity index), and $T=11$ denotes the total observation window. The index $s$ serves as a summation variable across the 11-day period. This normalization yields $\tilde{x}_t$, the relative share of mobility on a given day, allowing us to focus on temporal patterns and "shapes" of suppression rather than raw magnitudes.

\begin{figure}[!t]
    \centering
    \begin{subfigure}[b]{0.9\textwidth}
        \centering
        \includegraphics[width=\textwidth]{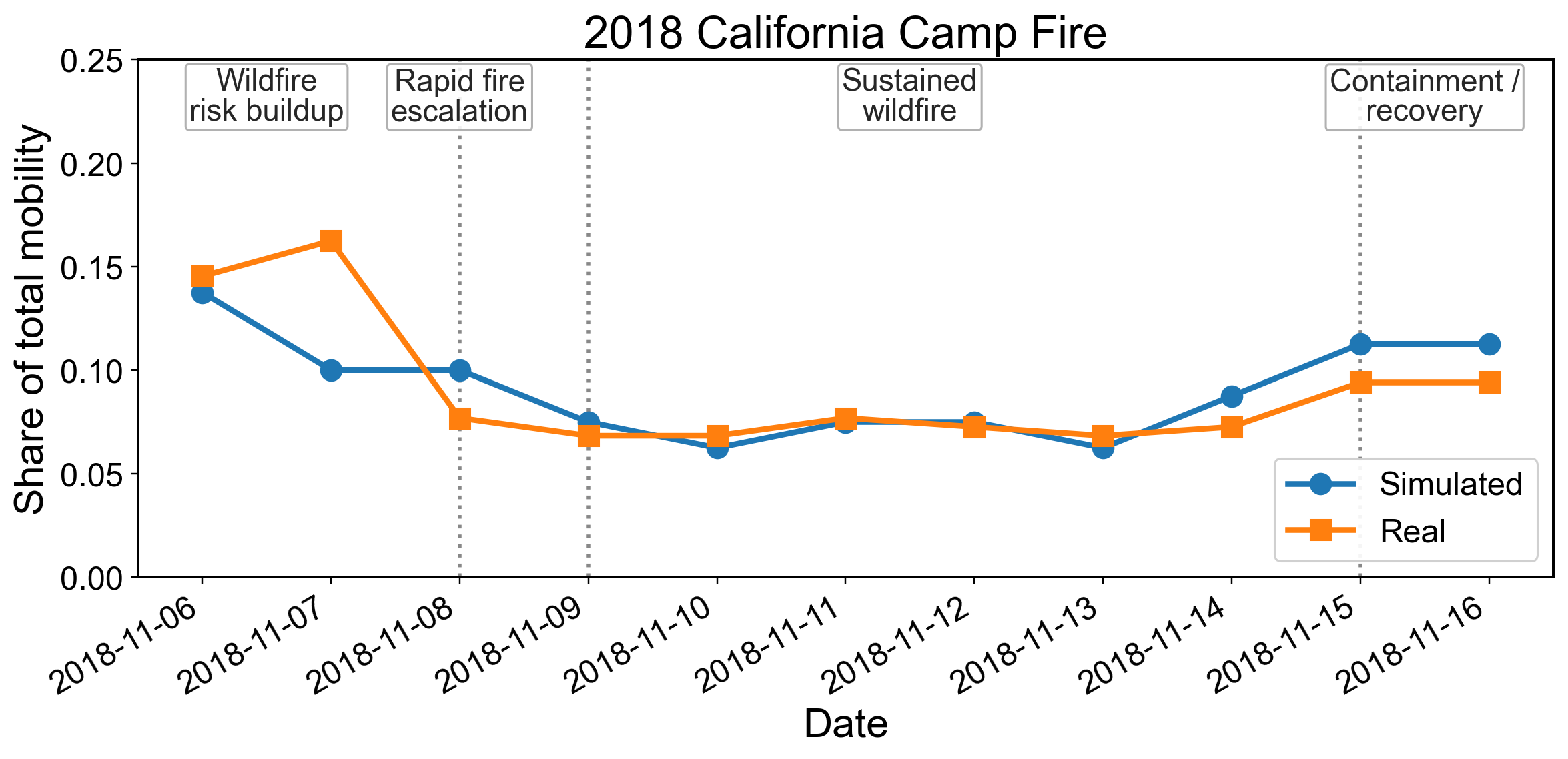}
        \caption{2018 California Camp Fire}
        \label{fig:camp}
    \end{subfigure}

    \vspace{0.5em}

    \begin{subfigure}[b]{0.9\textwidth}
        \centering
        \includegraphics[width=\textwidth]{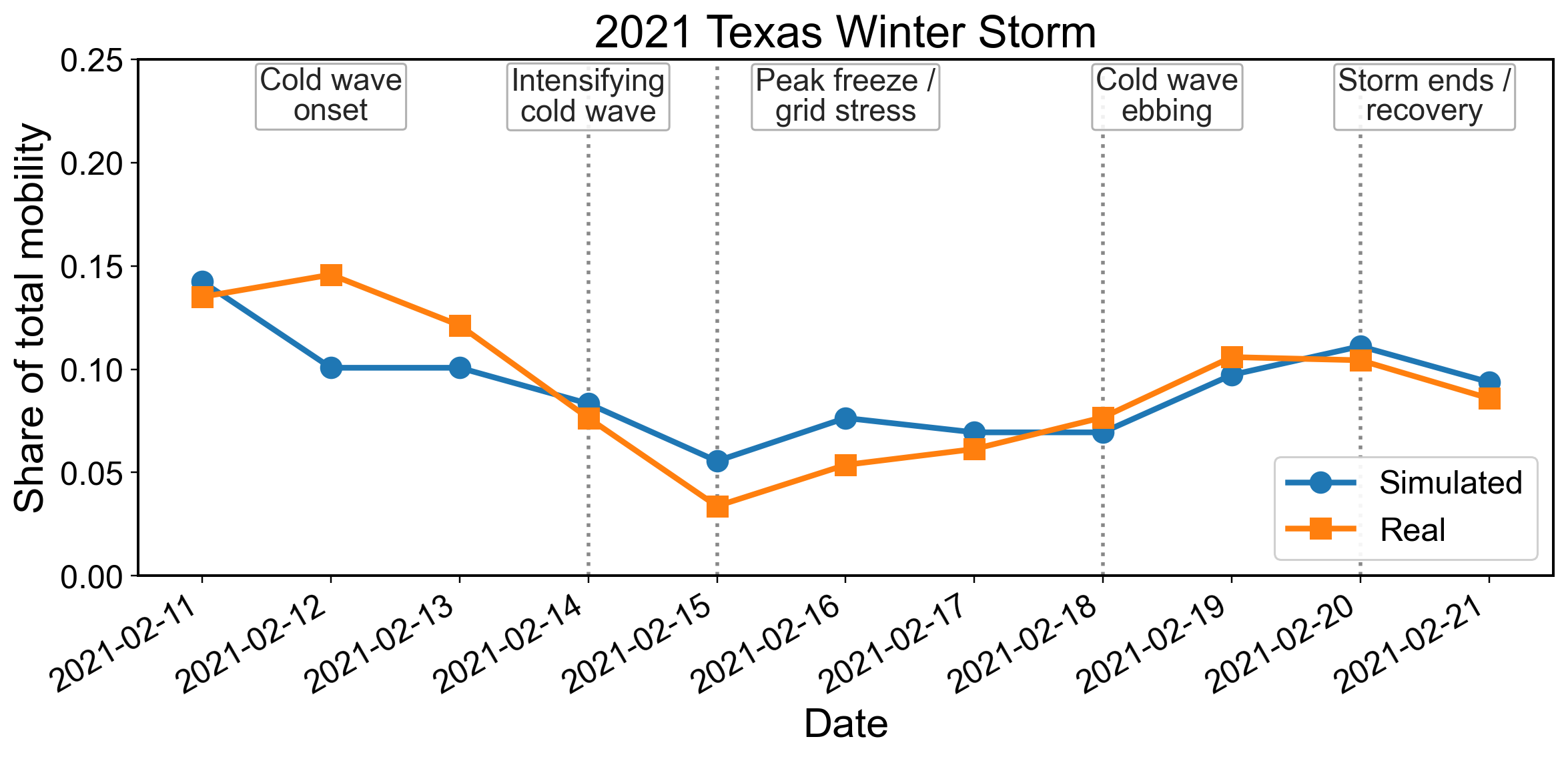}
        \caption{2021 Texas Winter Storm}
        \label{fig:winter}
    \end{subfigure}

    \caption{Comparison of empirical and simulated normalized mobility patterns for two disasters. The empirical curves are derived from SafeGraph mobility outflux indices adjusted for weekly seasonality, and the simulated curves are generated by 100 agents under historically reconstructed disaster timelines.}
    \label{fig:disaster}
\end{figure}

To further quantify the alignment between simulation and reality, we segment the timeline into distinct disaster phases based on the escalation of event conditions and the timing of broadcasted information (indicated by vertical dashed lines in Figure~\ref{fig:camp} and Figure~\ref{fig:winter}). For each phase, we calculate the Root Mean Square Error (RMSE) between the normalized sequences.

As shown in Figure~\ref{fig:camp} (California Camp Fire), the PersonAgent framework demonstrates high fidelity in capturing the rapid decline of mobility. During the "Sustained Wildfire" phase (Phase 3), the simulation achieves an exceptional fit with an RMSE of 0.0076, accurately reflecting the peak suppression of human activity. Similarly, in the Texas Winter Storm scenario (Figure~\ref{fig:winter}), the model precisely mirrors the empirical curve during the "Peak Freeze" (Phase 3, RMSE=0.0188) and the subsequent "Recovery" (Phase 5, RMSE=0.0073). These low error metrics across varying disaster stages indicate that by incorporating cognitive and social factors, PersonAgent can effectively reproduce the aggregate responsiveness of human populations to external shocks, demonstrating its potential for realistic social simulation in crisis management.

\section{Related Work}

\subsection{AI Scientist and Agentic Scientific Workflows}

AI Scientist systems extend LLM agents from isolated research assistance to staged scientific workflows. Recent systems differ in how they connect scientific reasoning with artifacts, tools, execution, validation, and writing. In computer science, AI Scientist-v2~\citep{lu2026towards} uses agentic tree search to propose research ideas, run code experiments, analyze results, write manuscripts, and support automated review for AI research. AutoSOTA~\citep{li2026autosota} reconstructs published AI-model papers into executable projects, repairs dependencies and environments, runs experiments, and searches for model improvements. AI-Researcher~\citep{tang2026ai} studies an automated pipeline that links literature review, hypothesis generation, implementation, evaluation, and manuscript preparation. Agent Laboratory~\citep{schmidgall2025agent} structures human-in-the-loop research assistance around literature review, experimentation, and report writing, while ResearchAgent~\citep{baek2025researchagent} focuses on retrieval-grounded idea generation over scientific literature. OmniScientist~\citep{shao2025omniscientist} studies a research ecosystem that links scientific knowledge infrastructure, collaboration, writing, review, and evaluation. DeepScientist~\citep{weng2025deepscientist} emphasizes persistent research execution and workspace state, ERA (Empirical Research Assistance)~\citep{aygun2026ai} formulates empirical scientific software generation as metric-driven search, and InternAgent-1.5~\citep{feng2026internagent} supports extended discovery workflows across scientific tasks. In chemistry, Coscientist~\citep{boiko2023autonomous} connects chemical reasoning with experimental planning, while ChemCrow~\citep{bran2023chemcrow} grounds chemistry assistance in external tools for molecular and reaction-related tasks. In biology and biomedical science, CellVoyager~\citep{alber2026cellvoyager} supports biological data analysis and hypothesis generation, Virtual Lab~\citep{swanson2025virtual} organizes biomedical design through multi-agent deliberation, Robin~\citep{ghareeb2026multi} links literature search, hypothesis generation, experimental proposal, and data analysis, Co-Scientist~\citep{gottweis2026accelerating} refines biomedical hypotheses through multi-agent review, and ToolUniverse~\citep{gao2025democratizing} provides biomedical tool access for scientific workflows. OpenLens~\citep{cheng2025openlens} targets health-informatics research with literature review, data analysis, code generation, visualization feedback, quality control, and manuscript preparation. Table~\ref{tab:ai_scientist} compares the systems across research stages. To the best of our knowledge, AgentSociety$^2$ is the first AI Scientist framework designed for computational social science, integrating silicon co-scientists, silicon subjects, and agentic social environments within an Integrated Research Environment. It supports literature search, hypothesis generation, experiment design, simulation execution, analysis, and manuscript preparation as a stage-structured research process for executable social experimentation.

\begin{table*}[t]
\centering
\caption{
Comparison of AI Scientist systems across scientific domains and research-stages.
}
\label{tab:ai_scientist}
\footnotesize
\setlength{\tabcolsep}{2.8pt}
\renewcommand{\arraystretch}{1.06}
\begin{tabular}{p{2.7cm} p{2.78cm} c c c c c c}
\toprule
\multirow{2}{*}{\textbf{Discipline}}
& \multirow{2}{*}{\textbf{System}}
& \textbf{Literature}
& \textbf{Hypothesis}
& \textbf{Experiment}
& \textbf{Experiment}
& \textbf{Result}
& \textbf{Manuscript} \\
& 
& \textbf{Search}
& \textbf{Generation}
& \textbf{Design}
& \textbf{Execution}
& \textbf{Analysis}
& \textbf{Writing} \\
\midrule

\multirow{9}{*}{Computer science}
& AI Scientist-v2~\citep{lu2026towards}
& \stars{2} & \stars{3} & \stars{3} & \stars{3} & \stars{3} & \stars{3} \\
& AutoSOTA~\citep{li2026autosota}
& \stars{2} & \stars{1} & \stars{2} & \stars{2} & \stars{2} & \stars{1} \\
& AI-Researcher~\citep{tang2026ai}
& \stars{3} & \stars{3} & \stars{2} & \stars{2} & \stars{2} & \stars{3} \\
& Agent Laboratory~\citep{schmidgall2025agent}
& \stars{2} & \stars{2} & \stars{2} & \stars{2} & \stars{2} & \stars{3} \\
& ResearchAgent~\citep{baek2025researchagent}
& \stars{3} & \stars{3} & \stars{2} & -- & \stars{1} & \stars{1} \\
& OmniScientist~\citep{shao2025omniscientist}
& \stars{1} & \stars{2} & \stars{2} & -- & -- & -- \\
& DeepScientist~\citep{weng2025deepscientist}
& \stars{1} & \stars{1} & \stars{2} & \stars{2} & \stars{2} & \stars{1} \\
& ERA~\citep{aygun2026ai}
& -- & \stars{1} & \stars{2} & \stars{3} & \stars{2} & -- \\
& InternAgent-1.5~\citep{feng2026internagent}
& \stars{3} & \stars{3} & \stars{3} & \stars{3} & \stars{3} & \stars{2} \\

\midrule

\multirow{2}{*}{Chemistry}
& Coscientist~\citep{boiko2023autonomous}
& \stars{1} & \stars{2} & \stars{3} & \stars{1} & \stars{1} & -- \\
& ChemCrow~\citep{bran2023chemcrow}
& \stars{1} & \stars{1} & \stars{2} & \stars{2} & \stars{1} & -- \\

\midrule

Biology
& CellVoyager~\citep{alber2026cellvoyager}
& \stars{2} & \stars{3} & \stars{2} & \stars{2} & \stars{3} & \stars{1} \\

\midrule

\multirow{4}{*}{Biomedical science}
& Virtual Lab~\citep{swanson2025virtual}
& \stars{1} & \stars{2} & \stars{3} & \stars{1} & \stars{2} & -- \\
& Robin~\citep{ghareeb2026multi}
& \stars{3} & \stars{3} & \stars{3} & \stars{2} & \stars{3} & \stars{1} \\
& Co-Scientist~\citep{gottweis2026accelerating}
& \stars{1} & \stars{3} & \stars{2} & -- & \stars{2} & \stars{1} \\
& ToolUniverse~\citep{gao2025democratizing}
& \stars{2} & \stars{1} & \stars{2} & \stars{2} & \stars{2} & \stars{1} \\

\midrule

Health informatics
& OpenLens~\citep{cheng2025openlens}
& \stars{3} & \stars{1} & \stars{3} & \stars{2} & \stars{3} & \stars{2} \\

\midrule

Social science
& \textbf{AgentSociety$^2$}
& \textbf{\stars{3}} & \textbf{\stars{3}} & \textbf{\stars{3}} & \textbf{\stars{3}} & \textbf{\stars{3}} & \textbf{\stars{3}} \\

\bottomrule
\end{tabular}

\vspace{2pt}
\begin{minipage}{\textwidth}
\footnotesize
\textit{Note.}
-- = unsupported;
$\star$ = assistive support;
$\star$$\star$ = modular support;
$\star$$\star$$\star$ = integrated autonomous support. 
The full rating rubric is provided in Appendix~\ref{app:rating_rubric_ai_scientist}.
\end{minipage}
\end{table*}

\subsection{LLM Agents as Social Participants}
Social simulation has evolved from classical agent-based models, where agents rely on manually specified rules and often face reproducibility and flexibility challenges in complex settings~\citep{epstein2012generative,macy2002factors,macal2005tutorial}, to LLM-driven frameworks for autonomous multi-agent interaction across social, online, and urban environments~\citep{piao2025agentsociety,wang2025yulan}. Related embodied and digital-twin simulators, including Megaverse~\citep{tang2025metaverse} and metaverse-style platforms~\citep{soliman2024artificial,tang2025metaverse}, develop high-performance interactive environments, but mainly target physical interaction, embodied control, or virtual spaces rather than social-science-oriented mechanism analysis. Generative Agents~\citep{park2023generative} show memory, planning, and autonomous interaction in a sandbox town. AgentScope~\citep{gao2024agentscope} and AgentVerse~\citep{chen2023agentverse} provide composable agent construction and multi-agent collaboration. OASIS~\citep{yang2024oasis} scales online social-network simulation for information diffusion and polarization. SocioVerse~\citep{zhang2025socioverse} grounds population simulation with a large real-user pool and alignment modules. YuLan-OneSim~\citep{wang2025yulan} and GenSim~\citep{tang2025gensim} support modular scenario construction and large-scale social simulation. Agent-Kernel~\citep{mao2025agent} studies adaptive social simulation with a microkernel multi-agent framework, and CitySim~\citep{bougie2025citysim} focuses on urban activity and mobility. AgentSociety~\citep{piao2025agentsociety} connects profile-based social agents with social, urban, and economic environments for large-scale behavioral simulation. Table~\ref{tab:social_sim_platform_comparison} compares AgentSociety$^2$ with representative agent-based social simulation platforms along four platform-level dimensions, including agent capability, environment capability, data use, and human-AI steering, assessing how each platform supports reusable agents, constructible environments, empirical data use, and researcher-guided design, execution, and interpretation. AgentSociety$^2$ advances this line from scalable simulation toward a research-facing simulation platform, combining configurable social agents, configurable agentic environments, multi-source data assets, study-level interventions, replayable artifacts, and evaluation support across individual, collective, and societal scales.

\begin{table*}[t]
\centering
\caption{Capability comparison of agent-based social simulation platforms.}
\label{tab:social_sim_platform_comparison}
\scriptsize
\setlength{\tabcolsep}{5pt}
\renewcommand{\arraystretch}{1.10}
\begin{tabularx}{\textwidth}{
p{3.0cm}
>{\centering\arraybackslash}X
>{\centering\arraybackslash}X
>{\centering\arraybackslash}X
>{\centering\arraybackslash}X
}
\toprule
\textbf{Platform}
& \textbf{Agent}
& \textbf{Environment}
& \textbf{Data}
& \textbf{Human-AI Steering} \\
\midrule

Generative Agents~\citep{park2023generative}
& \stars{2}
& \stars{1}
& \stars{1}
& \stars{1} \\

AgentScope~\citep{gao2024agentscope}
& \stars{3}
& \stars{3}
& \stars{3}
& \stars{3} \\

AgentVerse~\citep{chen2023agentverse}
& \stars{3}
& \stars{3}
& \stars{3}
& \stars{3} \\

OASIS~\citep{yang2024oasis}
& \stars{3}
& \stars{3}
& \stars{4}
& \stars{3} \\

SocioVerse~\citep{zhang2025socioverse}
& \stars{3}
& \stars{2}
& \stars{4}
& \stars{2} \\

YuLan-OneSim~\citep{wang2025yulan}
& \stars{4}
& \stars{4}
& \stars{3}
& \stars{4} \\

GenSim~\citep{tang2025gensim}
& \stars{2}
& \stars{3}
& \stars{2}
& \stars{2} \\

Agent-Kernel~\citep{mao2025agent}
& \stars{4}
& \stars{4}
& \stars{3}
& \stars{3} \\

CitySim~\citep{bougie2025citysim}
& \stars{3}
& \stars{3}
& \stars{3}
& \stars{2} \\

AgentSociety~\citep{piao2025agentsociety}
& \stars{4}
& \stars{4}
& \stars{3}
& \stars{3} \\

\textbf{AgentSociety$^2$}
& \stars{5}
& \stars{5}
& \stars{5}
& \stars{5} \\
\bottomrule
\end{tabularx}

\vspace{2pt}
\begin{minipage}{\textwidth}
\footnotesize
\textit{Note.}
$\star$ = minimal or case-specific support;
$\star\star$ = basic configurable support;
$\star\star\star$ = structured modular support;
$\star\star\star\star$ = advanced reusable platform-level support;
$\star\star\star\star\star$ = agentic research support across heterogeneous studies.
Higher scores reflect stronger support for generality, reusability, autonomy, empirical data use, and human-AI steering across design, execution, evaluation, and reporting. The full rating rubric is provided in Appendix~\ref{app:rating_rubric_social_simulation}.
\end{minipage}
\end{table*}

\subsection{LLM for Social Science}

LLMs are moving social-scientific inquiry from individual-level response simulation toward social interaction and society-level dynamics. Aher et al.~\citep{aher2023using} provide an early empirical test of LLMs as simulated participants, showing partial behavioral alignment but also sensitivity to prompts and task framing. Argyle et al.~\citep{argyle2023out} further show that demographic and political conditioning can approximate some public-opinion patterns, motivating LLMs as synthetic samples for survey research. This line also raises concerns about validity, bias, representativeness, and human-agent comparability~\citep{grossmann2023ai,ziems2024can,wang2024survey}. Another line evaluates LLMs through cognitive and decision-making tasks. Binz and Schulz~\citep{binz2023using} report both similarities and deviations between model and human behavior across psychological experiments, while Hagendorff et al.~\citep{hagendorff2023human} show that LLMs reproduce some human-like cognitive biases but differ in response patterns. Related studies examine theory of mind, analogy, text analysis, experimental simulation, and cognitive modeling, emphasizing validation against human evidence~\citep{kosinski2023theory,webb2023emergent,echterhoff2024cognitive,binz2025foundation}. Recent work extends from general respondents to socially situated actors and collective phenomena~\citep{horton2023large,zhao2023competeai,piao2025emergence,ashery2025emergent}. AgentSociety$^2$ transforms this line of work from task-specific LLM-based studies into an executable research process that connects study interpretation, protocol translation, simulation execution, and comparative analysis, making it possible to test social-scientific mechanisms across individual behavior, social interaction, and society-level outcomes with clearer control, traceability, and comparison with human evidence.

\section{Discussion}

\subsection{Design of Large-Scale and Realistic Social Agents}
AgentSociety$^2$ builds upon the social agents developed in AgentSociety-1 and advances large-scale social agent design in both efficiency and generality.
In terms of efficiency, skill-based incremental disclosure reduces the context and token consumption required for each LLM invocation.
In terms of generality, we decouple agent design from specific simulation environments through world descriptions provided by the environment and natural language interaction.
By enabling simulated agents to autonomously load Skills and maintain independent workspaces, AgentSociety$^2$ further enhances their adaptability to different environments and scenarios.
These improvements make agent simulation easier to extend to diverse social science research questions.

However, efficiency improvements still need to be pushed further.
In large-scale simulations, the number of agents, simulation steps, and environment interactions all amplify the total volume of LLM calls.
Therefore, even when token consumption per call is reduced, rate-limiting and cost issues can still restrict adoption by most social science researchers.
An important direction for the next stage is to improve the overall efficiency of large-scale agent simulation at the system level.
On the one hand, researchers can attempt to further reduce the number of ReAct cycles in a single simulation step or introduce on-the-fly code generation to replace multi-round calls.
On the other hand, fine-tuning ultra-small LLMs and deploying them locally may be another effective approach.

At the same time, greater generality also raises higher requirements for simulation realism.
When the same class of agents can be transferred to more environments and research tasks, researchers must ensure that this adaptability does not come at the expense of behavioral realism.
Realism first means that agent behavior should be consistent with profiles, memories, social contexts, and environmental constraints.
It also means that agent populations should generate plausible behavioral distributions, counterfactual responses, and group dynamics through interaction.
For social science research, realism determines whether simulation results can support mechanism exploration, experimental rehearsal, and hypothesis generation.

Therefore, future work needs to make realism evaluation part of general-purpose agent design.
The platform should build reusable calibration benchmarks for different classes of experiments, allowing researchers to judge whether a given agent population is appropriate for a specific research question.
These evaluations should also compare how different models, prompting strategies, memory structures, and environmental constraints affect simulation realism.
Only when efficiency optimization and realism evaluation advance together can large-scale general social agents more reliably support social science research.

\subsection{Community Building for Cumulative Social Simulation}
AgentSociety$^2$ improves platform extensibility through three key designs.
First, the decorator-based environment integration approach and CodeGenRouter transform agent inputs into environment operations, reducing the cost of connecting new environments to the platform.
Second, the generalized design of social agents allows the same class of agents to adapt to different scenarios and experimental tasks.
Third, AI social scientist supports research workflows such as hypothesis generation, experiment configuration, and result analysis.
These capabilities reduce usage and integration costs and make it easier for the platform to cover diverse areas of social science.

However, relying solely on our team's capabilities still cannot fully cover all research topics in social science and LLM Agent simulation.
If AgentSociety$^2$ is to become a sustainable research infrastructure, it needs an open and flexible community ecosystem.
Such a community can expand the domains covered by the platform and allow experiments, agents, environments, and evaluation methods to accumulate across researchers.
Therefore, community building will shape whether AgentSociety$^2$ and its methodology can meaningfully influence a paradigm shift in social science research.

The key to community building should not be merely sharing code, but sharing verifiable research components.
These components can include environment modules, experiment templates, reusable agent profiles, skills, benchmark tasks, evaluation protocols, and analysis pipelines.
Each component should be accompanied by clear metadata, version control, example configurations, and minimal validation tests.
In this way, community contributions can be compared, reproduced, and transferred to new research questions.

We think that one concrete approach is to draw inspiration from the mechanisms of skill marketplaces.
The platform can establish standardized conventions based on Git repository structures or Python packages, allowing researchers to publish, download, and install open-source components conveniently.
At the same time, existing LLM agent simulation code can be adapted into community components that satisfy the interfaces and validation requirements of AgentSociety$^2$.

\subsection{LLM Agent-Driven Co-Scientist and Accountable Social Research}
As part of the LLM agent co-scientist initiative that has garnered significant attention in recent years, our proposed AI social scientist has demonstrated promising results in assisting social science research, providing valuable insights based on simulation for social science studies.
This new form of scientific research organization will drive changes in the research paradigm of the social sciences and in the production relations between humans and machines within the field.

Throughout this process, we will encounter numerous challenges that require resolution.
First is the further enrichment and capability enhancement of agents and environment modules discussed above.
Furthermore, lowering the barrier to entry for social science experts using AI social scientist is a key challenge in ensuring that new technologies effectively support social science research and scientific discovery.
In this regard, an important success of AI social scientist is that it substantially lowers the barrier to using a complete agentic social science workflow.
Researchers no longer need to manually construct simulation pipelines, experiment configurations, and analysis workflows from scratch.
Instead, they can focus more on research questions, theoretical mechanisms, and experimental interpretation.
This capability makes AI social scientist not merely a local tool, but a collaborative system that spans the research workflow.

Precisely because it participates in multiple stages of the research workflow, scientific accountability and process auditability become more important.
AI social scientist becomes involved in hypothesis generation, experiment configuration, execution, result analysis, and manuscript preparation.
Therefore, researchers need to know which conclusions come from raw data, which come from model inference, and which come from automatic system organization.
If these sources cannot be traced, the convenience of automation may weaken the interpretability and credibility of research results.

Future work should therefore focus on auditable research workflows.
The platform needs to record the relationships among hypotheses, prompts, experiment configurations, generated code, execution logs, intermediate data, analysis scripts, and final reports.
It should also allow human researchers to inspect, revise, and confirm key steps.
Therefore, AI social scientist is better understood as a traceable research collaborator rather than a black-box automation system that replaces human judgment.
We believe that continuously improving the usability, transparency, and auditability of AI social scientist is essential for its role in supporting a paradigm shift in social science research.

\subsection{Risks and Methodological Safeguards}
AgentSociety$^2$ lowers the cost of constructing, executing, and analyzing large-scale agent-based simulation experiments.
However, the more fully the platform spans the research workflow, the more its conclusions require clear methodological safeguards.
Therefore, in this subsection, we discuss the risks of AgentSociety$^2$ and their corresponding methodological safeguards.
This discussion is organized around three levels: formal reliability, simulation realism, and experimental-mechanism validity.

The first level is formal reliability, namely whether research outputs can be traced and audited.
The previous subsection has discussed the auditability requirements introduced by the deep involvement of AI social scientist in the research workflow.
The key point is that formal traceability is the basis for all subsequent evaluation.
Researchers must be able to trace the source, generation process, and human confirmation points behind a claim.
This auditability does not by itself prove that a research conclusion is true, but it is a prerequisite for formal reliability.

The second level is the realism of the simulation design itself.
Even when the research workflow is traceable, the simulation design may still be biased by modeling assumptions and system configurations.
Therefore, realism evaluation needs to move beyond the plausibility of individual outputs toward systematic checks of behavior, interaction, and aggregate outcomes.
A simulation may match real-world data at the aggregate statistical level while still deviating in individual decision logic or interaction pathways.
Future work should turn such checks into reusable calibration benchmarks and evaluation protocols.

The third level is the validity of the overall simulated experimental mechanism.
A realistic simulation design does not automatically imply a valid experimental mechanism.
When the platform is used to test or explain social mechanisms, researchers still need to confirm a clear correspondence among mechanism implementation, agent response, and final outcome.
Such confirmation requires methods such as replication, ablation, sensitivity analysis, and external empirical comparison.
These checks help prevent researchers from mistaking prompt bias, environment implementation errors, or model contingency for social scientific findings.

In summary, we believe that these methodological safeguards are not only key challenges for AgentSociety$^2$ as it develops into infrastructure for social science research, but also important issues that the broader community must collectively address under this emerging research paradigm.
Establishing reliable methodological safeguards will therefore become an important direction for future research in this field.

\section{Conclusion}

The central challenge for AI scientists in social science is not simply to automate the researcher-facing side of science, but to make social inquiry executable. Social knowledge depends on a chain of linked objects: theories, data, participants, environments, interventions, measurements, and claims. When this chain is distributed across separate tools, stages, and researchers, the relation between assumptions, simulated behaviors, observed outcomes, and scientific claims becomes opaque. \textit{AgentSociety$^2$} addresses this challenge by placing this chain within an Integrated Research Environment, where it can be represented, executed, inspected, and revised through the coupling of scientific coordination and social simulation.

This coupling is realized through a dual-role design that brings AI social scientists and silicon participants into the same research runtime. AI social scientists coordinate empirical grounding, theory and hypothesis formulation, experiment design, simulation execution, result interpretation, and manuscript preparation, while silicon participants provide behavioral responses within configurable social environments. Together, these two roles turn social-science hypotheses into auditable agent behaviors, environment rules, interventions, measurements, and claims. Building on \textit{AgentSociety-1}, \textit{AgentSociety$^2$} extends social simulation from an experimental substrate into a closed-loop research environment in which scientific reasoning and executable social experimentation can inform one another.

Across seven illustrative studies spanning micro-level behavioral experiments, meso-level social-media dynamics, and macro-level urban scenarios, \textit{AgentSociety$^2$} demonstrates the capacity to support diverse research questions, reproduce major qualitative patterns from prior studies, identify informative deviations, and scale computational experiments through optimized agent–environment interactions. More broadly, the platform offers a step toward executable and accountable social science: a mode of inquiry in which human researchers retain conceptual authority, while agentic systems make the procedural links between theory, experiment, evidence, and claim more visible and controllable. This points to a broader future for agentic computational social science, where human researchers, AI social scientists, and silicon participants jointly support the production, testing, interpretation, and communication of social knowledge.

\bibliographystyle{unsrtnat}
\bibliography{ref}

@inproceedings{braunschweiler2025toolreagt,
  title={ToolReAGt: tool retrieval for LLM-based complex task solution via retrieval augmented generation},
  author={Braunschweiler, Norbert and Doddipatla, Rama and Zorila, Tudor-Catalin},
  booktitle={Proceedings of the 3rd Workshop on Towards Knowledgeable Foundation Models (KnowFM)},
  pages={75--83},
  year={2025}
}

@article{comanici2025gemini,
  title={Gemini 2.5: Pushing the frontier with advanced reasoning, multimodality, long context, and next generation agentic capabilities},
  author={Comanici, Gheorghe and Bieber, Eric and Schaekermann, Mike and Pasupat, Ice and Sachdeva, Noveen and Dhillon, Inderjit and Blistein, Marcel and Ram, Ori and Zhang, Dan and Rosen, Evan and others},
  journal={arXiv preprint arXiv:2507.06261},
  year={2025}
}

@article{team2025kimi,
  title={Kimi k2: Open agentic intelligence},
  author={Team, Kimi and Bai, Yifan and Bao, Yiping and Chen, Guanduo and Chen, Jiahao and Chen, Ningxin and Chen, Ruijue and Chen, Yanru and Chen, Yuankun and Chen, Yutian and others},
  journal={arXiv preprint arXiv:2507.20534},
  year={2025}
}

@article{yang2025qwen3,
  title={Qwen3 technical report},
  author={Yang, An and Li, Anfeng and Yang, Baosong and Zhang, Beichen and Hui, Binyuan and Zheng, Bo and Yu, Bowen and Gao, Chang and Huang, Chengen and Lv, Chenxu and others},
  journal={arXiv preprint arXiv:2505.09388},
  year={2025}
}

@article{glm2024chatglm,
  title={Chatglm: A family of large language models from glm-130b to glm-4 all tools},
  author={GLM, Team and Zeng, Aohan and Xu, Bin and Wang, Bowen and Zhang, Chenhui and Yin, Da and Zhang, Dan and Rojas, Diego and Feng, Guanyu and Zhao, Hanlin and others},
  journal={arXiv preprint arXiv:2406.12793},
  year={2024}
}

@inproceedings{patilberkeley,
  title={The Berkeley Function Calling Leaderboard (BFCL): From Tool Use to Agentic Evaluation of Large Language Models},
  author={Patil, Shishir G and Mao, Huanzhi and Yan, Fanjia and Ji, Charlie Cheng-Jie and Suresh, Vishnu and Stoica, Ion and Gonzalez, Joseph E},
  booktitle={Forty-second International Conference on Machine Learning}
}

@inproceedings{wang-etal-2023-plan,
    title = "Plan-and-Solve Prompting: Improving Zero-Shot Chain-of-Thought Reasoning by Large Language Models",
    author = "Wang, Lei  and
      Xu, Wanyu  and
      Lan, Yihuai  and
      Hu, Zhiqiang  and
      Lan, Yunshi  and
      Lee, Roy Ka-Wei  and
      Lim, Ee-Peng",
    editor = "Rogers, Anna  and
      Boyd-Graber, Jordan  and
      Okazaki, Naoaki",
    booktitle = "Proceedings of the 61st Annual Meeting of the Association for Computational Linguistics (Volume 1: Long Papers)",
    month = jul,
    year = "2023",
    address = "Toronto, Canada",
    publisher = "Association for Computational Linguistics",
    url = "https://aclanthology.org/2023.acl-long.147/",
    doi = "10.18653/v1/2023.acl-long.147",
    pages = "2609--2634"
}

@inproceedings{yao2022react,
  title={React: Synergizing reasoning and acting in language models},
  author={Yao, Shunyu and Zhao, Jeffrey and Yu, Dian and Du, Nan and Shafran, Izhak and Narasimhan, Karthik R and Cao, Yuan},
  booktitle={The eleventh international conference on learning representations},
  year={2022}
}

@article{qi2025comprehensive,
  title={A Comprehensive Dataset for Investigating the Structure of Self-Bias},
  author={Qi, Yuxuan and Zou, Fengjie and Chau, Xi Ying and Zhou, Michelle and Wang, Fei and Sui, Jie},
  journal={Scientific Data},
  volume={12},
  number={1},
  pages={1755},
  year={2025},
  publisher={Nature Publishing Group UK London}
}

@article{kahneman1990experimental,
  title={Experimental tests of the endowment effect and the Coase theorem},
  author={Kahneman, Daniel and Knetsch, Jack L and Thaler, Richard H},
  journal={Journal of political Economy},
  volume={98},
  number={6},
  pages={1325--1348},
  year={1990},
  publisher={The University of Chicago Press}
}

@article{heine1999there,
  title={Is there a universal need for positive self-regard?},
  author={Heine, Steven J and Lehman, Darrin R and Markus, Hazel Rose and Kitayama, Shinobu},
  journal={Psychological review},
  volume={106},
  number={4},
  pages={766},
  year={1999},
  publisher={American Psychological Association}
}

@article{greenwald2000using,
  title={Using the implicit association test to measure self-esteem and self-concept.},
  author={Greenwald, Anthony G and Farnham, Shelly D},
  journal={Journal of personality and social psychology},
  volume={79},
  number={6},
  pages={1022},
  year={2000},
  publisher={American Psychological Association}
}

@article{rogers1977self,
  title={Self-reference and the encoding of personal information.},
  author={Rogers, Timothy B and Kuiper, Nicholas A and Kirker, William S},
  journal={Journal of personality and social psychology},
  volume={35},
  number={9},
  pages={677},
  year={1977},
  publisher={American Psychological Association}
}

@article{taylor1988illusion,
  title={Illusion and well-being: a social psychological perspective on mental health.},
  author={Taylor, Shelley E and Brown, Jonathon D},
  journal={Psychological bulletin},
  volume={103},
  number={2},
  pages={193},
  year={1988},
  publisher={American Psychological Association}
}

@inproceedings{aher2023using,
  title={Using large language models to simulate multiple humans and replicate human subject studies},
  author={Aher, Gati V and Arriaga, Rosa I and Kalai, Adam Tauman},
  booktitle={International conference on machine learning},
  pages={337--371},
  year={2023},
  organization={PMLR}
}

@article{argyle2023out,
  title={Out of one, many: Using language models to simulate human samples},
  author={Argyle, Lisa P and Busby, Ethan C and Fulda, Nancy and Gubler, Joshua R and Rytting, Christopher and Wingate, David},
  journal={Political Analysis},
  volume={31},
  number={3},
  pages={337--351},
  year={2023},
  publisher={Cambridge University Press}
}

@article{binz2023using,
  title={Using cognitive psychology to understand GPT-3},
  author={Binz, Marcel and Schulz, Eric},
  journal={Proceedings of the National Academy of Sciences},
  volume={120},
  number={6},
  pages={e2218523120},
  year={2023},
  publisher={National Academy of Sciences}
}

@article{dillion2023can,
  title={Can AI language models replace human participants?},
  author={Dillion, Danica and Tandon, Niket and Gu, Yuling and Gray, Kurt},
  journal={Trends in Cognitive Sciences},
  volume={27},
  number={7},
  pages={597--600},
  year={2023},
  publisher={Elsevier}
}

@article{hagendorff2023human,
  title={Human-like intuitive behavior and reasoning biases emerged in large language models but disappeared in ChatGPT},
  author={Hagendorff, Thilo and Fabi, Sarah and Kosinski, Michal},
  journal={Nature Computational Science},
  volume={3},
  number={10},
  pages={833--838},
  year={2023},
  publisher={Nature Publishing Group US New York}
}

@article{kosinski2023theory,
  title={Theory of mind may have spontaneously emerged in large language models},
  author={Kosinski, Michal},
  journal={arXiv preprint arXiv:2302.02083},
  volume={4},
  number={169},
  pages={2},
  year={2023}
}

@techreport{horton2023large,
  title={Large language models as simulated economic agents: What can we learn from homo silicus?},
  author={Horton, John J and Filippas, Apostolos and Manning, Benjamin S},
  year={2023},
  institution={National Bureau of Economic Research}
}

@inproceedings{park2023generative,
  title={Generative agents: Interactive simulacra of human behavior},
  author={Park, Joon Sung and O'Brien, Joseph and Cai, Carrie Jun and Morris, Meredith Ringel and Liang, Percy and Bernstein, Michael S},
  booktitle={Proceedings of the 36th annual acm symposium on user interface software and technology},
  pages={1--22},
  year={2023}
}

@article{webb2023emergent,
  title={Emergent analogical reasoning in large language models},
  author={Webb, Taylor and Holyoak, Keith J and Lu, Hongjing},
  journal={Nature Human Behaviour},
  volume={7},
  number={9},
  pages={1526--1541},
  year={2023},
  publisher={Nature Publishing Group UK London}
}

@article{grossmann2023ai,
  title={AI and the transformation of social science research},
  author={Grossmann, Igor and Feinberg, Matthew and Parker, Dawn C and Christakis, Nicholas A and Tetlock, Philip E and Cunningham, William A},
  journal={Science},
  volume={380},
  number={6650},
  pages={1108--1109},
  year={2023},
  publisher={American Association for the Advancement of Science}
}

@article{wei2022chain,
  title={Chain-of-thought prompting elicits reasoning in large language models},
  author={Wei, Jason and Wang, Xuezhi and Schuurmans, Dale and Maiden, Maarten and Fei-Fei, Li and Chi, Ed and Le, Quoc V and Zhou, Denny},
  journal={Advances in Neural Information Processing Systems},
  volume={35},
  pages={24824--24837},
  year={2022}
}

@article{piao2025agentsociety,
  title={Agentsociety: Large-scale simulation of llm-driven generative agents advances understanding of human behaviors and society},
  author={Piao, Jinghua and Yan, Yuwei and Zhang, Jun and Li, Nian and Yan, Junbo and Lan, Xiaochong and Lu, Zhihong and Zheng, Zhiheng and Wang, Jing Yi and Zhou, Di and others},
  journal={arXiv preprint arXiv:2502.08691},
  year={2025}
}

@inproceedings{wang2025yulan,
  title={Yulan-onesim: Towards the next generation of social simulator with large language models},
  author={Wang, Lei and Gao, Heyang and Bo, Xiaohe and Chen, Xu and Wen, Ji-Rong},
  booktitle={Workshop on Scaling Environments for Agents},
  year={2025}
}

@article{gao2024agentscope,
  title={Agentscope: A flexible yet robust multi-agent platform},
  author={Gao, Dawei and Li, Zitao and Pan, Xuchen and Kuang, Weirui and Ma, Zhijian and Qian, Bingchen and Wei, Fei and Zhang, Wenhao and Xie, Yuexiang and Chen, Daoyuan and others},
  journal={arXiv preprint arXiv:2402.14034},
  year={2024}
}

@inproceedings{chen2023agentverse,
  title={Agentverse: Facilitating multi-agent collaboration and exploring emergent behaviors},
  author={Chen, Weize and Su, Yusheng and Zuo, Jingwei and Yang, Cheng and Yuan, Chenfei and Chan, Chi-Min and Yu, Heyang and Lu, Yaxi and Hung, Yi-Hsin and Qian, Chen and others},
  booktitle={The Twelfth International Conference on Learning Representations},
  year={2023}
}

@article{zhang2025socioverse,
  title={Socioverse: A world model for social simulation powered by llm agents and a pool of 10 million real-world users},
  author={Zhang, Xinnong and Lin, Jiayu and Mou, Xinyi and Yang, Shiyue and Liu, Xiawei and Sun, Libo and Lyu, Hanjia and Yang, Yihang and Qi, Weihong and Chen, Yue and others},
  journal={arXiv preprint arXiv:2504.10157},
  year={2025}
}

@inproceedings{bougie2025citysim,
  title={Citysim: Modeling urban behaviors and city dynamics with large-scale llm-driven agent simulation},
  author={Bougie, Nicolas and Watanabe, Narimawa},
  booktitle={Proceedings of the 2025 Conference on Empirical Methods in Natural Language Processing: Industry Track},
  pages={215--229},
  year={2025}
}

@article{yang2024oasis,
  title={Oasis: Open agent social interaction simulations with one million agents},
  author={Yang, Ziyi and Zhang, Zaibin and Zheng, Zirui and Jiang, Yuxian and Gan, Ziyue and Wang, Zhiyu and Ling, Zijian and Chen, Jinsong and Ma, Martz and Dong, Bowen and others},
  journal={arXiv preprint arXiv:2411.11581},
  year={2024}
}

@article{mao2025agent,
  title={Agent-Kernel: A MicroKernel Multi-Agent System Framework for Adaptive Social Simulation Powered by LLMs},
  author={Mao, Yuren and Liu, Peigen and Wang, Xinjian and Ding, Rui and Miao, Jing and Zou, Hui and Qi, Mingjie and Luo, Wanxiang and Lai, Longbin and Wang, Kai and others},
  journal={arXiv preprint arXiv:2512.01610},
  year={2025}
}

@article{zhao2023competeai,
  title={Competeai: Understanding the competition dynamics in large language model-based agents},
  author={Zhao, Qinlin and Wang, Jindong and Zhang, Yixuan and Jin, Yiqiao and Zhu, Kaijie and Chen, Hao and Xie, Xing},
  journal={arXiv preprint arXiv:2310.17512},
  year={2023}
}

@article{levy2021social,
  author = {Levy, Ro'ee},
  title = {Social Media, News Consumption, and Polarization: Evidence from a Field Experiment},
  journal = {American Economic Review},
  volume = {111},
  number = {3},
  year = {2021},
  month = {March},
  pages = {831--870},
  doi = {10.1257/aer.20191777},
  url = {https://www.aeaweb.org/articles?id=10.1257/aer.20191777}
}

@article{piao2025emergence,
  title={Emergence of human-like polarization among large language model agents}, 
  author={Jinghua Piao and Zhihong Lu and Chen Gao and Fengli Xu and Qinghua Hu and Fernando P. Santos and Yong Li and James Evans},
  year={2025},
  journal={arXiv preprint arXiv:2501.05171}
}

@article{lu2024ai,
  title={The ai scientist: Towards fully automated open-ended scientific discovery},
  author={Lu, Chris and Lu, Cong and Lange, Robert Tjarko and Foerster, Jakob and Clune, Jeff and Ha, David},
  journal={arXiv preprint arXiv:2408.06292},
  year={2024}
}

@article{boiko2023autonomous,
  title={Autonomous chemical research with large language models},
  author={Boiko, Daniil A and MacKnight, Robert and Kline, Ben and Gomes, Gabe},
  journal={Nature},
  volume={624},
  number={7992},
  pages={570--578},
  year={2023},
  publisher={Nature Publishing Group UK London}
}

@article{bran2023chemcrow,
  title={Chemcrow: Augmenting large-language models with chemistry tools},
  author={Bran, Andres M and Cox, Sam and Schilter, Oliver and Baldassari, Carlo and White, Andrew D and Schwaller, Philippe},
  journal={arXiv preprint arXiv:2304.05376},
  year={2023}
}

@inproceedings{baek2025researchagent,
  title={Researchagent: Iterative research idea generation over scientific literature with large language models},
  author={Baek, Jinheon and Jauhar, Sujay Kumar and Cucerzan, Silviu and Hwang, Sung Ju},
  booktitle={Proceedings of the 2025 Conference of the Nations of the Americas Chapter of the Association for Computational Linguistics: Human Language Technologies (Volume 1: Long Papers)},
  pages={6709--6738},
  year={2025}
}

@article{ziems2024can,
  title={Can large language models transform computational social science?},
  author={Ziems, Caleb and Held, William and Shaikh, Omar and Chen, Jiaao and Zhang, Zhehao and Yang, Diyi},
  journal={Computational Linguistics},
  volume={50},
  number={1},
  pages={237--291},
  year={2024},
  publisher={MIT Press One Broadway, 12th Floor, Cambridge, Massachusetts 02142, USA~…}
}

@article{zhou2025autonomous,
  title={Autonomous agents for scientific discovery: Orchestrating scientists, language, code, and physics},
  author={Zhou, Lianhao and Ling, Hongyi and Fu, Cong and Huang, Yepeng and Sun, Michael and Yu, Wendi and Wang, Xiaoxuan and Li, Xiner and Su, Xingyu and Zhang, Junkai and others},
  journal={arXiv preprint arXiv:2510.09901},
  year={2025}
}

@book{hey2009fourth,
  title={The fourth paradigm: data-intensive scientific discovery},
  author={Hey, Anthony JG and Tansley, Stewart and Tolle, Kristin Michele and others},
  volume={1},
  year={2009},
  publisher={Microsoft research Redmond, WA}
}

@book{kuhn1970structure,
  title={The structure of scientific revolutions},
  author={Kuhn, Thomas S and Hacking, Ian},
  volume={2},
  number={2},
  year={1970},
  publisher={University of Chicago press Chicago}
}

@book{winsberg2019science,
  title={Science in the age of computer simulation},
  author={Winsberg, Eric},
  year={2019},
  publisher={University of Chicago Press}
}

@book{epstein2012generative,
  title={Generative social science: Studies in agent-based computational modeling},
  author={Epstein, Joshua M},
  year={2012},
  publisher={Princeton University Press}
}

@article{macy2002factors,
  title={From factors to actors: Computational sociology and agent-based modeling},
  author={Macy, Michael W and Willer, Robert},
  journal={Annual review of sociology},
  volume={28},
  number={1},
  pages={143--166},
  year={2002},
  publisher={Annual Reviews 4139 El Camino Way, PO Box 10139, Palo Alto, CA 94303-0139, USA}
}

@article{macal2005tutorial,
  title={Tutorial on agent-based modeling and simulation},
  author={Macal, Charles M and North, Michael J},
  journal={Proceedings of the Winter Simulation Conference},
  volume={2005},
  pages={2--15},
  year={2005}
}

@article{zhang2026equipping,
  title={Equipping agents for the real world with Agent Skills, October 2025},
  author={Zhang, Barry and Lazuka, Keith and Murag, Mahesh},
  journal={URL https://www. anthropic. com/engineering/equipping-agents-for-the-real-world-with-agent-skills. Accessed},
  pages={01--28},
  year={2026}
}

@article{cohen2017scientific,
  title={Scientific workflows for computational reproducibility in the life sciences: Status, challenges and opportunities},
  author={Cohen-Boulakia, Sarah and Belhajjame, Khalid and Collin, Olivier and Chopard, J{\'e}r{\^o}me and Froidevaux, Christine and Gaignard, Alban and Hinsen, Konrad and Larmande, Pierre and Le Bras, Yvan and Lemoine, Fr{\'e}d{\'e}ric and others},
  journal={Future Generation Computer Systems},
  volume={75},
  pages={284--298},
  year={2017},
  publisher={Elsevier}
}

@misc{chirigati2013es,
  title={{\'e}s J. Freire,„ReproZip: Using Provenance to Support Computational Reproducibility.”},
  author={Chirigati, FS and Shasha, D},
  year={2013},
  publisher={TaPP}
}

@article{lee2015complexities,
  title={The complexities of agent-based modeling output analysis},
  author={Lee, Ju-Sung and Filatova, Tatiana and Ligmann-Zielinska, Arika and Hassani-Mahmooei, Behrooz and Stonedahl, Forrest and Lorscheid, Iris and Voinov, Alexey and Polhill, J Gareth and Sun, Zhanli and Parker, Dawn C},
  journal={Journal of Artificial Societies and Social Simulation},
  volume={18},
  number={4},
  year={2015},
  publisher={Guildford: University of Surrey}
}

@article{1986An,
  title={An Evolutionary Approach to Norms},
  author={ Axelrod, Robert },
  journal={American Political Science Review},
  volume={80},
  number={4},
  pages={1095-1111},
  year={1986},
}

@article{fischbacher2010social,
  title={Social preferences, beliefs, and the dynamics of free riding in public goods experiments},
  author={Fischbacher, Urs and G{\"a}chter, Simon},
  journal={American economic review},
  volume={100},
  number={1},
  pages={541--556},
  year={2010},
  publisher={American Economic Association}
}

@book{hacking1983representing,
  title={Representing and intervening: Introductory topics in the philosophy of natural science},
  author={Hacking, Ian},
  year={1983},
  publisher={Cambridge university press}
}

@article{lazer2009computational,
  title={Computational social science},
  author={Lazer, David and Pentland, Alex and Adamic, Lada and Aral, Sinan and Barab{\'a}si, Albert-L{\'a}szl{\'o} and Brewer, Devon and Christakis, Nicholas and Contractor, Noshir and Fowler, James and Gutmann, Myron and others},
  journal={Science},
  volume={323},
  number={5915},
  pages={721--723},
  year={2009},
  publisher={American Association for the Advancement of Science}
}

@book{salganik2019bit,
  title={Bit by bit: Social research in the digital age},
  author={Salganik, Matthew J},
  year={2019},
  publisher={Princeton University Press}
}

@article{brown2020language,
  title={Language models are few-shot learners},
  author={Brown, Tom and Mann, Benjamin and Ryder, Nick and Subbiah, Melanie and Kaplan, Jared D and Dhariwal, Prafulla and Neelakantan, Arvind and Shyam, Pranav and Sastry, Girish and Askell, Amanda and others},
  journal={Advances in neural information processing systems},
  volume={33},
  pages={1877--1901},
  year={2020}
}

@article{achiam2023gpt,
  title={Gpt-4 technical report},
  author={Achiam, Josh and Adler, Steven and Agarwal, Sandhini and Ahmad, Lama and Akkaya, Ilge and Aleman, Florencia Leoni and Almeida, Diogo and Altenschmidt, Janko and Altman, Sam and Anadkat, Shyamal and others},
  journal={arXiv preprint arXiv:2303.08774},
  year={2023}
}

@article{deng2023mind2web,
  title={Mind2web: Towards a generalist agent for the web},
  author={Deng, Xiang and Gu, Yu and Zheng, Boyuan and Chen, Shijie and Stevens, Sam and Wang, Boshi and Sun, Huan and Su, Yu},
  journal={Advances in Neural Information Processing Systems},
  volume={36},
  pages={28091--28114},
  year={2023}
}

@inproceedings{shi2022natural,
  title={Natural language to code translation with execution},
  author={Shi, Freda and Fried, Daniel and Ghazvininejad, Marjan and Zettlemoyer, Luke and Wang, Sida I},
  booktitle={Proceedings of the 2022 Conference on Empirical Methods in Natural Language Processing},
  pages={3533--3546},
  year={2022}
}

@article{chen2021evaluating,
  title={Evaluating large language models trained on code},
  author={Chen, Mark and Tworek, Jerry and Jun, Heewoo and Yuan, Qiming and Pinto, Henrique Ponde De Oliveira and Kaplan, Jared and Edwards, Harri and Burda, Yuri and Joseph, Nicholas and Brockman, Greg and others},
  journal={arXiv preprint arXiv:2107.03374},
  year={2021}
}

@article{github2022quantifying,
  title={quantifying GitHub copilot’s impact on developer productivity and happiness},
  author={GitHub},
  year={2022}
}

@misc{steinberger2026openclaw,
  author       = {Steinberger, Peter and OpenClaw Contributors},
  title        = {{OpenClaw}: Personal AI Assistant},
  year         = {2026},
  howpublished = {\url{https://github.com/openclaw/openclaw}},
  note         = {GitHub repository},
}

@article{lu2026towards,
  title={Towards end-to-end automation of AI research},
  author={Lu, Chris and Lu, Cong and Lange, Robert Tjarko and Yamada, Yutaro and Hu, Shengran and Foerster, Jakob and Ha, David and Clune, Jeff},
  journal={Nature},
  volume={651},
  number={8107},
  pages={914--919},
  year={2026},
  publisher={Nature Publishing Group UK London}
}

@article{shao2025omniscientist,
  title={OmniScientist: Toward a Co-evolving Ecosystem of Human and AI Scientists},
  author={Shao, Chenyang and Huang, Dehao and Li, Yu and Zhao, Keyu and Lin, Weiquan and Zhang, Yining and Zeng, Qingbin and Chen, Zhiyu and Li, Tianxing and Huang, Yifei and others},
  journal={arXiv preprint arXiv:2511.16931},
  year={2025}
}

@article{li2026autosota,
  title={AutoSOTA: An End-to-End Automated Research System for State-of-the-Art AI Model Discovery},
  author={Li, Yu and Shao, Chenyang and Liu, Xinyang and Zhao, Ruotong and Liu, Peijie and Su, Hongyuan and Chen, Zhibin and Yang, Qinglong and Xu, Anjie and Fang, Yi and others},
  journal={arXiv preprint arXiv:2604.05550},
  year={2026}
}

@article{feng2026internagent,
  title={Internagent-1.5: A unified agentic framework for long-horizon autonomous scientific discovery},
  author={Feng, Shiyang and Ma, Runmin and Yan, Xiangchao and Fan, Yue and Hu, Yusong and Huang, Songtao and Zhang, Shuaiyu and Cao, Zongsheng and Peng, Tianshuo and Yuan, Jiakang and others},
  journal={arXiv preprint arXiv:2602.08990},
  year={2026}
}

@article{jiang2016timegeo,
  author    = {Jiang, Shan and Yang, Yingxiang and Gupta, Siddharth and Venevsky, Sergey and Athavale, Prashant and Gonz{\'a}lez, Marta C.},
  title     = {TimeGeo modeling framework for urban mobility without tracking data},
  journal   = {Proceedings of the National Academy of Sciences (PNAS)},
  volume    = {113},
  number    = {37},
  pages     = {E5370--E5378},
  year      = {2016},
  publisher = {National Acad Sciences}
}

@inproceedings{feng2018predicting,
  author    = {Feng, Jie and Li, Yong and Zhang, Chao and Sun, Pengshuai and Zhou, Mengting and Meng, Jiazhen and Jin, Depeng},
  title     = {Predicting human mobility with semantic motivation via deep generative model},
  booktitle = {Proceedings of the 24th ACM SIGKDD International Conference on Knowledge Discovery \& Data Mining (KDD)},
  pages     = {1388--1397},
  year      = {2018}
}

@article{volunteer_baseline,
  author    = {Song, Chaoming and Qu, Zehui and Blumm, Nicholas and Barab{\'a}si, Albert-L{\'a}szl{\'o}},
  title     = {Limits of predictability in human mobility},
  journal   = {Science},
  volume    = {327},
  number    = {5968},
  pages     = {1018--1021},
  year      = {2010},
  publisher = {American Association for the Advancement of Science}
}

@inproceedings{zhu2023difftraj,
  author    = {Zhu, Shifu and Zhao, Hongjian and Semanjski, Ivana and Zhang, Xi and Mladenovi{\'c}, Milo{\v{s}} N. and Feng, Jiyuan and Cheng, Weiwei},
  title     = {DiffTraj: A novel approach for trajectory generation using diffusion models},
  booktitle = {Proceedings of the 29th ACM SIGKDD International Conference on Knowledge Discovery \& Data Mining (KDD)},
  pages     = {3645--3656},
  year      = {2023}
}

@inproceedings{zheng2024act2loc,
  author    = {Zheng, Zhiheng and Li, Yong and Piao, Jinghua and Zhang, Jun and Jin, Depeng},
  title     = {Act2Loc: A Generative Framework for Microscopic Human Mobility Simulation with Activity-driven Motivations},
  booktitle = {Proceedings of the 30th ACM SIGKDD International Conference on Knowledge Discovery \& Data Mining (KDD)},
  year      = {2024}
}

@inproceedings{shang2025large,
  title={A large-scale dataset with behavior, attributes, and content of mobile short-video platform},
  author={Shang, Yu and Gao, Chen and Li, Nian and Li, Yong},
  booktitle={Companion Proceedings of the ACM on Web Conference 2025},
  pages={793--796},
  year={2025}
}

@article{shao2024chain,
  title={Chain-of-planned-behaviour workflow elicits few-shot mobility generation in llms},
  author={Shao, Chenyang and Xu, Fengli and Fan, Bingbing and Ding, Jingtao and Yuan, Yuan and Wang, Meng and Li, Yong},
  journal={arXiv preprint arXiv:2402.09836},
  year={2024}
}

@misc{safegraph2019core,
  author       = {{SafeGraph}},
  title        = {{SafeGraph Core Places and Patterns Data}},
  year         = {2019},
  howpublished = {Data set},
  url          = {https://www.safegraph.com/}
}

@misc{safegraph2021core,
  author       = {{SafeGraph}},
  title        = {{SafeGraph Core Places and Patterns Data}},
  year         = {2021},
  howpublished = {Data set},
  url          = {https://www.safegraph.com/}
}

@misc{uscensus2020acs,
  author       = {{U.S. Census Bureau}},
  title        = {{American Community Survey 5-Year Estimates, 2015--2019}},
  year         = {2020},
  publisher    = {{U.S. Department of Commerce}},
  howpublished = {Data set},
  url          = {https://data.census.gov/}
}

@article{soliman2024artificial,
  title={Artificial intelligence powered Metaverse: analysis, challenges and future perspectives},
  author={Soliman, Mona M and Ahmed, Eman and Darwish, Ashraf and Hassanien, Aboul Ella},
  journal={Artificial Intelligence Review},
  volume={57},
  number={2},
  pages={36},
  year={2024},
  publisher={Springer}
}

@article{tang2025metaverse,
  title={Metaverse and digital twins in the age of AI and extended reality},
  author={Tang, Ming and Nikolaenko, Mikhail and Alrefai, Ahmad and Kumar, Aayush},
  journal={Architecture},
  volume={5},
  number={2},
  pages={36},
  year={2025},
  publisher={MDPI}
}

@article{swanson2025virtual,
  title={The Virtual Lab of AI agents designs new SARS-CoV-2 nanobodies},
  author={Swanson, Kyle and Wu, Wesley and Bulaong, Nash L and Pak, John E and Zou, James},
  journal={Nature},
  volume={646},
  number={8085},
  pages={716--723},
  year={2025},
  publisher={Nature Publishing Group UK London}
}

@article{ghareeb2026multi,
  title={A multi-agent system for automating scientific discovery},
  author={Ghareeb, Ali Essam and Chang, Benjamin and Mitchener, Ludovico and Yiu, Angela and Szostkiewicz, Caralyn J and Shved, Dmytro and Gyimesi, Gavin J and Laurent, Jon M and Wright, Samantha M and Razzak, Muhammed T and others},
  journal={Nature},
  pages={1--3},
  year={2026},
  publisher={Nature Publishing Group}
}

@article{tang2026ai,
  title={Ai-researcher: Autonomous scientific innovation},
  author={Tang, Jiabin and Xia, Lianghao and Li, Zhonghang and Huang, Chao},
  journal={Advances in Neural Information Processing Systems},
  volume={38},
  pages={9481--9520},
  year={2026}
}

@article{gao2025democratizing,
  title={Democratizing AI scientists using ToolUniverse},
  author={Gao, Shanghua and Zhu, Richard and Sui, Pengwei and Kong, Zhenglun and Aldogom, Sufian and Huang, Yepeng and Noori, Ayush and Shamji, Reza and Parvataneni, Krishna and Tsiligkaridis, Theodoros and others},
  journal={arXiv preprint arXiv:2509.23426},
  year={2025}
}

@article{weng2025deepscientist,
  title={Deepscientist: Advancing frontier-pushing scientific findings progressively},
  author={Weng, Yixuan and Zhu, Minjun and Xie, Qiujie and Sun, Qiyao and Lin, Zhen and Liu, Sifan and Zhang, Yue},
  journal={arXiv preprint arXiv:2509.26603},
  year={2025}
}

@article{alber2026cellvoyager,
  title={Cellvoyager: Ai compbio agent generates new insights by autonomously analyzing biological data},
  author={Alber, Samuel and Chen, Bowen and Sun, Eric and Isakova, Alina and Wilk, Aaron J and Zou, James},
  journal={Nature Methods},
  pages={1--11},
  year={2026},
  publisher={Nature Publishing Group US New York}
}

@article{wang2024survey,
  title={A survey on human-centric llms},
  author={Wang, Jing Yi and Sukiennik, Nicholas and Li, Tong and Su, Weikang and Hao, Qianyue and Xu, Jingbo and Huang, Zihan and Xu, Fengli and Li, Yong},
  journal={arXiv preprint arXiv:2411.14491},
  year={2024}
}

@misc{census2023acs_pums,
  author       = {{U.S. Census Bureau}},
  title        = {{2023 American Community Survey: 1-Year Estimates, Public Use Microdata Sample}},
  year         = {2023},
  howpublished = {Public Use Microdata Sample (PUMS)},
  url          = {https://catalog.data.gov/dataset/2023-american-community-survey-1-year-estimates-public-use-microdata-sample},
  note         = {Accessed: 2026-05-26}
}

@misc{bls2023ce_pumd,
  author       = {{U.S. Bureau of Labor Statistics}},
  title        = {{Consumer Expenditure Surveys Public Use Microdata, 2023}},
  year         = {2023},
  howpublished = {Consumer Expenditure Surveys Public Use Microdata},
  url          = {https://www.bls.gov/cex/pumd_data.htm},
  note         = {Accessed: 2026-05-26}
}

@misc{psid2023_family,
  author       = {{Panel Study of Income Dynamics}},
  title        = {{Panel Study of Income Dynamics, 2023 Family Survey: Public-Use Dataset}},
  year         = {2023},
  howpublished = {Survey Research Center, Institute for Social Research, University of Michigan},
  url          = {https://psidonline.isr.umich.edu/},
  note         = {Accessed: 2026-05-26}
}

@misc{fed2023shed,
  author       = {{Board of Governors of the Federal Reserve System}},
  title        = {{2023 Survey of Household Economics and Decisionmaking: Survey Data and Codebook}},
  year         = {2023},
  howpublished = {Public-use survey data},
  url          = {https://www.federalreserve.gov/consumerscommunities/shed_data.htm},
  note         = {Accessed: 2026-05-26}
}

@misc{census2023sipp,
  author       = {{U.S. Census Bureau}},
  title        = {{2023 Survey of Income and Program Participation}},
  year         = {2023},
  howpublished = {Public-use microdata and API dataset},
  url          = {https://catalog.data.gov/dataset/2023-survey-of-income-and-program-participation-sipp},
  note         = {Accessed: 2026-05-26}
}

@misc{census2019acs5year,
  author       = {{U.S. Census Bureau}},
  title        = {{American Community Survey 5-Year Data, 2019}},
  year         = {2019},
  howpublished = {ACS 5-year estimates, including block-group-level detailed tables},
  url          = {https://www.census.gov/data/developers/data-sets/acs-5year.html},
  note         = {Used as source data for census block-group features. Accessed: 2026-05-26}
}

@misc{factsetOwnership,
  author       = {{FactSet}},
  title        = {{FactSet Ownership}},
  year         = {2026},
  howpublished = {Institutional ownership data product},
  url          = {https://www.factset.com/marketplace/catalog/product/factset-ownership},
  note         = {Accessed: 2026-05-26}
}

@misc{finnhubCompanyNews,
  author       = {{Finnhub}},
  title        = {{Company News API}},
  year         = {2026},
  howpublished = {Finnhub API Documentation},
  url          = {https://finnhub.io/docs/api/company-news},
  note         = {Accessed: 2026-05-26}
}

@misc{finnhubStockCandles,
  author       = {{Finnhub}},
  title        = {{Stock Candles API}},
  year         = {2026},
  howpublished = {Finnhub API Documentation},
  url          = {https://finnhub.io/docs/api/stock-candles},
  note         = {Accessed: 2026-05-26}
}

@article{liu2024lost,
  title={Lost in the middle: How language models use long contexts},
  author={Liu, Nelson F and Lin, Kevin and Hewitt, John and Paranjape, Ashwin and Bevilacqua, Michele and Petroni, Fabio and Liang, Percy},
  journal={Transactions of the association for computational linguistics},
  volume={12},
  pages={157--173},
  year={2024}
}

@inproceedings{liu2024longgenbench,
  title={Longgenbench: Long-context generation benchmark},
  author={Liu, Xiang and Dong, Peijie and Hu, Xuming and Chu, Xiaowen},
  booktitle={Findings of the Association for Computational Linguistics: EMNLP 2024},
  pages={865--883},
  year={2024}
}

@inproceedings{jin2025long,
  title={Long-context llms meet rag: Overcoming challenges for long inputs in rag},
  author={Jin, Bowen and Yoon, Jinsung and Han, Jiawei and Arik, Sercan},
  booktitle={International Conference on Learning Representations},
  volume={2025},
  pages={37784--37822},
  year={2025}
}

@article{mou2026individual,
  title={From individual to society: A survey on social simulation driven by large language model-based agents},
  author={Mou, Xinyi and Ding, Xuanwen and He, Qi and Wang, Liang and Liang, Jingcong and Zhang, Xinnong and Sun, Libo and Lin, Jiayu and Zhou, Jie and Xuanjing, Huang and others},
  journal={ACM Computing Surveys},
  volume={58},
  number={11},
  pages={1--41},
  year={2026},
  publisher={ACM New York, NY}
}

@article{gao2024large,
  title={Large language models empowered agent-based modeling and simulation: A survey and perspectives},
  author={Gao, Chen and Lan, Xiaochong and Li, Nian and Yuan, Yuan and Ding, Jingtao and Zhou, Zhilun and Xu, Fengli and Li, Yong},
  journal={Humanities and Social Sciences Communications},
  volume={11},
  number={1},
  pages={1--24},
  year={2024},
  publisher={Palgrave}
}

@misc{anthropic2025agentskills,
  title={Agent Skills},
  author={{Anthropic}},
  year={2025},
  howpublished={\url{https://platform.claude.com/docs/en/agents-and-tools/agent-skills/overview}}
}

@misc{openai2026codexskills,
  title={Agent Skills},
  author={{OpenAI}},
  year={2026},
  howpublished={\url{https://developers.openai.com/codex/skills}}
}

@inproceedings{tang2025gensim,
  title={Gensim: A general social simulation platform with large language model based agents},
  author={Tang, Jiakai and Gao, Heyang and Pan, Xuchen and Wang, Lei and Tan, Haoran and Gao, Dawei and Chen, Yushuo and Chen, Xu and Lin, Yankai and Li, Yaliang and others},
  booktitle={Proceedings of the 2025 Conference of the Nations of the Americas Chapter of the Association for Computational Linguistics: Human Language Technologies (System Demonstrations)},
  pages={143--150},
  year={2025}
}

@article{ashery2025emergent,
  title={Emergent social conventions and collective bias in LLM populations},
  author={Ashery, Ariel Flint and Aiello, Luca Maria and Baronchelli, Andrea},
  journal={Science Advances},
  volume={11},
  number={20},
  pages={eadu9368},
  year={2025},
  publisher={American Association for the Advancement of Science}
}

@article{binz2025foundation,
  title={A foundation model to predict and capture human cognition},
  author={Binz, Marcel and Akata, Elif and Bethge, Matthias and Br{\"a}ndle, Franziska and Callaway, Fred and Coda-Forno, Julian and Dayan, Peter and Demircan, Can and Eckstein, Maria K and {\'E}ltet{\H{o}}, No{\'e}mi and others},
  journal={Nature},
  volume={644},
  number={8078},
  pages={1002--1009},
  year={2025},
  publisher={Nature Publishing Group UK London}
}

@article{demszky2023using,
  title={Using large language models in psychology},
  author={Demszky, Dorottya and Yang, Diyi and Yeager, David S and Bryan, Christopher J and Clapper, Margarett and Chandhok, Susannah and Eichstaedt, Johannes C and Hecht, Cameron and Jamieson, Jeremy and Johnson, Meghann and others},
  journal={Nature Reviews Psychology},
  volume={2},
  number={11},
  pages={688--701},
  year={2023},
  publisher={Nature Publishing Group US New York}
}

@inproceedings{echterhoff2024cognitive,
  title={Cognitive bias in decision-making with LLMs},
  author={Echterhoff, Jessica Maria and Liu, Yao and Alessa, Abeer and McAuley, Julian and He, Zexue},
  booktitle={Findings of the association for computational linguistics: EMNLP 2024},
  pages={12640--12653},
  year={2024}
}

@misc{anthropic_skill_creator,
  title        = {Skill Creator},
  author       = {{Anthropic}},
  year         = {2026},
  howpublished = {\url{https://claude.com/plugins/skill-creator}},
  note         = {Accessed: 2026-05-28}
}

@article{aygun2026ai,
  title={An AI system to help scientists write expert-level empirical software},
  author={Ayg{\"u}n, Eser and Belyaeva, Anastasiya and Comanici, Gheorghe and Coram, Marc and Cui, Hao and Garrison, Jake and Johnston, Renee and Kast, Anton and McLean, Cory Y and Norgaard, Peter and others},
  journal={Nature},
  pages={1--3},
  year={2026},
  publisher={Nature Publishing Group UK London}
}

@article{gottweis2026accelerating,
  title={Accelerating scientific discovery with Co-Scientist},
  author={Gottweis, Juraj and Weng, Wei-Hung and Daryin, Alexander and Tu, Tao and Sirkovic, Petar and Myaskovsky, Artiom and Glowaty, Grzegorz and Weissenberger, Felix and Orlandi, Alessio and Popovici, Dan and others},
  journal={Nature},
  pages={1--3},
  year={2026},
  publisher={Nature Publishing Group UK London}
}

@article{schmidgall2025agent,
  title={Agent laboratory: Using llm agents as research assistants},
  author={Schmidgall, Samuel and Su, Yusheng and Wang, Ze and Sun, Ximeng and Wu, Jialian and Yu, Xiaodong and Liu, Jiang and Moor, Michael and Liu, Zicheng and Barsoum, Emad},
  journal={Findings of the Association for Computational Linguistics: EMNLP 2025},
  pages={5977--6043},
  year={2025},
  publisher={Association for Computational Linguistics}
}

@article{cheng2025openlens,
  title={Openlens ai: Fully autonomous research agent for health infomatics},
  author={Cheng, Yuxiao and Suo, Jinli},
  journal={arXiv preprint arXiv:2509.14778},
  year={2025}
}
\clearpage
\appendix

\section{Appendix}
\label{s:appendix}

\subsection{Rating Guidance for Agent-Based Social Simulation Platform Comparison}
\label{app:rating_rubric_social_simulation}

\paragraph{Rating protocol.}
We rate representative agent-based social simulation platforms using a five-star maturity scale.
The ratings reflect platform-level capabilities that are explicitly described or demonstrated in the corresponding papers, documents, or released systems, rather than capabilities that could be implemented with substantial additional engineering.
Higher scores indicate stronger support for generality, reusability, autonomy, empirical data use, and human-AI steering across the social-science research process.
We rate each platform along four dimensions: agent capability, environment capability, data use, and human-AI steering.
Table~\ref{tab:social_sim_rating_rubric} defines the detailed scoring criteria for these four dimensions.

\paragraph{Overall scale.}
A score of $\star$ indicates minimal support that is mostly fixed, manual, or case-specific.
A score of $\star$$\star$ indicates basic support through simple configuration, state, logging, or scenario-level use.
A score of $\star$$\star$$\star$ indicates structured support through modular components, structured inputs and outputs, tool interfaces, standard logs, metrics, or reusable runtime support.
A score of $\star$$\star$$\star$$\star$ indicates advanced support through generalizable modules, platform-level data interfaces, interactive control, or automated analysis support.
A score of $\star$$\star$$\star$$\star$$\star$ indicates agentic research support across heterogeneous social-science studies, including assistance for design, execution, evaluation, interpretation, or reporting.

\begin{table*}[t]
\centering
\scriptsize
\caption{Rating rubric for social simulation platform comparison.}
\label{tab:social_sim_rating_rubric}
\setlength{\tabcolsep}{3pt}
\renewcommand{\arraystretch}{1.00}
\begin{tabularx}{\textwidth}{
p{2.0cm}
>{\raggedright\arraybackslash}p{2.75cm}
>{\raggedright\arraybackslash}p{2.75cm}
>{\raggedright\arraybackslash}p{2.75cm}
>{\raggedright\arraybackslash}p{2.75cm}
>{\raggedright\arraybackslash}X
}
\toprule
\textbf{Dimension}
& \textbf{\stars{1}}
& \textbf{\stars{2}}
& \textbf{\stars{3}}
& \textbf{\stars{4}}
& \textbf{\stars{5}} \\
\midrule

\textbf{Agent}
& Prompt-only agents based on static roles, personas, or manually specified rules.
& Agents with profiles, memory, planning, or fixed workflows, but mainly tied to one scenario.
& ReAct or tool-use agents with local autonomy for observation, reasoning, tool use, state update, and action.
& Workflow-capable agents that complete multi-step tasks through reusable workflows, modules, or role structures.
& Skill-based social agents with reusable skills, tool interfaces, persistent states, and cross-study adaptability. \\
\midrule

\textbf{Environment}
& A fixed sandbox, task setting, or hard-coded scenario.
& A specialized environment with configurable parameters, rules, or populations.
& A programmable environment that users can define or modify through code, modules, APIs, or structured interfaces.
& Modular multi-domain environments with standard interfaces and tool access.
& Agentically constructible environments that can be generated, adapted, validated, and executed from study protocols. \\
\midrule

\textbf{Data}
& Synthetic data, simulation-only data, or prompt-described context.
& External data injected into prompts, profiles, initial states, or scenario descriptions.
& Structured records, logs, configuration files, or datasets loaded for simulation and evaluation.
& Platform-level data assets, dataset interfaces, benchmarks, registries, or evaluation datasets.
& Agentic data use for data retrieval, preparation, simulation grounding, analysis, and validation across the research workflow. \\
\midrule

\textbf{Human-AI Steering}
& Manual setup through prompts, scripts, parameters, or configuration files, followed by external inspection.
& Config-level steering over agents, scenarios, rules, or parameters, with fixed output inspection.
& Interactive steering through log inspection, data interfaces, setting modification, reruns, metrics, or execution control.
& Artifact-level steering over hypotheses, protocols, configs, code, logs, results, figures, and analysis outputs.
& Co-scientist steering for hypothesis refinement, experiment design, execution monitoring, intervention, debugging, evaluation, interpretation, and reporting. \\

\bottomrule
\end{tabularx}
\end{table*}

\paragraph{Dimension definitions.}
\textbf{Agent} measures whether agents move from static prompt-based personas to reusable, tool-using, and skill-based social agents.
\textbf{Environment} measures whether environments move from fixed scenarios to configurable, programmable, modular, and agentically constructible experimental spaces.
\textbf{Data} measures whether empirical data are absent, manually injected, structurally loaded, integrated as platform assets, or autonomously used by agentic workflows.
\textbf{Human-AI Steering} measures how deeply human researchers can steer, inspect, revise, evaluate, interpret, and report agent-based social simulation studies with AI assistance.
Table~\ref{tab:social_sim_rating_rationale} provides the per-platform evidence used to assign the ratings in Table~\ref{tab:social_sim_platform_comparison} of the main manuscript.

\begin{table*}[t]
\centering
\caption{
Rationale for ratings in the agent-based social simulation platform comparison.
}
\label{tab:social_sim_rating_rationale}
\scriptsize
\setlength{\tabcolsep}{3.0pt}
\renewcommand{\arraystretch}{0.96}
\resizebox{\textwidth}{!}{%
\begin{tabularx}{\textwidth}{
p{2.45cm}
X
X
X
X
}
\toprule
\textbf{Platform}
& \textbf{Agent}
& \textbf{Environment}
& \textbf{Data}
& \textbf{Human-AI Steering} \\
\midrule

Generative Agents~\citep{park2023generative}
& Supports memory, reflection, and planning personas, but agent behavior is mainly tied to a single sandbox-town scenario.
& Uses a fixed town sandbox rather than a configurable or reusable environment construction platform.
& Data mainly consist of manually specified profiles, memories, and simulation-produced traces.
& Researchers mainly set up the scenario, run the simulation, and inspect traces after execution. \\
\midrule

AgentScope~\citep{gao2024agentscope}
& Supports structured autonomous agents with tools, memory, planning, and workflow-based multi-agent coordination.
& Provides programmable runtime interfaces for building agent applications and task environments.
& Supports structured logs, configurations, evaluation hooks, tool inputs, and user-defined data through code.
& Users steer runs through code, APIs, configuration, execution control, and inspection tools. \\
\midrule

AgentVerse~\citep{chen2023agentverse}
& Supports role-based and task-oriented agents with reusable interaction structures.
& Provides programmable task and simulation environments through rules, configurations, and custom environments.
& Uses structured traces, task inputs, benchmark outputs, and user-defined resources.
& Human steering mainly occurs through task setup, configuration edits, reruns, and trace inspection. \\
\midrule

OASIS~\citep{yang2024oasis}
& Supports structured social-media agents with platform-specific actions and interaction behavior.
& Provides a programmable online social-network simulator, but the environment remains domain-specific rather than broadly multi-domain.
& Uses platform-level posts, networks, interaction logs, recommendation signals, and evaluation targets for social-media phenomena.
& Supports configurable interventions, reruns, metrics, scenario control, and downstream analysis, while the broader research workflow remains external. \\
\midrule

SocioVerse~\citep{zhang2025socioverse}
& Supports population-level agents grounded in user characteristics and alignment modules, but not a broad reusable skill or tool agent architecture.
& Provides specialized world-model and scenario support, but environment construction is less openly programmable under the released evidence.
& Strong data support comes from real-user grounding, user pools, alignment resources, and evaluation assets.
& Steering mainly depends on scenario, population, question setup, and result inspection rather than artifact-level research control. \\
\midrule

YuLan-OneSim~\citep{wang2025yulan}
& Supports workflow-capable agents with memory, planning, role structures, and reusable components across scenarios.
& Provides modular multi-domain scenario construction and broad environment support.
& Uses structured configurations, logs, and evaluation outputs, but is less centered on a full platform-level data asset lifecycle.
& Provides strong user steering for scenario construction, execution, analysis, and reporting artifacts. \\
\midrule

GenSim~\citep{tang2025gensim}
& Supports LLM-based agents for generated simulations, but reusable tool-using autonomy is limited.
& Provides a general programming framework for customized and large-scale social simulations.
& Data grounding mainly comes from external initialization and simulation outputs.
& Human steering mainly occurs through scenario setup, scripts, correction mechanisms, and output inspection. \\
\midrule

Agent-Kernel~\citep{mao2025agent}
& Provides reusable and adaptive agent infrastructure through a microkernel-style architecture.
& Supports modular and extensible environment, action, tool, and system components.
& Supports structured runtime records, interfaces, and experiment artifacts, but not a full platform-level data asset system.
& Supports interactive control and structured intervention, but does not provide a full research-facing IRE. \\
\midrule

CitySim~\citep{bougie2025citysim}
& Supports structured urban agents with planning, memory, goals, and adaptive activity behavior.
& Provides a programmable urban-scale simulation environment, but the domain remains relatively specific.
& Uses structured urban, mobility, activity, and simulation records for city-dynamics modeling.
& Researcher steering mainly occurs through scenario or policy configuration and downstream analysis. \\
\midrule

AgentSociety~\citep{piao2025agentsociety}
& Supports reusable profile-based social agents with cognition, memory, social relations, and activity modeling.
& Connects agents with modular social, urban, and economic environments for large-scale behavioral simulation.
& Uses structured profiles, logs, trajectories, and benchmarks, but does not yet provide a full agentic data workflow.
& Supports experiment configuration, reruns, logs, and evaluation, while the broader research cycle remains outside the simulator. \\
\midrule

\textbf{AgentSociety$^2$}
& Supports skill-based social generative agents, ReAct-style execution, persistent workspaces, reusable skills, and AI social co-scientists across heterogeneous studies.
& Provides agentic environment construction through unified interfaces, module integration, CodeGenRouter, and agent-assisted environment generation across different social-science scenarios.
& Provides dataset assets, online registry, metadata-aware discovery, documentation-guided extraction, dataset skills, replayable artifacts, benchmark support, and validation tools.
& Provides an IRE where researchers steer hypothesis refinement, experiment design, intervention, execution monitoring, analysis, validation, interpretation, manuscript drafting, and revision. \\

\bottomrule
\end{tabularx}
}
\end{table*}

\subsection{Rating Guidance for AI Scientist System Comparison}
\label{app:rating_rubric_ai_scientist}

\paragraph{Rating protocol.}
We rate representative AI Scientist systems using a three-star capability scale with an unsupported marker.
The ratings reflect capabilities explicitly described or demonstrated in the corresponding papers, documents, or released systems, rather than capabilities that would require substantial additional engineering.
Because the comparison is code-grounded, we prioritize released code paths and lightweight local run checks over paper-level claims.
We rate each system along six research-stage dimensions: literature search, hypothesis generation, experiment design, experiment execution, result analysis, and manuscript writing.
Human-AI steering is reflected within each stage, ranging from prompt-level assistance to modular human-in-the-loop workflows and integrated autonomous support.
Table~\ref{tab:ai_scientist_rating_rubric} defines the scoring criteria used for these six research-stage dimensions.

\paragraph{Overall scale.}
A score of -- indicates that the system does not explicitly support the stage.
A score of $\star$ indicates assistive support, where the system provides prompt-level help, isolated outputs, suggestions, summaries, or human-led support for the stage.
A score of $\star\star$ indicates modular support, where the system includes a dedicated module, tool, agent, or constrained workflow for the stage.
A score of $\star\star\star$ indicates integrated autonomous support, where the stage is natively connected to other stages in an automated or closed-loop research workflow.

\begin{table*}[t]
\centering
\scriptsize
\caption{Rating rubric for AI Scientist system comparison.}
\label{tab:ai_scientist_rating_rubric}
\setlength{\tabcolsep}{3pt}
\renewcommand{\arraystretch}{1.12}
\begin{tabularx}{\textwidth}{
p{2.35cm}
>{\raggedright\arraybackslash}p{3.1cm}
>{\raggedright\arraybackslash}p{3.3cm}
>{\raggedright\arraybackslash}p{3.5cm}
>{\raggedright\arraybackslash}X
}
\toprule
\textbf{Dimension}
& \textbf{--}
& \textbf{$\star$}
& \textbf{$\star\star$}
& \textbf{$\star\star\star$} \\
\midrule

\textbf{Literature Search}
& No explicit literature-search support.
& Uses user-provided papers, citations, or prompt context.
& Retrieves, summarizes, or ranks papers through a module, tool, or agent.
& Searches, filters, and synthesizes literature as part of the research workflow. \\

\midrule

\textbf{Hypothesis Generation}
& No hypothesis-generation support.
& Produces simple ideas, questions, or suggestions from prompts.
& Generates multiple hypotheses with ranking, critique, novelty checking, or domain constraints.
& Iteratively generates, evaluates, and refines hypotheses within an automated research loop. \\

\midrule

\textbf{Experiment Design}
& No experiment-design support.
& Gives high-level suggestions or informal plans.
& Produces structured plans, protocols, code plans, task configurations, or evaluation settings.
& Produces executable designs with variables, controls, metrics, and validation logic connected to execution. \\

\midrule

\textbf{Experiment Execution}
& No execution support.
& Provides instructions, code snippets, or external execution guidance.
& Executes code, tools, simulations, or lab-adjacent steps in constrained settings.
& Runs experiments through an integrated loop with monitoring, correction, reruns, or search. \\

\midrule

\textbf{Result Analysis}
& No analysis support.
& Gives qualitative interpretation, summaries, or simple score reading.
& Computes metrics, generates plots, tracks outcomes, or performs structured data analysis.
& Performs autonomous analysis and feeds findings back into later research stages. \\

\midrule

\textbf{Manuscript Writing}
& No writing support.
& Drafts short summaries, notes, or isolated sections.
& Produces structured reports, partial manuscripts, figures, or citation-aware drafts.
& Generates full research outputs, including methods, results, figures, citations, and review-style revision. \\

\bottomrule
\end{tabularx}
\end{table*}

\paragraph{Dimension definitions.}
\textbf{Literature Search} measures whether the system can retrieve and synthesize scientific evidence.
\textbf{Hypothesis Generation} measures whether the system can generate, rank, critique, and refine research hypotheses.
\textbf{Experiment Design} measures whether the system can translate hypotheses into structured or executable studies.
\textbf{Experiment Execution} measures whether the system can run code, tools, simulations, or lab-adjacent procedures.
\textbf{Result Analysis} measures whether the system can compute metrics, interpret results, and update the research process.
\textbf{Manuscript Writing} measures whether the system can produce reports, manuscripts, figures, citations, or review outputs.
Table~\ref{tab:ai_scientist_rating_rationale} provides the per-system evidence used to assign the ratings in Table~\ref{tab:ai_scientist} of the main manuscript.

\begin{table*}[t]
\centering
\caption{
Rationale for ratings in the AI Scientist system comparison.
}
\label{tab:ai_scientist_rating_rationale}
\tiny
\setlength{\tabcolsep}{2.0pt}
\renewcommand{\arraystretch}{1.12}
\resizebox{0.96\textwidth}{!}{%
\begin{tabularx}{\textwidth}{
p{1.5cm}
X X X X X X
}
\toprule
\textbf{System}
& \textbf{Literature Search}
& \textbf{Hypothesis Generation}
& \textbf{Experiment Design}
& \textbf{Experiment Execution}
& \textbf{Result Analysis}
& \textbf{Manuscript Writing} \\
\midrule

AI Scientist-v2~\citep{lu2026towards}
& Uses dedicated literature grounding and novelty checking, but literature search remains a supporting component of the experiment loop.
& Treats iterative ideation as a first-class stage in the research workflow.
& Translates research ideas into executable experiment plans.
& Runs experiments through an integrated tree-search loop with debugging and retries.
& Supports plotting, aggregation, and interpretation of experiment results.
& Includes full writeup and review stages. \\
\midrule

AutoSOTA~\citep{li2026autosota}
& Uses paper understanding and research-code grounding during optimization setup.
& Generates optimization strategies, but does not center broad scientific hypothesis refinement.
& Produces structured optimization plans and candidate model modifications.
& Runs model-improvement experiments, but the loop is narrower than a full autonomous research workflow.
& Tracks scores, curves, and improvement summaries.
& Provides reporting support rather than full manuscript writing. \\
\midrule

AI-Researcher~\citep{tang2026ai}
& Makes literature review a central stage of the research pipeline.
& Supports idea generation and refinement as native functions.
& Produces structured implementation and experiment plans, but design remains modular rather than fully closed-loop.
& Supports implementation and evaluation, but execution is constrained by runtime and environment requirements.
& Collects and reports results within the pipeline, but analysis remains modular.
& Provides a dedicated manuscript-writing pipeline. \\
\midrule

Agent Laboratory~\citep{schmidgall2025agent}
& Provides literature review as a named workflow stage.
& Supports idea and plan formation with human feedback.
& Produces structured experiment plans.
& Supports experiment execution through workflow modules.
& Provides analysis support over experiment outputs.
& Treats report writing as a major native output. \\
\midrule

ResearchAgent~\citep{baek2025researchagent}
& Centers retrieval-grounded scientific literature use.
& Uses retrieved papers for iterative problem and method generation.
& Provides proposal-level experiment design and validation modules.
& --
& Provides limited evaluation or feedback on proposed content.
& Produces structured outputs, but not full manuscript-native writing. \\
\midrule

OmniScientist~\citep{shao2025omniscientist}
& Provides assistive use of scientific knowledge resources in the released evidence.
& Describes idea-generation functions through ecosystem modules.
& Describes experiment-design capabilities through linked modules.
& --
& --
& -- \\
\midrule

DeepScientist~\citep{weng2025deepscientist}
& Provides limited literature support in the released workflow.
& Supports research work broadly, but hypothesis generation is not a strong native stage.
& Provides structured quest or workflow setup.
& Supports persistent execution through long-lived runners and workspace state.
& Uses statefulness and tracking to support analysis.
& Provides limited direct writing support. \\
\midrule

ERA~\citep{aygun2026ai}
& --
& Can frame candidate solutions, but not strong scientific hypotheses.
& Structures search and evaluation tasks.
& Makes code generation, execution, and metric-driven search the core workflow.
& Uses score history and feedback for candidate selection.
& -- \\
\midrule

InternAgent-1.5~\citep{feng2026internagent}
& Supports deep-research and literature-grounded QA within discovery workflows.
& Integrates idea generation and task decomposition.
& Supports planning across multiple scientific task types.
& Provides discovery-loop execution and scientific task execution.
& Integrates evaluation, memory, reproduction tasks, and iteration over outcomes.
& Supports reporting and downstream outputs, but is less manuscript-native than writing-centered systems. \\
\midrule

Coscientist~\citep{boiko2023autonomous}
& Uses literature mainly as assistive context.
& Supports chemistry reasoning and proposal formation.
& Provides strong protocol-level experiment planning.
& Provides limited execution support under the code-grounded rubric.
& Provides limited downstream analysis support.
& -- \\
\midrule

ChemCrow~\citep{bran2023chemcrow}
& Provides assistive chemistry knowledge access rather than strong literature synthesis.
& Provides limited scientific hypothesis iteration.
& Uses chemistry tools to support structured task and protocol planning.
& Executes constrained chemistry-related tool calls.
& Provides limited post-execution analysis.
& -- \\
\midrule

CellVoyager~\citep{alber2026cellvoyager}
& Uses biological background and research resources to support analysis.
& Makes biological hypothesis generation a central function.
& Plans analyses over biological datasets in a structured way.
& Executes constrained notebook-based analysis workflows.
& Makes biological data analysis the main native strength.
& Provides summary-level writing support rather than full manuscripts. \\
\midrule

Virtual Lab~\citep{swanson2025virtual}
& Uses literature mainly as assistive context.
& Develops biomedical ideas through agent discussion.
& Provides strong multi-agent biomedical design.
& Relies on external validation or scientific toolchains for execution.
& Supports comparison and interpretation of candidate outputs.
& -- \\
\midrule

Robin~\citep{ghareeb2026multi}
& Integrates literature review as a core discovery function.
& Generates and compares disease or therapy hypotheses in structured workflows.
& Produces assay and therapeutic-candidate designs.
& Supports execution through external experimental infrastructure rather than a lightweight integrated loop.
& Provides strong ranking, comparison, and structured synthesis.
& Provides detailed outputs, but not full manuscript-native writing. \\
\midrule

Co-Scientist~\citep{gottweis2026accelerating}
& Uses literature mainly as assistive context in the released workflow.
& Centers multi-agent hypothesis generation, review, ranking, and refinement.
& Refines hypotheses toward actionable research directions.
& --
& Uses peer review, ranking, and meta-review for structured analysis.
& Synthesizes outputs, but does not provide a full paper-writing pipeline. \\
\midrule

ToolUniverse~\citep{gao2025democratizing}
& Provides biomedical knowledge retrieval through tool access.
& Serves as infrastructure, so hypothesis generation depends on external agent design.
& Supports planning through tool composition.
& Runs biomedical tools in constrained workflows.
& Supports tool-backed analysis, but orchestration is external.
& Provides limited direct writing support. \\
\midrule

OpenLens~\citep{cheng2025openlens}
& Strongly supports literature and evidence gathering for health-informatics research.
& Focuses more on planning, execution, and analysis than hypothesis-native automation.
& Makes experiment-plan generation a major component.
& Supports real workflow execution, but setup remains dependency and environment heavy.
& Emphasizes result generation, validation, visualization feedback, and completion audits.
& Provides structured reporting and manuscript-support functions, but not fully autonomous manuscript generation. \\
\midrule

\textbf{AgentSociety$^2$}
& Integrates literature-grounded research skills into the computational social-science workflow.
& Supports explicit social-science hypothesis generation and refinement.
& Treats social experiments, interventions, agent design, environment design, and measurement design as first-class stages.
& Integrates simulation-based execution into the platform workflow.
& Provides replay, measurement, validation, and simulation-output analysis.
& Supports evidence-linked manuscript drafting, revision, and preparation within the same research workflow. \\

\bottomrule
\end{tabularx}
}
\end{table*}

\end{document}